\documentclass[twocolumn,amsmath,trackchanges]{aastex702}
\usepackage{makecell}
\usepackage{longtable}
\usepackage{graphicx}
\usepackage{booktabs}
\usepackage{enumitem}
\usepackage{placeins}
\usepackage{float}
\usepackage{xcolor}
\usepackage{caption}

\begin{document}

\title{Confirming HSC strong lens candidates with DESI Spectroscopy. II. Analysis of 80 galaxy-scale systems with lens redshifts above 0.8}

\author[orcid=0009-0007-5455-020X]{Shen Li}
\affiliation{Purple Mountain Observatory, Chinese Academy of Sciences, Nanjing, Jiangsu, 210023, China}
\affiliation{School of Astronomy and Space Sciences, University of Science and Technology of China, Hefei 230026, China}
\email{shenli@pmo.ac.cn}  

\author[orcid=0000-0002-9063-698X]{Yiping Shu} 
\affiliation{Purple Mountain Observatory, Chinese Academy of Sciences, Nanjing, Jiangsu, 210023, China}
\email[show]{yiping.shu@pmo.ac.cn}

\author[orcid=0000-0001-6495-1890]{Lin Nie} 
\affiliation{Shanghai Astronomical Observatory (SHAO), Nandan Road 80, Shanghai 200030, China}
\email{}

\begin{abstract}

The total mass distribution of galaxies provides a sensitive probe of the fundamental physical processes governing galaxy evolution, including gas inflow/star formation, AGN/SN feedback, galaxy merging, and the nature of dark matter. State-of-the-art cosmological simulations have therefore made extensive predictions for the evolution of the total-mass density slope. Interestingly, their predictions diverge significantly at high redshift, primarily due to differences in how these key physical mechanisms are implemented. However, measurements of the total-mass structure evolution have been largely confined to low and intermediate redshifts, leaving the regime where model predictions differ most significantly ($z>0.8$) largely unexplored. In the first paper of this series, we reported the discovery of 84 candidate strong-lens systems with lens redshifts above 0.8. In this work, we perform lens modeling of the 80 galaxy-scale systems to assess their lensing nature. Of these, 71 systems ($\approx 90\%$) exhibit clear lensing features that are well reproduced by our lens models and are therefore classified as secure strong lenses. Another eight systems are classified as probable (5) or possible (3) strong lenses, based on the prominence of their lensing features and the quality of the lens modeling. One system lacking reliable lensing features in current imaging data is classified as a non-lens. Our analysis nearly triples the number of known galaxy-scale strong lenses at $z_{\rm lens} > 0.8$, making statistical studies of the total mass structure of high-redshift galaxies possible for the first time. This expanded sample provides an unprecedented opportunity to test the predictions of cosmological simulations and discriminate between competing physical prescriptions.

\end{abstract}



\section{Introduction}\label{section1}

Under the $\Lambda$CDM paradigm, massive early-type galaxies (ETGs) assemble hierarchically through mergers of lower-mass progenitors (e.g., \citealt{Toomre1977, White1978, White1991, Cole1994, Cole2000}). Their total mass distribution, comprising both dark and baryonic matter, imprints the competing among the multiple physical mechanisms that regulate mass and energy exchange throughout the galaxy ecosystem. Crucially, the redshift evolution of their total mass distribution serves as a sensitive probe to constrain key processes that drive galaxy evolution such as gas inflow, star formation, AGN/stellar feedback, merging, as well as to test galaxy evolution models and the nature of dark matter (e.g., \citealt{Ma2004, Naab2009, Duffy2010, Velliscig2014}).

Motivated by this diagnostic power, over the past decade, state-of-the-art cosmological simulations have made predictions for the total-mass distribution of ETGs, parameterized by a power-law profile $\rho(r) \propto r^{\gamma}$, and its redshift evolution (e.g., \citealt{Remus2017, Peirani2019, Wang2019}). The predictions across simulations are in close agreement toward low redshifts ($z \lesssim 0.2$). This is presumably due to the simulations being calibrated against local scaling relations \citep{Kaviraj2017, Pillepich2018}, rendering the low-redshift regime less effective for discriminating between models. In contrast, the high-redshift regime, where such calibrations are not imposed, remains sensitive to the differences in the sub-grid physics adopted in different simulations. Indeed, the spread in the predicted average $\gamma$ values across IllustrisTNG, Magneticum, and Horizon-AGN simulations is below 0.1 at $z = 0.1$ but increases to $\approx 0.3$ by $z=1$. This divergence is suggested to be mainly driven by the varying implementations of AGN feedback in those simulations (e.g., \citealt{Wang2019, Derkenne2021}).

Measurements of ETGs total mass density profiles at high redshifts would therefore provide greater leverage for distinguishing among simulations and constraining the physical prescriptions responsible for their divergent predictions. Existing strong lensing and dynamical measurements, however, remain concentrated at low-to-intermediate redshifts (e.g., \citealt{Ruff2011, Bolton2012, Etherington2023, Poci2017, Derkenne2021, Derkenne2023}). Within that redshift range, the intrinsic scatter in $\gamma$, approximately 0.15, is comparable to the predicated differences and limits discriminatory power of current samples. Beyond $z=0.8$, measurements are only available for a few systems (e.g. \citealt{Treu2004, Sahu2024, Huang2025}), insufficient for providing any statistically significant constraint. A substantially sizable sample at $z>0.8$ is therefore needed to rigorously test simulations and, in turn, constrain the poorly understood physical mechanisms driving galaxy evolution.

The unprecedented expansion of wide-field imaging surveys (e.g., \citealt{DES, HSC, DECaLS, Pan-STARRS}) coupled with large-scale spectroscopic campaigns (e.g., \citealt{SDSS, DESI}) has spurred the discovery of tens of thousands of strong lens candidates(e.g., \citealt{sugohi_I, holismokesII, Nightingale2025, Bolton2006}), of which several hundred have been securely confirmed to date. In particular, prior to \citet{ShuLi2025}, 27 galaxy-galaxy strong lenses with ETG lenses at $z_{\text{lens}} > 0.8$ had been spectroscopically confirmed (i.e., with spectroscopic redshifts for both lens and source) in the literature (e.g., \citealt{Crampton2002,  Cabanac2005, Faure2011, Dye2014, Sonnenfeld2015, Canameras2017, Jaelani2020, LiRui2025, Euclid2025B}). In the first paper of this series (\citealp{ShuLi2025}, hereafter Paper~I), we cross-matched imaging-selected strong lens candidates from the Hyper Suprime-Cam Subaru Strategic Program (e.g., \citealt{sugohi_II, sugohi_V, sugohi_VI, sugohi_VII, sugohi_VIII, sugohi_X, holismokesVI, holismokesVIII, holismokesXIII, holismokesXVI}) with the Data Release 1 of the Dark Energy Spectroscopic Instrument (DESI DR1, \citealp{DESI_DR1}), identifying 84 systems whose putative lens galaxies are at $z_{\text{lens}} > 0.8$. Although the majority of these systems exhibit compelling lensing features, as many as 74 lack spectroscopic source redshifts and is conservatively treated as unconfirmed. In this work, we perform lens modeling analyses to validate their lensing nature, assuming that bona fide strong lenses can be well described by a physically motivated model. The resulting high-fidelity catalog of high-redshift strong lenses will serve as an indispensable observational benchmark for statistical studies of ETG evolution.

This paper is structured as follows. In Section 2, we outline our sample selection criteria and imaging datasets used for lens modeling. Section 3 describes the lens modeling methodology, and the modeling results are summarized in Section 4. The kinematic analysis and stellar mass measurements of the lens galaxies are given in Section 5. Finally, we provide a discussion and conclusions in Sections 6 and 7, respectively. Throughout this work, we assume a flat $\Lambda$CDM cosmology with $\Omega_m = 0.3$, $\Omega_\Lambda = 0.7$, and $H_0 = 70 \text{ km s}^{-1} \text{ Mpc}^{-1}$.



\section{Sample and imaging}\label{sec:data}
\subsection{Sample selection}

In Paper~I, we identified 84 strong lens candidates with spectroscopically confirmed lens redshifts at $z_{\text{lens}} > 0.8$, which constitute the parent sample of this work. Among these, ten systems also have secure spectroscopic source redshifts, including eight measured from DESI spectra in Paper~I and two reported by \citet{sugohi_IV} and \citet{Jaelani2020}. Since this work focuses specifically on galaxy-scale lenses, we excluded from the parent sample four systems with putative lensing features separated from the lens center by more than 4\farcs0, namely 023331$-$032801, 084632$-$015416, 090404$+$012516, and \allowbreak223316$-$010409. The subsequent analyses were performed on the remaining 80 systems. Table~\ref{Mass_para} lists these systems and their redshift measurements.

\subsection{Imaging data}

For lens modeling, we used $g$, $r$, $i$, $z$, and $y$-band imaging data from the Wide layer (\texttt{pdr3\_wide}) of the Third Public Data Release (PDR3) of the Hyper Suprime-Cam Subaru Strategic Program (HSC-SSP; \citealt{PDR3}), along with associated pixel-level uncertainty maps. Covering approximately $1200\,\text{deg}^2$, the PDR3 Wide layer observations achieve median $5\sigma$ point-source depths of $26.5$, $26.5$, $26.2$, $25.2$, and $24.4$ mag in the $grizy$ bands, respectively, with corresponding median seeing of 0\farcs79, 0\farcs75, 0\farcs61, 0\farcs68, and 0\farcs68. For each strong lens system, we extracted 12$^{\prime\prime}$$\times$12$^{\prime \prime}$ image cutouts centered on the lens galaxy in $grizy$ bands from the HSC DAS Cutout Service\footnote{\url{https://hsc-release.mtk.nao.ac.jp/das_cutout/pdr3/}}.

We constructed the PSF models using field stars. In each band, we selected stars from the size-magnitude locus \citep{HybPSF}, retaining only unblended stars with $\mathrm{S/N} > 20$ to reconstruct the PSF models. The local PSFs were then constructed using the Expectation-Maximization-PCA (EMPCA) method (e.g., \citealt{Li2016, Nie2021}). To capture spatial variations, we fitted third-order bivariate polynomials to the principal component coefficients (e.g., \citealt{Jee2007, Nie2021b}). The final PSF models were rendered on a $41 \times 41$ pixel grid at the native HSC pixel scale.

\section{Lens modeling methodology}\label{sec:lens_modeling}

We employed a parametric forward-modeling approach to simultaneously reconstruct the lens mass distribution and the surface brightness profiles of both the foreground lens and the background source. For each system, we jointly modeled five-band ($g$, $r$, $i$, $z$, and $y$) images utilizing \texttt{lenstronomy}\footnote{\url{https://lenstronomy.readthedocs.io}}, an open-source, Astropy-affiliated gravitational lens modeling package \citep{lenstronomy}. All systems were initially modeled using a uniform baseline configuration consisting of a single mass component and one light component each for the lens and source. We also included a uniform light component in each band to fit the residual sky background using the \texttt{UNIFORM} profile. The initial batch modeling of all 80 systems with the baseline configuration yielded satisfactory models for approximately 40\% of the sample. 

When significant residuals indicated an inadequate fit to the foreground galaxy light or lensed images, we added an additional lens-light or source-light component, respectively. For a subset of the remaining systems with complex or ambiguous image configurations, we deployed a two-stage modeling strategy. We first performed rapid single-band modeling using the open-source lens modeling tool \texttt{lfit\_gui} \citep{lfitgui} with a graphical user interface (GUI). For each system, the single-band modeling was performed in the $g$, $r$, or $i$ band in which the lensed features were most clearly discernible. We then utilized the best-fit mass parameters and source position to initialize the joint five-band modeling run with \texttt{lenstronomy}, resulting in a final model constrained by all five bands.

\subsection{Lens light model}\label{subsec:lens_lignt}

To reconstruct the surface brightness distribution of lens galaxies, we adopt an elliptical S\'{e}rsic profile \citep{Sersic1963, Sersic1968}, parameterized as:
\begin{equation}
I(R) = I_e \exp\left\{ -b_n \left[ \left( \frac{R}{R_e} \right)^{1/n} - 1 \right] \right\}, \label{eq:sersic}
\end{equation}
where $R=\sqrt{q x^2 + y^2 / q}$ is the circularized elliptical radius, with $q$ denoting the projected axis ratio of the isophotes. Here, $n$ is the S\'ersic index, $ R_e$ is the circularized effective (half-light) radius, and $I_e$ is the surface brightness at $R = R_e$. The dimensionless constant $b_n$ can be approximated as $b_n \approx 1.9992n - 0.3271$ for $0.5<n<10$ \citep{Corwin1989}. For each lens galaxy, the center coordinates and ellipticity of the S\'{e}rsic profile are tied across all five bands. 
During modeling, we found that the S\'{e}rsic indices of the lens galaxies were poorly constrained and tended to approach the upper bound of the prior in many cases. We therefore fixed it to the fiducial value for ETGs, namely $n = 4$, for all lenses in all bands. As shown in Figure~\ref{fig:1}, for most systems, a single S\'ersic component with $n = 4$ was sufficient to describe the lens light, with no noticeable residuals. Nevertheless, for systems 120849$+$012845 and HSC\,J1600$+$4345, it was necessary to add a concentric $n=1$ S\'ersic component to the $n=4$ component in order to reproduce the observed lens light distribution down to the noise level. Other objects within the cutout were either modeled explicitly with S\'{e}rsic profiles, if they significantly contaminated the lens system, or masked out otherwise.

\subsection{Source light model}\label{subsec:source_lignt}

We modeled the background source light using elliptical S\'{e}rsic profiles. For each system, the center and ellipticities of the S\'{e}rsic components were fixed to the same values across all five imaging bands, while the effective radii and S\'{e}rsic indices were allowed to vary. For 74 systems, using a single S\'{e}rsic component was sufficient to reproduce the observed lensing features. For five other systems, a second S\'{e}rsic component was needed to achieve satisfactory residuals. In addition, for system 220405$+$043454, which was suggested to be a lensed QSO \citep{sugohi_IX}, we adopted a point source model (\texttt{SOURCE\_POSITION}) instead of the S\'{e}rsic model to characterize the background source. 

\subsection{Lens mass model}\label{subsec:lens_mass_model}

The primary goal of this work is to confirm the lensing nature of these candidates using a minimal parametric model. For all 80 systems, we modeled the lens galaxy's projected total mass distribution with a Singular Isothermal Ellipsoid (SIE) profile, which has been shown to provide a close approximation for ETGs (e.g., \citealt{Koopmans2006, Gavazzi2007}). The surface mass density of the SIE model is given by
\begin{equation}
\Sigma(x, y) = \Sigma_{\mathrm{crit}} \frac{\sqrt{q}}{2} \frac{b_{\mathrm{SIE}}}{\sqrt{x^2 + q^2 y^2}},
\label{eq:sie_surface_density}
\end{equation}
where $\Sigma_{\mathrm{crit}}$ is the critical surface mass density,
\begin{equation}
\Sigma_{\mathrm{crit}} = \frac{c^2}{4\pi G} \frac{D_{\mathrm{S}}}{D_{\mathrm{L}} D_{\mathrm{LS}}},
\label{eq:sigma_crit}
\end{equation}
and $D_{\mathrm{L}}$, $D_{\mathrm{S}}$, and $D_{\mathrm{LS}}$ denote the angular diameter distances from the observer to the lens, from the observer to the source, and from the lens to the source, respectively. The parameter $b_{\mathrm{SIE}}$ corresponds to the Einstein radius under the intermediate-axis convention. The coordinates $x$ and $y$ in the formula are defined in a lens-centered coordinate frame aligned with the principal axis of the mass distribution, where the $x$- and $y$-axis correspond to the major and the minor axis, respectively. The parameter $q$ denotes the axis ratio of the elliptical isodensity contours, defined as the ratio of the minor to major axis ($0 < q \leq 1$). The SIE centroid was uniformly tied to the light center of the primary (brightest) lens galaxy. The sole exception is system 155625$+$432413, where  enforcing strict concentricity significantly degraded the modeling, suggesting an intricate central mass distribution. We therefore allowed the mass centroid to vary within the region between the two dominant lens galaxies. 
Following common practice, we added an external shear term to the lensing potential to account for deviations from the SIE model as 
\begin{equation}
\psi_{\mathrm{shear}}(r, \phi) = \frac{\gamma}{2} r^2 \cos\!\left[2(\phi - \phi_\gamma)\right],
\label{eq:shear_potential}
\end{equation}
where $\gamma$ is the shear strength and $\phi_\gamma$ is its position angle measured counterclockwise from the $x$-axis. 

\subsection{Parameter inference}\label{subsec:para_inference}
We adopted a two-step procedure for parameter inference. 
A Particle Swarm Optimization (PSO) algorithm (e.g., \citealt{Kennedy1995, Eberhart1995}) was first used to efficiently navigate the high-dimensional parameter space and locate the high-likelihood region, yielding an approximate maximum likelihood estimate (MLE). This solution then initialized the walkers for Markov Chain Monte Carlo (MCMC) sampling using the affine-invariant ensemble sampler in the \texttt{emcee} package \citep{Foreman-Mackey2013}.
For all models presented here, the PSO stage was configured with 800 particles and up to 2,000 iterations with convergence-based early stopping. In the subsequent MCMC sampling, we allocated 15 walkers per non-linear free parameter, with the initial 1,000 steps discarded as burn-in and the subsequent 1,000 steps retained for posterior evaluation. The lens-mass, lens-light, and source-light parameters are given in Tables~\ref{Mass_para}--\ref{source_para}, while the best-fit modeling results for the 80 systems are shown in Figure~\ref{fig:1}.

\section{Modeling Results}\label{Results}

\begin{figure*}[t]
    \centering \includegraphics[width=0.98\textwidth]{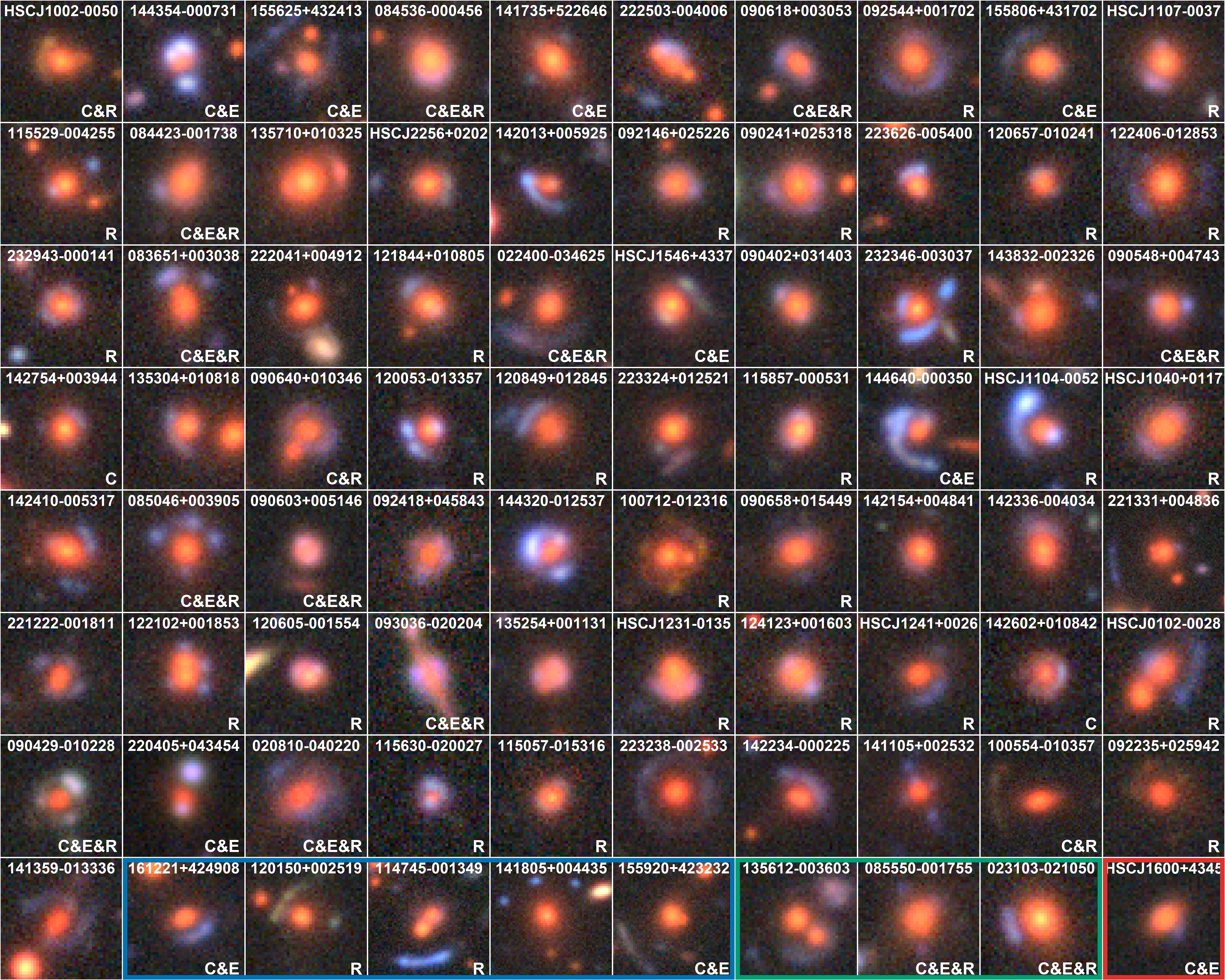}
    \caption{Montage of the 80 systems analyzed in this work, displayed as five-band ($grizy$) composite color cutouts. Blue, green and red frames enclose probable lenses, possible lenses, and non-lenses, respectively, whereas unframed panels denote secure lenses. Within each classification, systems are ordered by increasing lens redshift. White labels in the lower-right corners indicate planned coverage by CSST (C), Euclid (E) and Roman (R) as explained in Section~\ref{sec:discussions}. 
    }
    \label{fig:sample_lenses}
\end{figure*}

Based on the modeling results, we classified the 80 systems into four categories (Figure~\ref{fig:sample_lenses}):
\begin{enumerate}[topsep=1pt, partopsep=1pt, itemsep=3pt, parsep=0pt]
    \item \textbf{Secure lens (71 systems):} Systems exhibiting robust lensing features with clear counter-images that are successfully reproduced with no noticeable residuals (Section~\ref{sec:4.1}).
    \item \textbf{Probable lens (5 systems):} Systems displaying prominent lensing features that can be reasonably reproduced, but lacking identifiable counter-images in the foreground-subtracted residuals (Section~\ref{sec:4.2}).
    \item \textbf{Possible lens (3 systems):} Systems with atypical image configurations and no discernible counter-images after lens light subtraction (Section~\ref{sec:4.3}).
    \item \textbf{Non-lens (1 system):} Systems lacking reliable lensing features in the current imaging data (Section~\ref{sec:4.4}).
\end{enumerate}

\vspace{-0.25cm}
\vspace{-0.25cm}

\subsection{Secure lens}\label{sec:4.1}

Of the 71 secure lenses, 53 systems each contain only one lens galaxy and exhibit typical strong-lensing configurations, including nine with both lens and source redshifts \citep{ShuLi2025, Jaelani2020, sugohi_IV}. These systems are particularly well suited for high-resolution imaging and spectroscopic follow-up. The remaining 18 systems appear to contain multiple lens galaxies. Nevertheless, a single SIE model still effectively reproduces the lensing features, further corroborating their classifications as secure strong lenses.

System 141105$+$002532 exhibits six blue clumps forming a ``claymore''-like structure that can be successfully reproduced using two source components. The two source components are separated by 0\farcs69 (5.33 kpc at an assumed $z_{\text{source}} = 3$) in the source plane. The brighter component lies inside the caustic and forms an Einstein cross, whereas the fainter one lies outside the caustic and forms a double.

\subsection{Probable lens}\label{sec:4.2}

The five systems classified as probable lenses are 114745$-$001349, 120150$+$002519, 141805$+$004435, 155920$+$423232, and 161221$+$424908. In particular, 120150$+$002519 exhibits a prominent arc alongside a southwestern clump that was initially treated as a counter-image. However, a single-source model failed to reproduce both features. The spectral energy distributions (SEDs) of the arc and the clump also appear to be different after subtracting the lens light, suggesting they were not multiple images of the same source. We therefore proceeded to only model the arc.

\subsection{Possible lens}\label{sec:4.3}

Although the observed features are adequately reproduced by the current models, the following three systems are classified as possible lenses because they lack unambiguous lensing features and identifiable counter-images. For system 085550$-$001755, our model roughly reproduces the three clumps around the central galaxy, but the non-detection of the fourth image, predicted to be faint, is somewhat concerning. System 135612$-$003603 contains two lens galaxies of comparable brightness. While our model can account for the observed blue arcs as a fold configuration, no counter-image was detected after foreground-light subtraction. Regarding 023103$-$021050, the blue extended feature to the east of the central galaxy appears less curved than a typical lensing arc, and no identifiable counter-image is detected after the subtraction of the lens light.

\subsection{Non-lens}\label{sec:4.4}

HSC\,J1600$+$4345 is the only one among the 80 systems that we consider as a non-lens based on our lens modeling results. A purple clump to the southwest of the central galaxy exhibits no arc-like morphology, and no counter-image is discernible after foreground-light-subtraction. These signatures favor a chance alignment over strong lensing.

\section{Lens galaxy properties}\label{Properties}

To further characterize the physical properties of the lens galaxies, we measured their central line-of-sight stellar velocity dispersions from DESI spectra using full-spectrum fitting (Section~\ref{velocity_dispersion}) and estimated the stellar masses from SED fitting to the five-band ($g, r, i, z, y$) photometry (Section~\ref{Stellar_mass}). 

\subsection{Stellar velocity dispersion}\label{velocity_dispersion}

We employed the penalized pixel-fitting (PPXF) method \citep{Cappellari2004, Cappellari2017, Cappellari2023} to determine the line-of-sight stellar velocity dispersion of the lens galaxies. PPXF performs full-spectrum fitting in pixel space, constructing model spectra from a weighted linear combination of broadened stellar templates and overlaying a series of orthogonal polynomials to adjust the overall continuum, thereby measuring the stellar and gas kinematics as well as the star formation history of the galaxies. In our analysis, we fitted only the first two moments (velocity and velocity dispersion) of the line-of-sight velocity distribution (LOSVD). To mitigate template mismatch and account for imperfect sky subtraction or scattered light \citep{Cappellari2017}, we included 6th-order additive Legendre polynomials in the fits.

We adopted the Indo-US stellar spectral library \citep{Valdes2004}, which contains 1,273 spectra covering 3460--9464 \AA\ at a resolution of FWHM $\approx$ 1.35 \AA, as the empirical templates.  For all lens galaxy spectra, we masked the telluric absorption features, namely the A-band (7590--7720 \AA), B-band (6860--6950 \AA) \citep{Smette2015}, and the water vapor absorption band near 9360 \AA\ (9310--9480 \AA) \citep{Widemann1994}, whenever they fell inside the fitting window and did not overlap with the crucial $\mathrm{Ca\,\textsc{II}}\;\mathrm{H\,\&\,K}$ absorption lines. For lens galaxies with multiple DESI spectra, we prioritized the fitting results from the spectrum with the highest signal-to-noise rations (S/Ns). However, full-spectrum fitting failed for system 115630$-$020027, whose spectrum has a mean S/N of only 0.91. We therefore restricted the fit to the $\mathrm{Ca\,\textsc{II}}\;\mathrm{H\,\&\,K}$ absorption lines at observed frame wavelengths of 7800--8000 \AA, where the mean S/N is 1.44. Furthermore, for five dual-lens systems, namely 114745$-$001349, 122102$+$001853, 135612$-$003603, 142234$-$000225, and 221222$-$001811, the reported velocity dispersions are for reference only, because the DESI fibers were actually centered between the two lens galaxies rather than on the primary lens. 

The best-fit PPXF models reproduced the principal absorption features and continua well in general. We obtained velocity dispersion estimates for all 80 systems, with reduced $\chi^2$ ranging from 0.58 to 1.10. The velocity dispersions range from 91 to 515 \,$\mathrm{km\,s^{-1}}$, with a median of 232\,$\mathrm{km\,s^{-1}}$. The median fractional uncertainty across this high-redshift sample is $14.9\%$, comparable to the typical uncertainty of $14\%$ reported for BOSS galaxies, predominantly at $0.15 \lesssim z\lesssim0.7$ \citep{Thomas2013}. The individual velocity dispersion measurements and their uncertainties are listed in the final column of Table~\ref{Mass_para}. 

\subsection{Stellar mass}\label{Stellar_mass}

Stellar masses of the lens galaxies were estimated through SED fittings to the $g, r, i, z, y$ five-band photometry using the Code Investigating GALaxy Emission (\texttt{CIGALE}) \citep{boquien2019}, a Python-based SED fitting tool that has been widely used for high-redshift galaxies (e.g., \citealt{lofaro2017, Alvarez-Marquez2019, Fang2026}). In the fitting, we adopted the spectroscopic lens redshifts reported in Paper~I, assuming that any companion lens galaxies without spectroscopic measurements share the confirmed redshift within the same system. We corrected for the Milky Way extinction using the Galactic dust map from \citet{Schlegel1998}, assuming the \citet{Fitzpatrick2007} extinction law with $R_V = 3.1$. For all galaxies, we assumed a delayed star formation history (SFH; \texttt{sfhdelayed}), where the SFR is parameterized as:
\begin{equation}
    \text{SFR}(t) \propto \frac{t}{\tau^2} \exp\left(-\frac{t}{\tau}\right), \quad t \in [0, t_0]
    \label{eq:sfh_delayed}
\end{equation}
where $t_0$ denotes the galaxy age since the onset of star formation and $\tau$ is the $e$-folding timescale that corresponds to the time at which the SFR peaks. To avoid unphysical age solutions, the maximum stellar age was capped at the age of the Universe corresponding to the lowest lens redshift in this sample. Although the \texttt{sfhdelayed} module allows for an additional recent burst component, its contribution to all target galaxies was found to be negligible in preliminary fits. Therefore, it was fixed to zero in the final fitting. Stellar populations were modeled using the \citet{Bruzual2003} single stellar population library (\texttt{bc03}), assuming a Chabrier \citep{Chabrier2003} initial mass function (IMF) with a metallicity range of Z = 0.008--0.05. The age boundary between young and old stellar populations was set at 10 Myr \citep{Charlot2000}. Nebular emission was modeled using the templates of \citet{Inoue2011}, which were generated by \texttt{CLOUDY 13.01} \citep{Ferland1998, Ferland2013}. Dust attenuation was modeled using the \texttt{dustatt\_modified\_CF00} module of \texttt{CIGALE}, which implements the two-component dust attenuation model proposed by \citet{Charlot2000}. In this model, young stars are subject to attenuation from both their birth clouds (BC) and the diffuse interstellar medium (ISM), whereas older stars are only attenuated by the ISM. The attenuation curves for the BC and ISM components are described by power laws $A(\lambda) \propto \lambda^\delta$ with different slopes. The relative contribution of these components is parameterized by $\mu = A_{V,\mathrm{ISM}} / (A_{V,\mathrm{BC}} + A_{V,\mathrm{ISM}})$, which quantifies the differential dust attenuation experienced by young and old stellar populations.

The SED analysis yielded stellar mass estimates for 105 foreground galaxies in the 80 systems. The best-fitting models matched the observed five-band photometry within $3\sigma$ for 67 galaxies. The median reduced $\chi^2$ is 1.63. The estimated stellar masses spanned $9.08 \leq \log_{10}(M_\star/M_\odot) \leq 12.10$, with a median of 11.36 and a median fractional uncertainty of $20.9\%$. The inferred ages of the main stellar population have a sample median of 4.52\,Gyr. The stellar masses and uncertainties are given in Table~\ref{lens_light_para}.

\section{Discussions}\label{sec:discussions}

\begin{figure}[htbp]
    \centering    
    \includegraphics[width=0.48\textwidth]{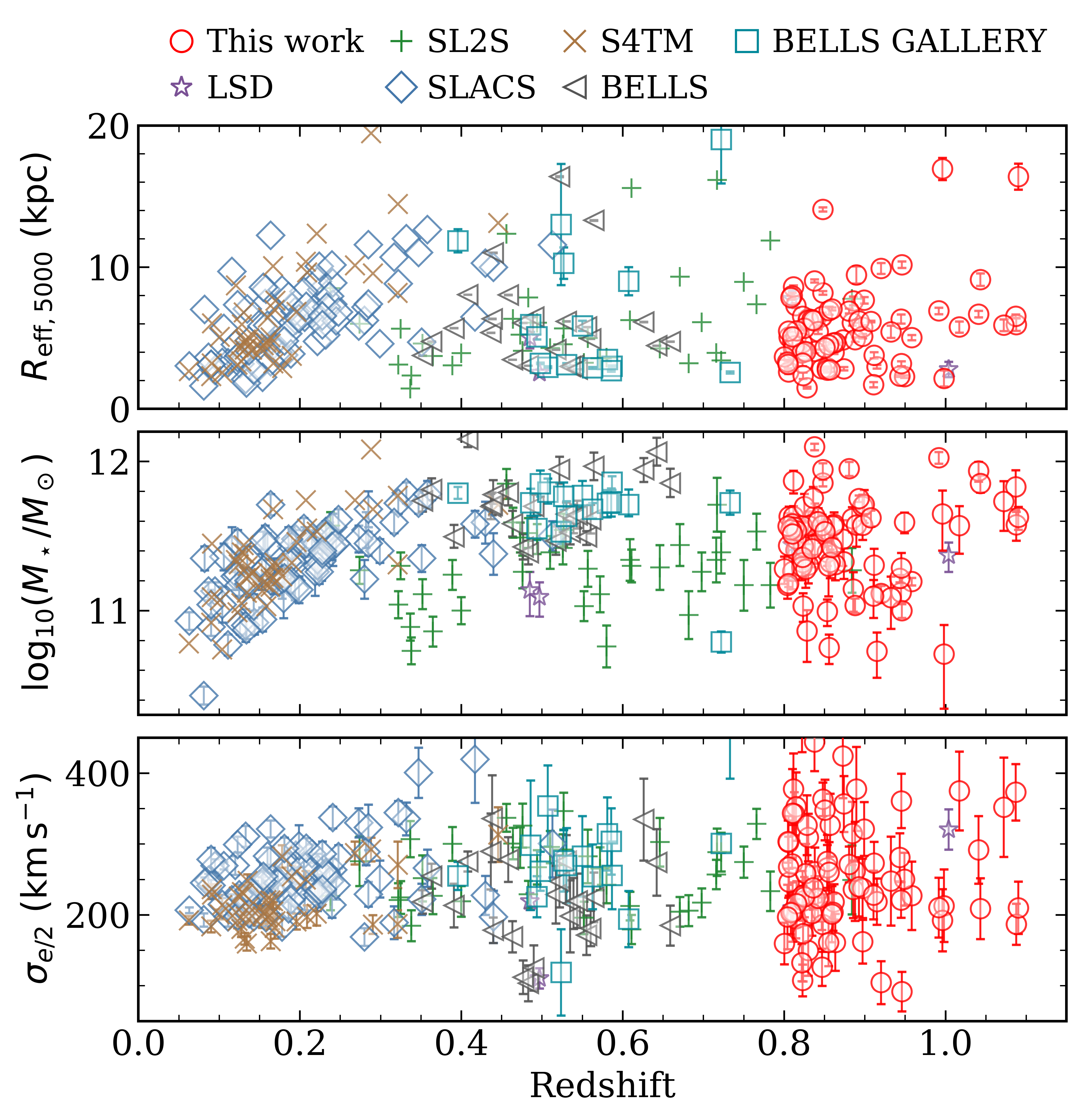}
    \caption{Rest-frame $5000\,\text{\AA}$ effective radius (top), stellar mass (middle), and stellar velocity dispersion (bottom) as a function of lens redshift. Red circles mark lens galaxies from this work with spectroscopic $z_{\mathrm{lens}}$. Literature samples include LSD \citep{Ruff2011, Koopmans2002, Koopmans2003, Treu2004}, SL2S \citep{Sonnenfeld2013, Sonnenfeld13, Sonnenfeld2015}, SLACS \citep{Auger2009}, S4TM \citep{Shu2017}, BELLS \citep{Brownstein2012}, and BELLS GALLERY \citep{lfitgui}. }
    \label{fig:stellar_population_z}
\end{figure}

\begin{figure*}[htbp]
    \centering
    \includegraphics[width=0.75\textwidth]{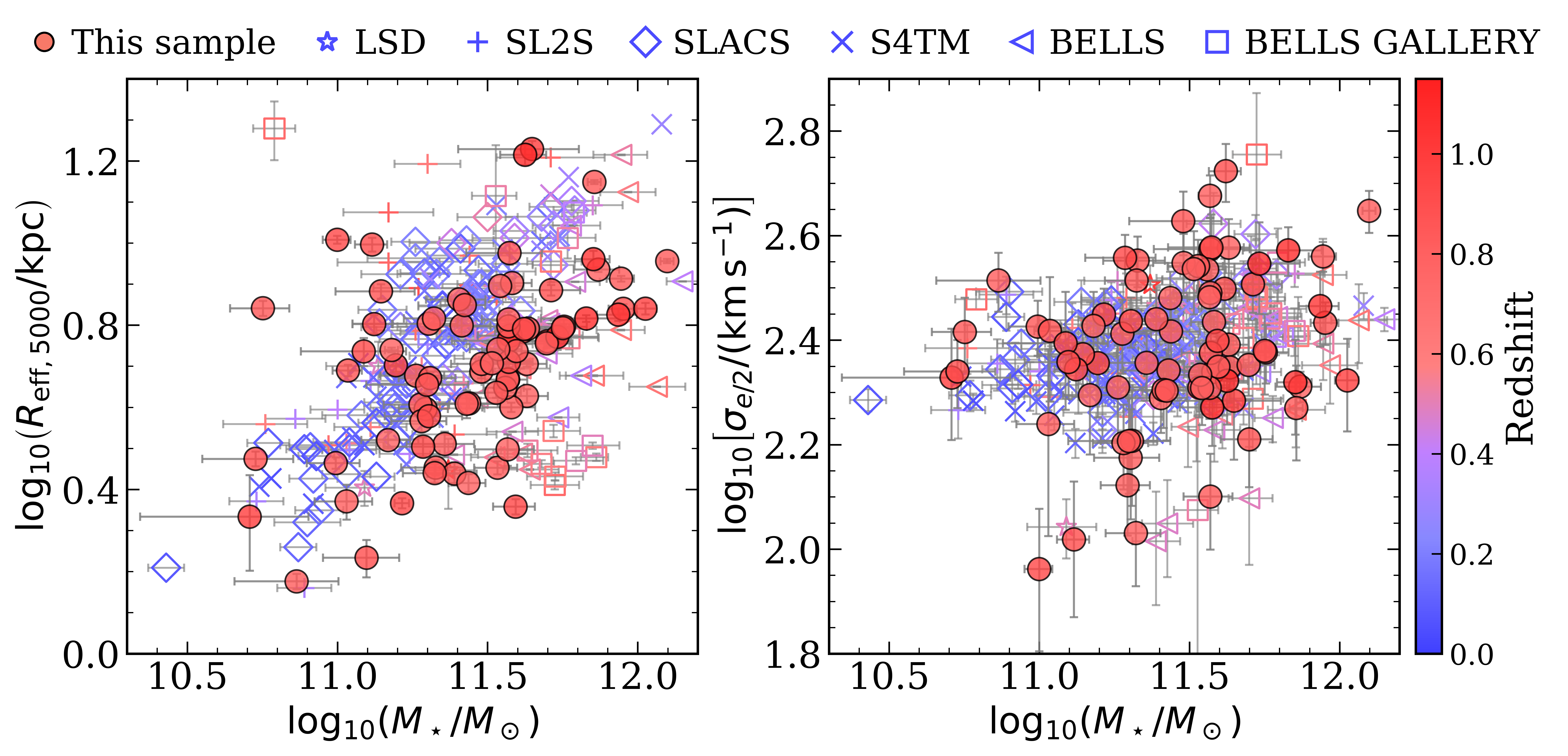}
    \vspace{-0.2cm}
    \caption{Size--stellar mass (left) and velocity dispersion--stellar mass (right) relations for our sample (circles) and literature comparison samples, color-coded by lens redshift.
    }    \label{fig:mass_size_dispersion_relations}
\end{figure*}

To place our sample into context, we examine the distributions of several basic properties of the lens galaxies in our sample and compare with previously known lens galaxies. Figure~\ref{fig:stellar_population_z} shows the effective radius $R_{\rm eff, 5000}$, stellar mass $\log_{10} M_*$, and central velocity dispersion $\sigma_{e/2}$ as functions of lens redshift for our sample and literature samples at lower redshifts. Figure~\ref{fig:mass_size_dispersion_relations} shows the corresponding size--mass and velocity dispersion--mass relations. Given the broad redshift range, the reported effective radii are converted to values at rest-frame 5000\AA\, using the relation $R_{\mathrm{eff}} \propto \lambda^{-0.25}$ \citep{vanderWel2014}. All stellar masses correspond to values derived assuming a Chabrier IMF. For stellar masses reported assuming a Salpeter IMF, we applied a constant shift of $-0.26\,\mathrm{dex}$, the typical stellar mass difference between the Salpeter and Chabrier IMFs for ETGs \citep{Longhetti2009, Deng2025}. Velocity dispersions are corrected to a circular aperture of radius $R_{\mathrm{eff},\,5000\,\text{\AA}}/2$ following \citet{Jorgensen1995}. We find that all the samples span similar ranges in effective radius, stellar mass, and central velocity dispersion. At fixed stellar mass, the high-redshift lens galaxies in our sample tend to have smaller sizes than lower-redshift counterparts on average, consistent with the redshift evolution trends of the size-mass relation (e.g., \citealt{Trujillo2007, Newman2012, vanderWel2014}). The central velocity dispersion is positively correlated with stellar mass for all the samples. Our sample exhibits somewhat larger scatter, potentially reflecting the generally low S/Ns of the DESI spectra.

We next highlight several systems with noteworthy features. HSC imaging data of HSC\,J1231$-$0135 shows an elongated pink arc to the south of the lens galaxy ($z_{\rm{lens}} = 0.94332$, $M_* = (1.64 \pm 0.09) \times 10^{11}\,M_\odot$). Our initial modeling of the lens galaxy using a single S\'{e}rsic component revealed another compact galaxy projected right on the arc. Subtracting the light from this compact galaxy reveals two sub-arcs around it, particularly in the $i$ and $z$ bands. We used \texttt{CIGALE} to derive its photometric redshift based on the five-band photometry and found $z_{\text{phot}} = 0.98 \pm 0.05$, consistent with the redshift of the central lens galaxy. The stellar mass of the compact galaxy is estimated to be $M_* = (0.94 \pm 0.25) \times 10^{11}\,M_\odot$. Given its mass and proximity to the arc, we consider it highly likely that this compact galaxy acts as a secondary lens. Nevertheless, our current model with a single lens mass component is able to reproduce the overall structure of the arc. We defer a detailed analysis of this system to a future paper. 

Some systems in this sample consist of multiple lens galaxies. In particular, system 221222$-$001811 shows a pair of lens galaxies with similar brightness surrounded by four lensed images in a cross configuration. Although the DESI fiber was centered between the two lens galaxies, no distinct second set of redshift features was detected. The velocity dispersion measured from the DESI spectrum is $211 \pm 32 \text{ km s}^{-1}$, further suggesting that the two lens galaxies have very similar redshifts. Adopting the DESI redshift of $z_{\rm{lens}} = 0.91083$ for both lens galaxies gives a projected separation of $3.39^{+0.06}_{-0.09}$ kpc, the smallest among this sample. Their $z$-band effective radii are $6.39^{+0.25}_{-0.28}$ kpc and $1.74^{+0.19}_{-0.17}$ kpc, with corresponding stellar masses of $(1.51 \pm 0.24) \times 10^{11}\,M_\odot$ and $(1.25 \pm 0.36) \times 10^{11}\,M_\odot$. Notably, the projected separation in 221222$-$001811 is smaller than that in HSC\,J0233$-$0205 ($3.596 \pm 0.046$ kpc), which was recently reported by \citet{He2026} as a strong-lens galaxy pair with the smallest projected separation. This type of strong-lens systems provides a valuable probe of galaxy interactions and dark matter physics (e.g., \citealt{Suyu2009, Bussmann2012, Shu2016}). 

The lens galaxy of system 093036$-$020204 exhibits a highly flattened ($q \sim 0.14$), edge-on disk morphology, flanked by two petal-shaped purple arcs on either side of the central bulge. Our lens model reproduces the HSC imaging data well. The measured velocity dispersion of $104 \pm 30 \text{ km s}^{-1}$ is also consistent with a disk lens galaxy nature. Known disk lenses are predominantly at $z_{\text{lens}} \lesssim 0.4$ (e.g., \citealt{Treu2011, Dutton2013}), and the highest-redshift disk lens to date is PKS 1830$-$211, a face-on disk at $z_{\text{lens}} = 0.886$ \citep{Winn2002}. The disk lens in system 093036$-$020204 is at a higher redshift $z_{\text{lens}} = 0.92016 \pm 0.00001$ and viewed edge on, providing a rare opportunity to probe the mass distribution of disk galaxies at high redshifts using strong lensing and potentially kinematics measurements.

\begin{figure*}[t]
    \centering
    \includegraphics[width=0.9\textwidth]{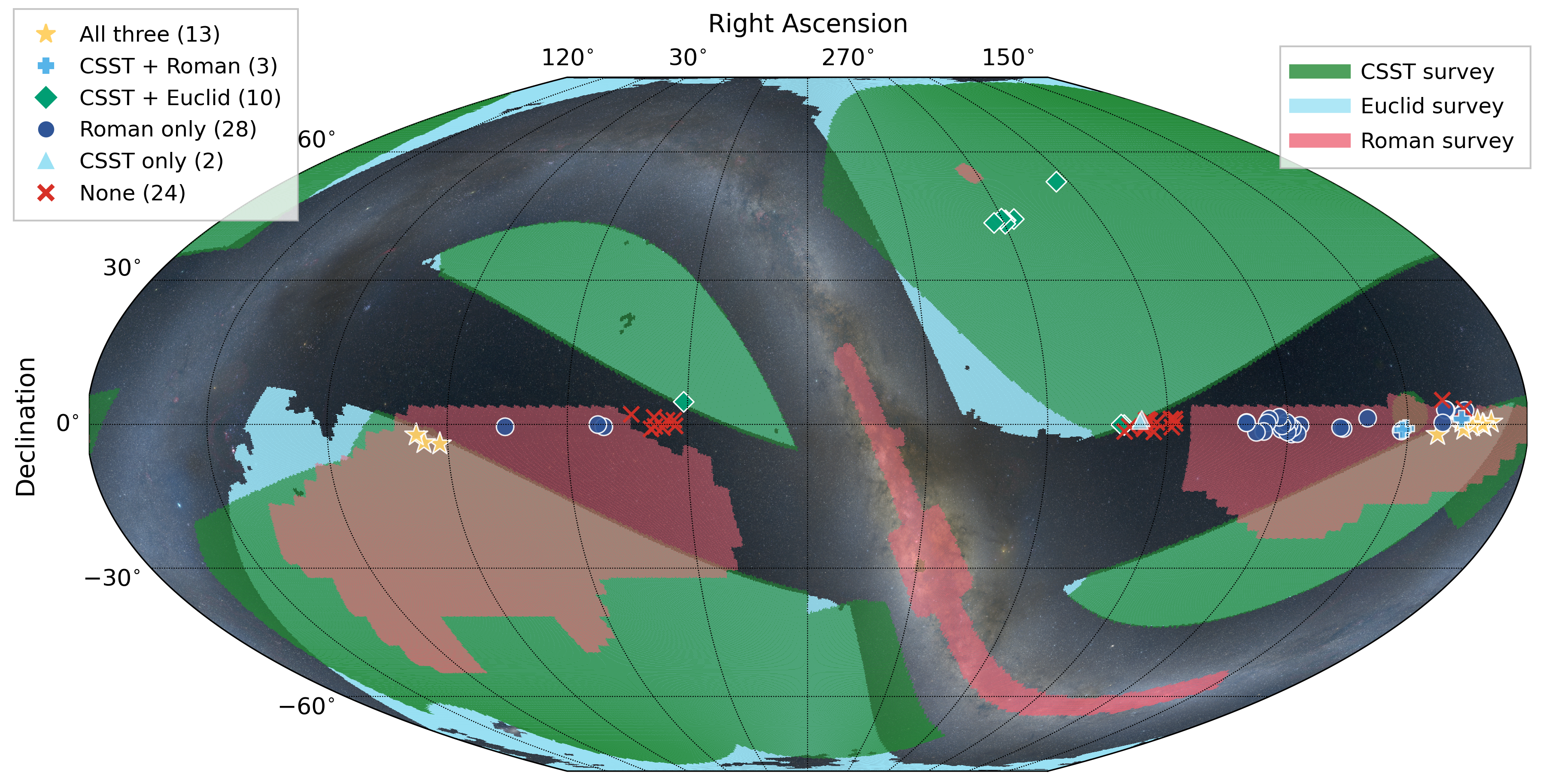}
    \vspace{-0.2cm}
    \caption{Planed CSST (green), Euclid (cyan) and Roman (pink) survey coverages. Symbols distinguish systems covered by individual surveys and their intersections, with category counts given in the legend. Of the 80 systems, CSST, Euclid and Roman will cover 28, 21 and 44 targets, respectively. Thirteen are accessible to all three surveys. The background is the optical RGB all-sky panorama of Mellinger, accessed through the CDS HiPS survey \texttt{CDS/P/Mellinger/color}. 
    }
    \label{fig:survey_coverage}
\end{figure*}

Five systems in our sample (092235$+$025942, HSC\,J1002$-$0050, 100554$-$010357, 100712$-$012316 and 135710$+$010325) display orange/red lensed images, suggesting their sources are quiescent galaxies. Massive quiescent galaxies at $z\sim2$ are found to be 3–6 times more compact than local ellipticals of similar mass, i.e., ``red nugget'' (e.g., \citealt{vanDokkum2008, Toft2009}), and are widely attributed to dry minor mergers that build up outer stellar envelopes via inside-out growth \citep{Hilz2013}. Leveraging lensing magnification and distortion, this type of ETG–ETG lens systems will allow to decompose the core–envelope structures and measure color gradients of the lensed quiescent galaxies (e.g., \citealt{Oldham2017a}), thereby testing the inside-out growth picture and probing their Fundamental Plane offsets relative to the local relation (e.g., \citealt{Oldham2017b}).

Although the HSC imaging is of sufficient quality for initial lens modeling aimed primarily at verifying the lensing nature of these systems, higher-resolution data remain essential for detailed follow-up analyses. To assess which systems in our sample could benefit from future surveys, we cross-matched our targets with the planned footprints of three ongoing or upcoming space-based wide-field missions: the Chinese Space-station Survey Telescope (CSST, \citealt{CSST2026}), \text{Euclid} \citep{Euclid}, and the Nancy Grace Roman Space Telescope (\text{Roman}, \citealt{Roman2025}). Together, these three facilities are expected to cover a combined area of approximately $23{,}000\,\text{deg}^2$ ($\sim 56\%$ of the full sky). Figure~\ref{fig:survey_coverage} depicts the planned sky coverages of these three missions and the locations of the 80 systems in this work. We found that 56 systems ($70\%$) reside within the footprint of at least one mission. \text{CSST}, \text{Euclid}, and \text{Roman} each will cover 28, 21 and 44 systems, respectively. Thirteen systems are expected to be covered by all three surveys. In particular, HSC\,J1600$+$4345, classified as a non-lens in this work, falls within both the CSST and \textit{Euclid} footprints, which will provide a definitive examination of its classification.

\section{Conclusions}\label{sec:conclusions}

The total mass distribution of ETGs and its redshift evolution encode key physical mechanisms governing galaxy assembly, yet such measurements remain limited at high redshifts where predictions from cosmological simulations diverge markedly. To date, 37 galaxy-galaxy strong lenses with ETG lenses at $z_{\text{lens}} > 0.8$ have been spectroscopically confirmed (e.g., \citealt{Hewitt1988, Lagattuta2012, Wong2014, Spilker2016, sugohi_IV, Tran2022, Lin2025, Barone2026, Jullo2026}), with the measurements of total mass density slope ($\gamma$) available for only ten systems (i.e., \citeauthor{Chen1995, Treu2004, Lagattuta2012, Dye2014, Wong2014, Filipp2023, Sahu2024, Stacey2024, LiRui2025, Huang2025}). 
In this work, we perform joint five-band lens modeling for 80 galaxy-scale systems at $z_{\text{lens}} > 0.8$ selected from the parent sample reported in Paper~I. Based on the modeling results, we classify 71 systems as secure lenses which exhibit typical lensing features with clear counter-images. Five systems that display clear features are classified as probable lenses due to the lack of counter-images, and another three systems with atypical lensing features are marked as possible lenses. The remaining system is classified as a non-lens because of the absence of convincing lensing features. This classification demonstrates the value of detailed lens modeling based on ground-based imaging for assessing the reliability of strong-lens candidates. Our modeling also reveals several systems with unusual lensing configurations that merit further investigation.

We measured stellar velocity dispersions from DESI spectra and estimated stellar masses based on modeled HSC five-band photometry for the lens galaxies. Overall, the effective radii and stellar masses of lens galaxies in our sample are comparable to those of known lenses. At fixed stellar mass, these high-redshift lens galaxies tend to have smaller size on average than lower-redshift counterparts. Across all samples, the central velocity dispersion positively correlates with stellar mass.

A total of 56 systems ($70\%$) in our sample fall within the scheduled footprints of \text{CSST}, \text{Euclid}, or \text{Roman}. Future higher-resolution imaging delivered by these missions will resolve their lensing features more clearly, enabling more precise constraints on the lens mass distributions and background source reconstructions. Among the 71 secure systems, 53 each contain a single lens galaxy, of which only 9 have spectroscopic source redshifts. Spectroscopic follow-up is urgently needed for the remaining 44 systems to determine the source redshifts and refine the stellar velocity dispersion measurements. Positioned at the intersection of next-generation space-based imaging and ground-based spectroscopy, this sample will help shed light on the long-standing questions surrounding the mass assembly and structural evolution of ETGs across cosmic time.

\begin{acknowledgments}

We thank Drs. Guang Yang and Denis Burgarella for helpful discussions on SED fitting with \text{CIGALE}. We are grateful to Drs. Youhua Xu and Tian Li for providing the planned footprints of CSST and \text{Euclid}, respectively, and we thank the \text{Roman} Science Operations Center for making the planned footprint publicly available, along with Drs. Javier S\'{a}nchez and Rosa Diaz for the accompanying visualization. This work was supported by the China Manned Space Project (No. CMS-CSST-2025-A20) and the National Natural Science Foundation of China (Nos. 12573115 and 12533008). This study is based on HSC-SSP PDR3 imaging and DESI DR1 spectroscopy, and we gratefully acknowledge the HSC and DESI collaborations, their participating institutions and respective funding agencies.
\end{acknowledgments}

\begin{contribution}
 
S. Li performed the lens modeling, measured the stellar velocity dispersions, estimated the stellar masses, and wrote the initial draft of the manuscript. Y. Shu conceived and supervised the study and participated in the writing and revision of the manuscript. L. Nie contributed to the construction of the PSF models.


\end{contribution}

%

\facilities{Subaru (HSC), Mayall (DESI)}


\software{
lenstronomy \citep{lenstronomy},
lfit\_gui \citep{lfitgui},
PPXF \citep{Cappellari2004, Cappellari2017, Cappellari2023},
CIGALE \citep{boquien2019},
emcee \citep{Foreman-Mackey2013}
}


\appendix

\section{Lens model atlas}\label{sec:figures}

\setcounter{figure}{0}
\renewcommand{\thefigure}{A\arabic{figure}}
\begin{figure*}[htbp]
\centering
\setlength{\parskip}{-15pt} 
\includegraphics[width=0.98\textwidth]{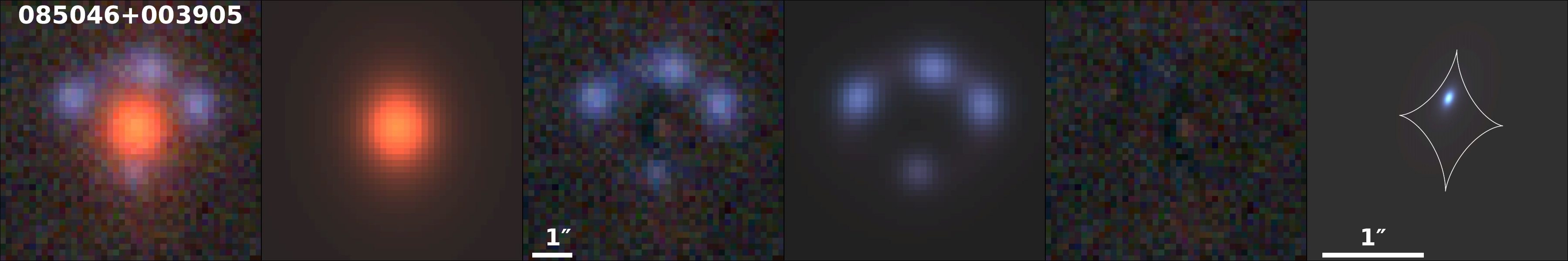}
\includegraphics[width=0.98\textwidth]{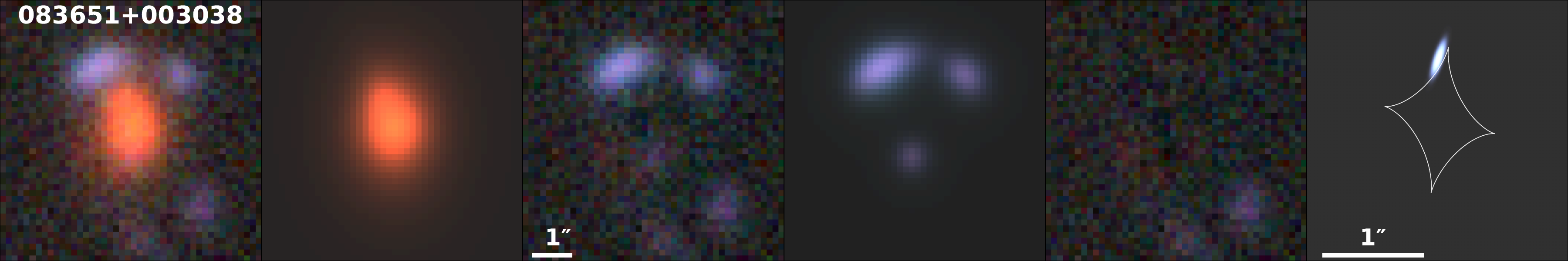}
\includegraphics[width=0.98\textwidth]{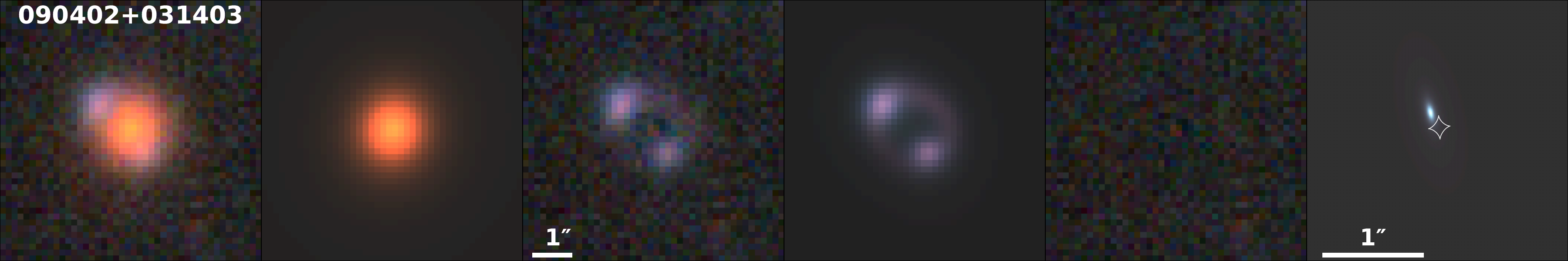}
\includegraphics[width=0.98\textwidth]{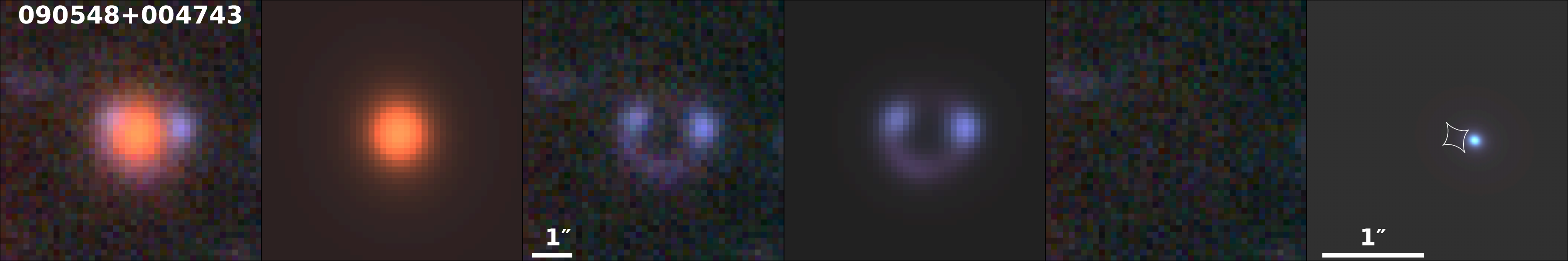}
\includegraphics[width=0.98\textwidth]{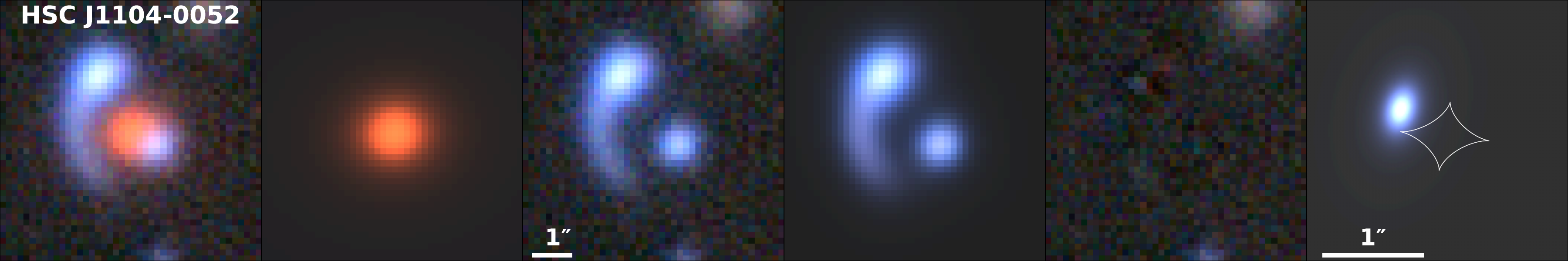}
\includegraphics[width=0.98\textwidth]{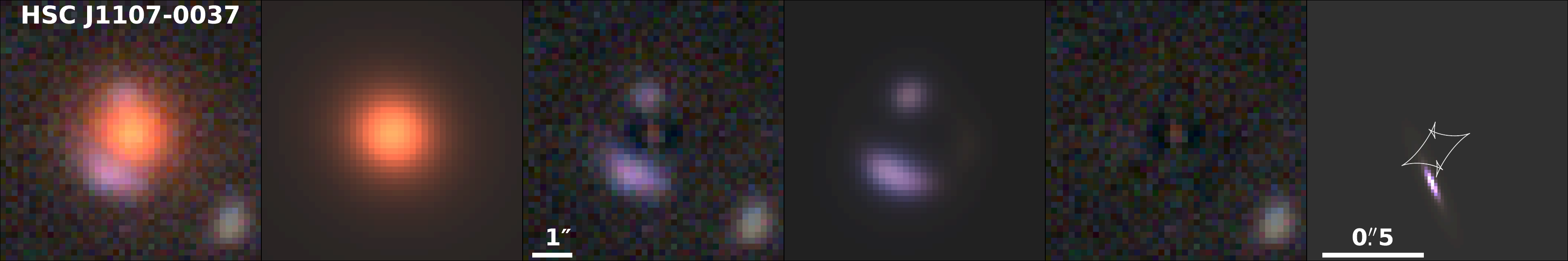}
\includegraphics[width=0.98\textwidth]{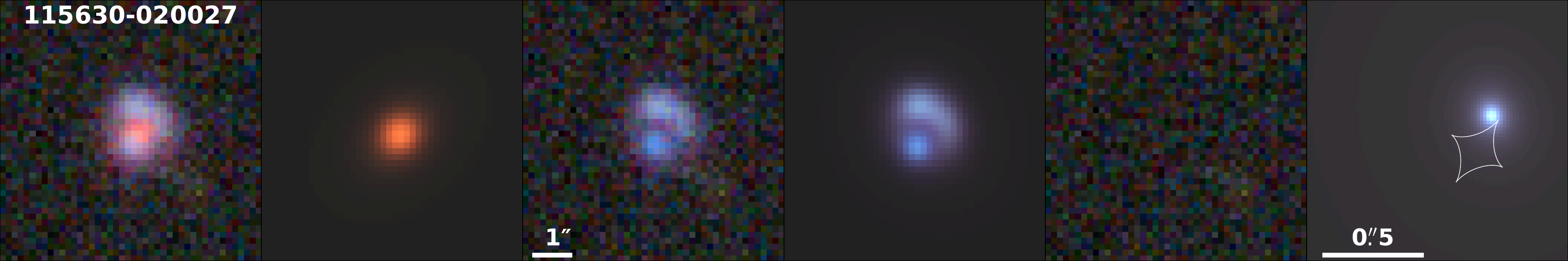}
\includegraphics[width=0.98\textwidth]{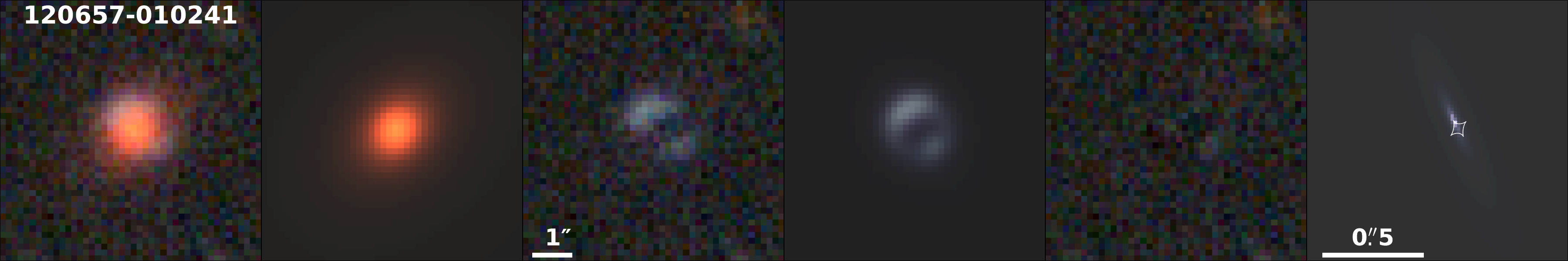}
\vspace{-0.1cm}
\caption{Best-fit lens models for the 80 systems, shown as $grizy$ color composites. From left to right, columns show the observed data, lens light model, lens light-subtracted data, lensed-source model, residuals, and source-plane reconstruction. Unless otherwise noted, image-plane cutouts are 7\farcs4 $\times$ 7\farcs4 with north up and east to the left. Scale bars for the image plane and the source plane are indicated in the lower-left corners of the third and sixth panels, respectively.}
\label{fig:1}
\end{figure*}

\begin{figure*}[htbp]
\ContinuedFloat
\centering
\includegraphics[width=0.98\textwidth]{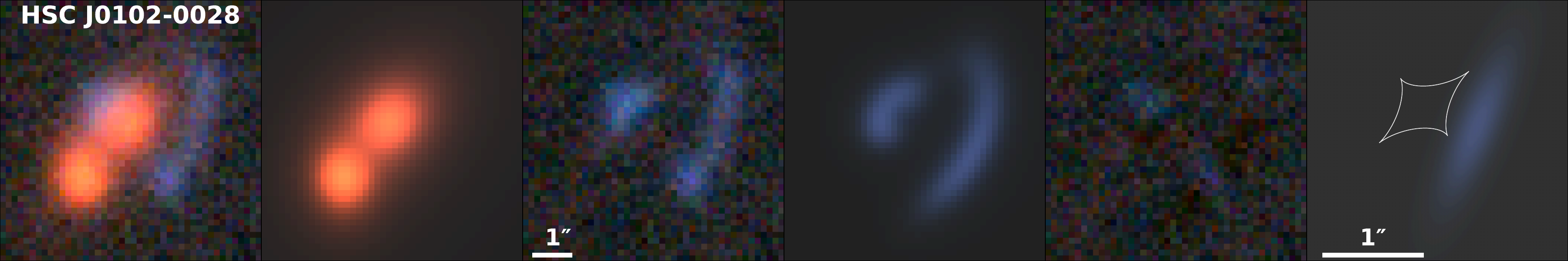}
\includegraphics[width=0.98\textwidth]{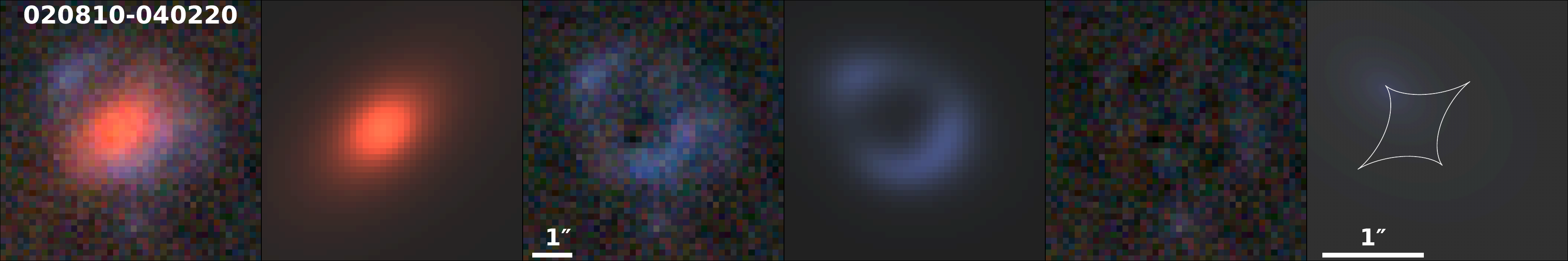}
\includegraphics[width=0.98\textwidth]{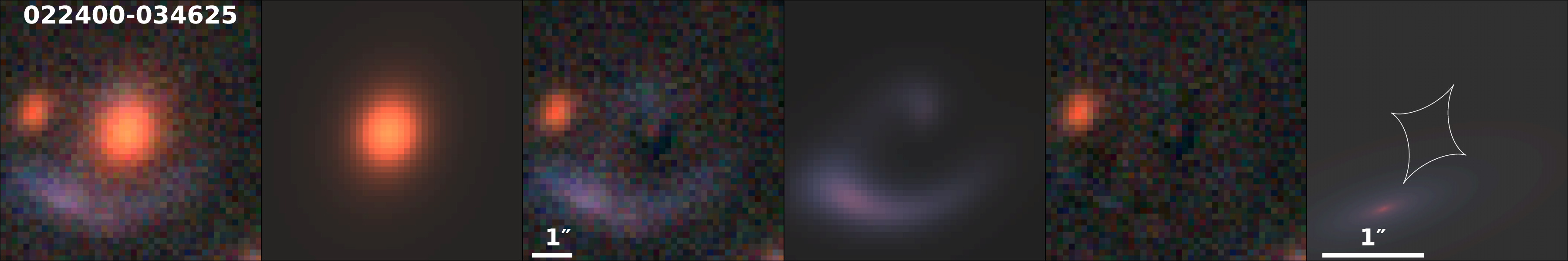}
\includegraphics[width=0.98\textwidth]{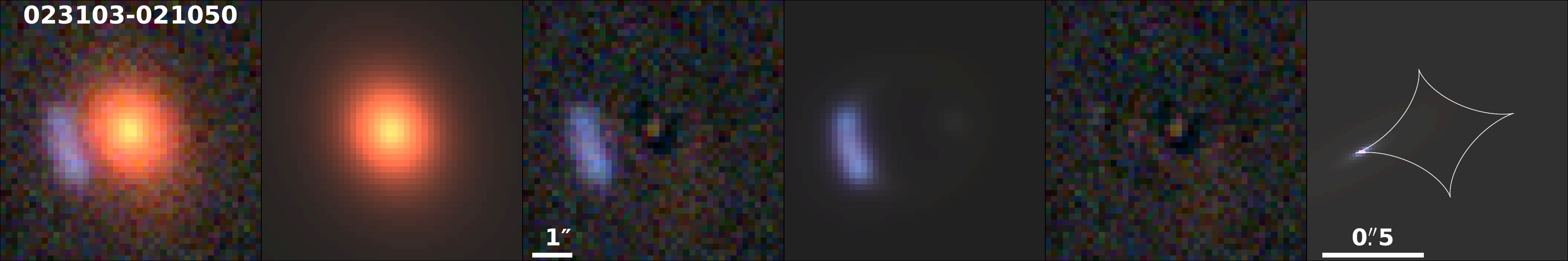}
\includegraphics[width=0.98\textwidth]{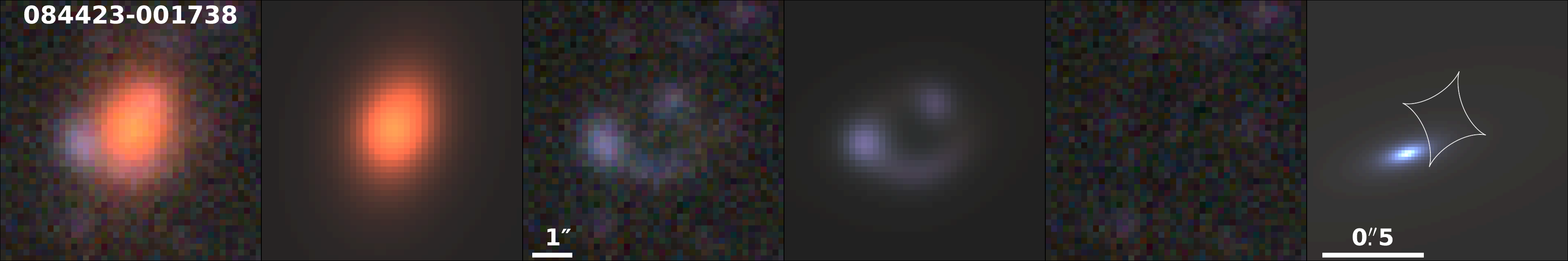}
\includegraphics[width=0.98\textwidth]{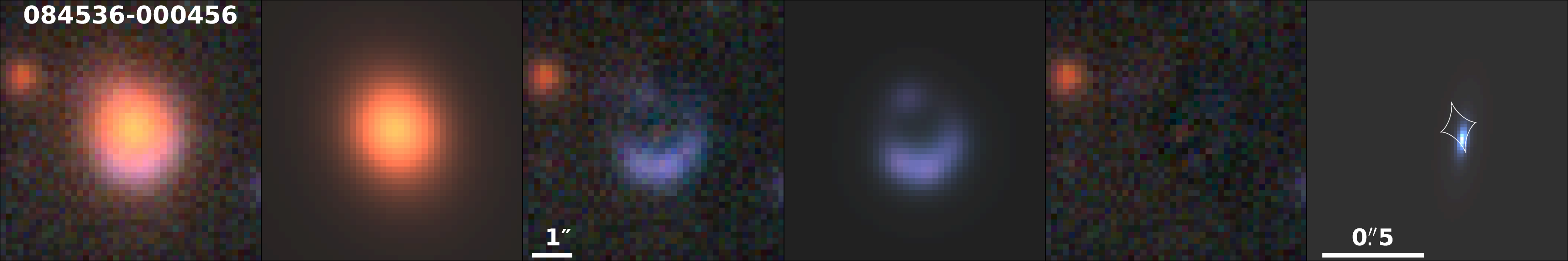}
\includegraphics[width=0.98\textwidth]{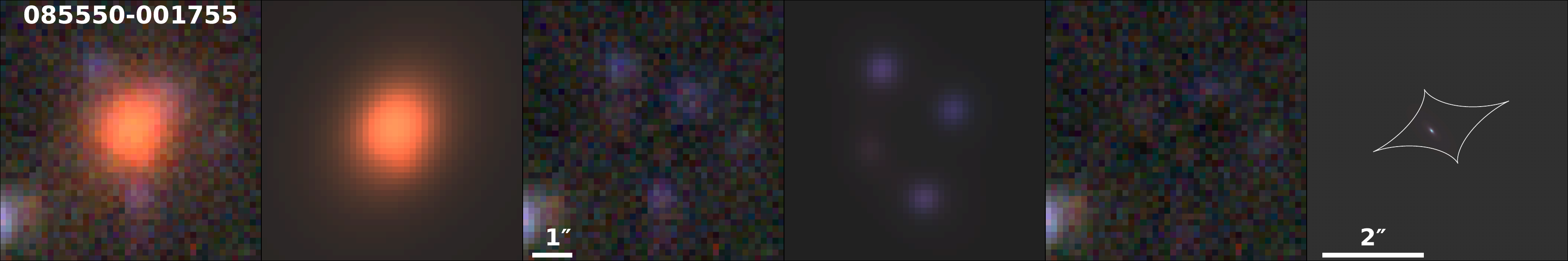}
\includegraphics[width=0.98\textwidth]{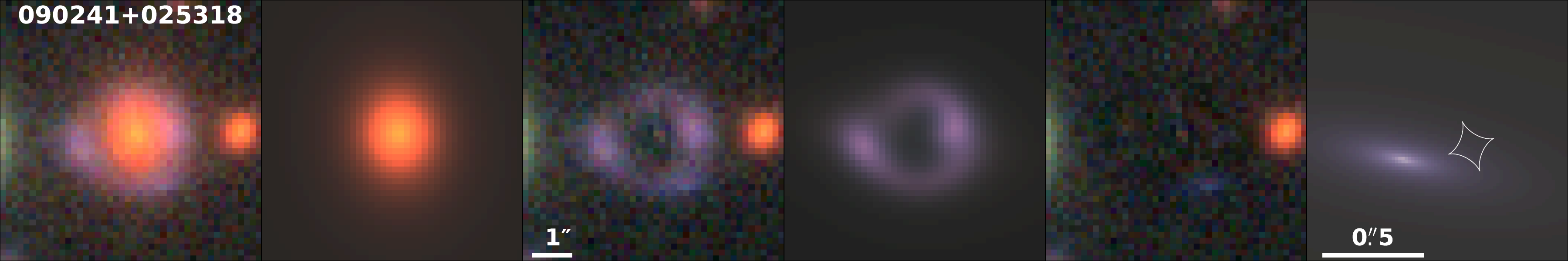}
\makeatletter
\long\def\@makecaption#1#2{%
  \vskip\abovecaptionskip
  {\small\raggedright\noindent\textbf{#1.} #2\par}%
  \vskip\belowcaptionskip
}
\makeatother
\caption{\textit{Continued}.}
\label{fig:continued_figure1_1}
\end{figure*}

\begin{figure*}[htbp]
\ContinuedFloat
\centering
\includegraphics[width=0.98\textwidth]{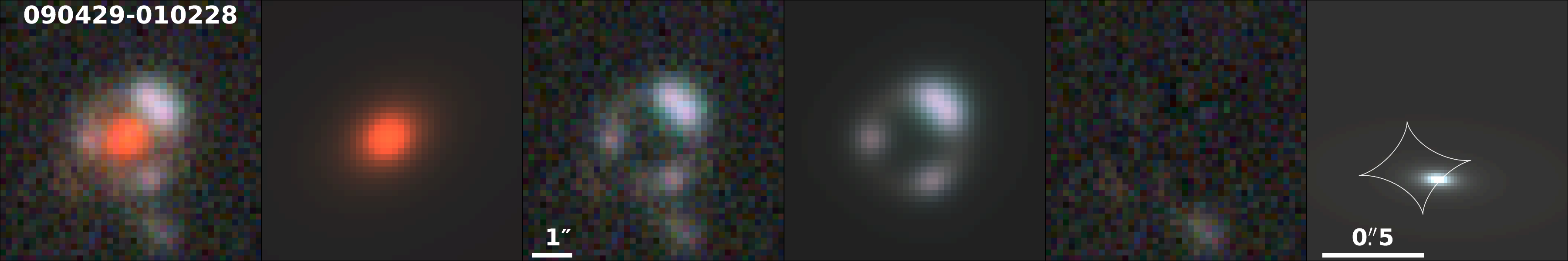}
\includegraphics[width=0.98\textwidth]{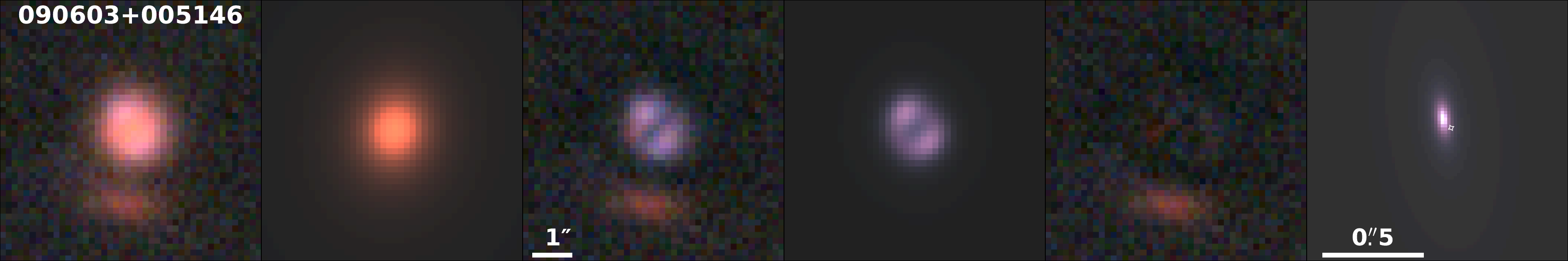}
\includegraphics[width=0.98\textwidth]{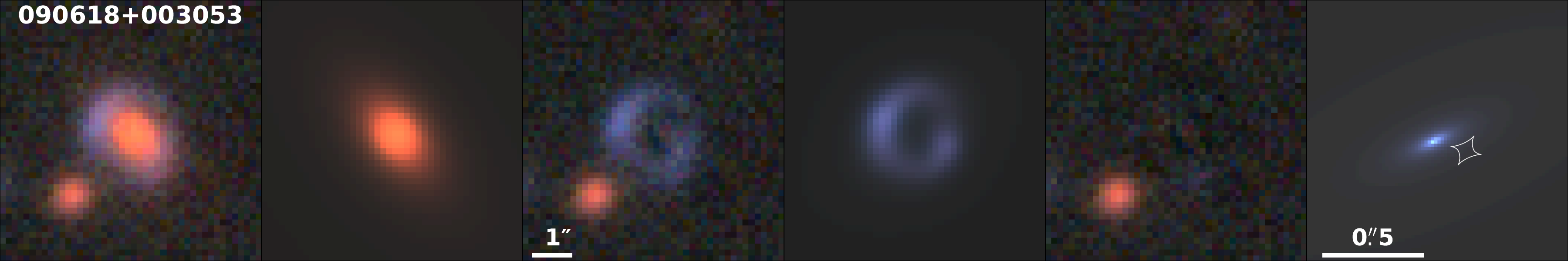}
\includegraphics[width=0.98\textwidth]{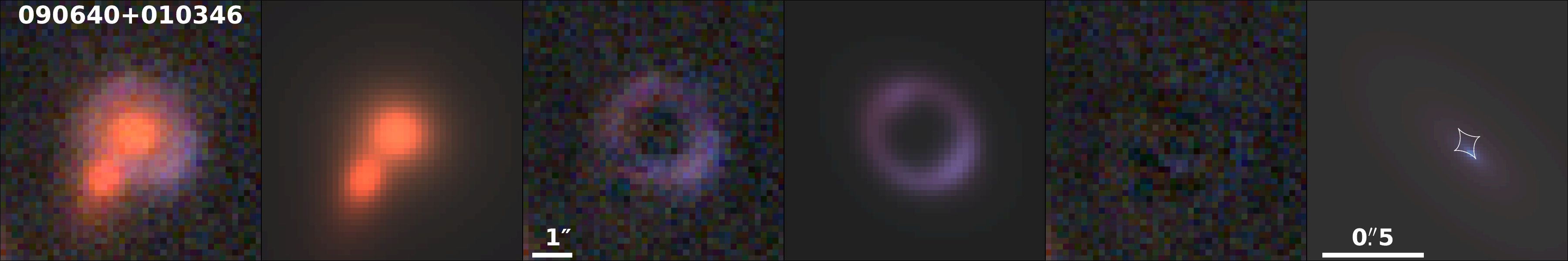}
\includegraphics[width=0.98\textwidth]{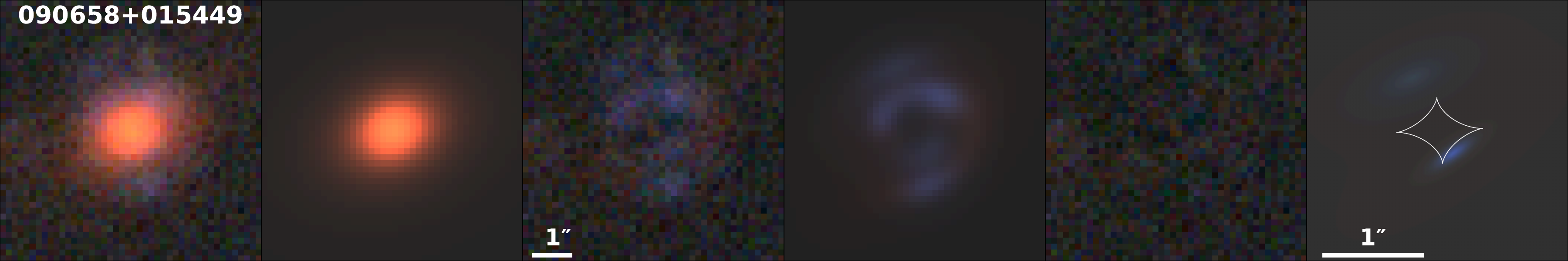}
\includegraphics[width=0.98\textwidth]{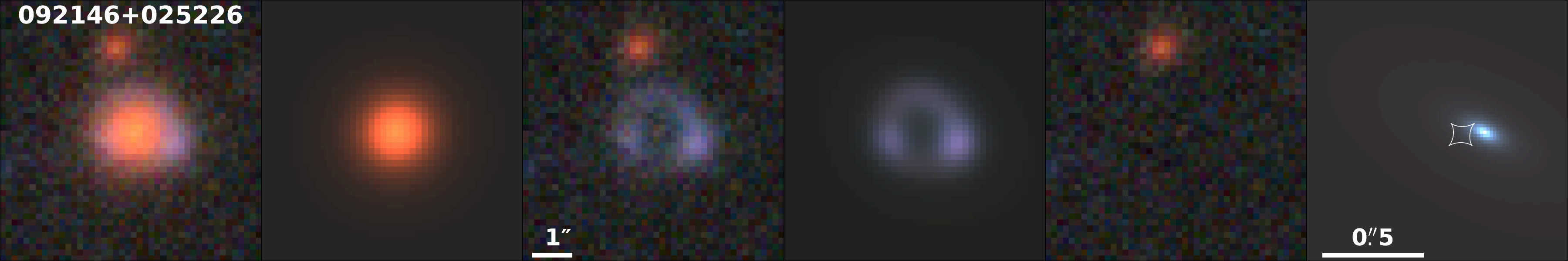}
\includegraphics[width=0.98\textwidth]{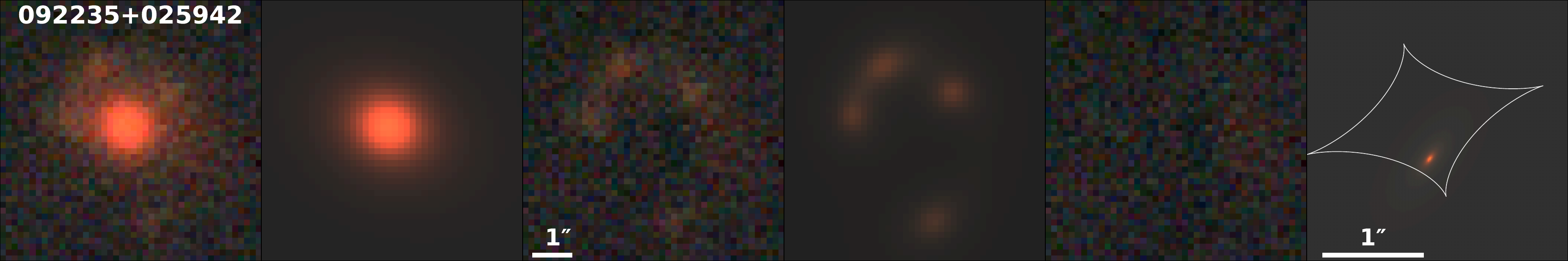}
\includegraphics[width=0.98\textwidth]{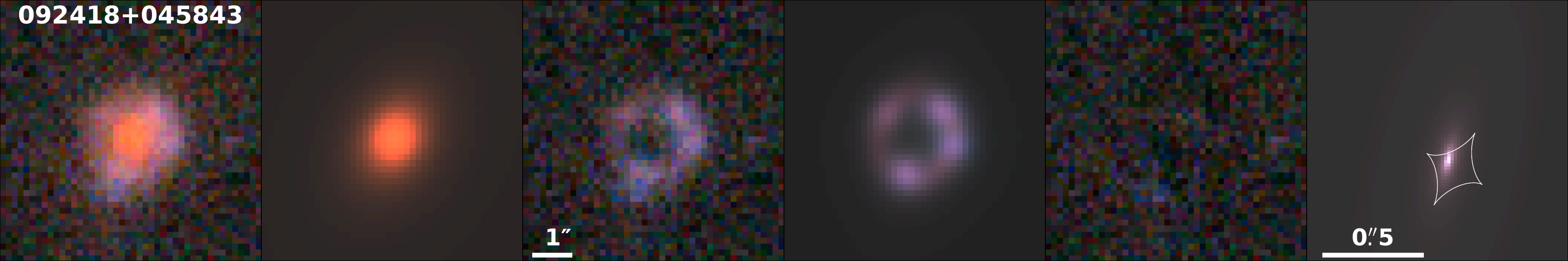}
\caption{\textit{Continued.} System 092235$+$025942 exhibits four faint, orange images that are well reproduced by our lens model.}
\label{fig:continued_figure1_2}
\end{figure*}

\begin{figure*}[htbp]
\ContinuedFloat
\centering
\includegraphics[width=0.98\textwidth]{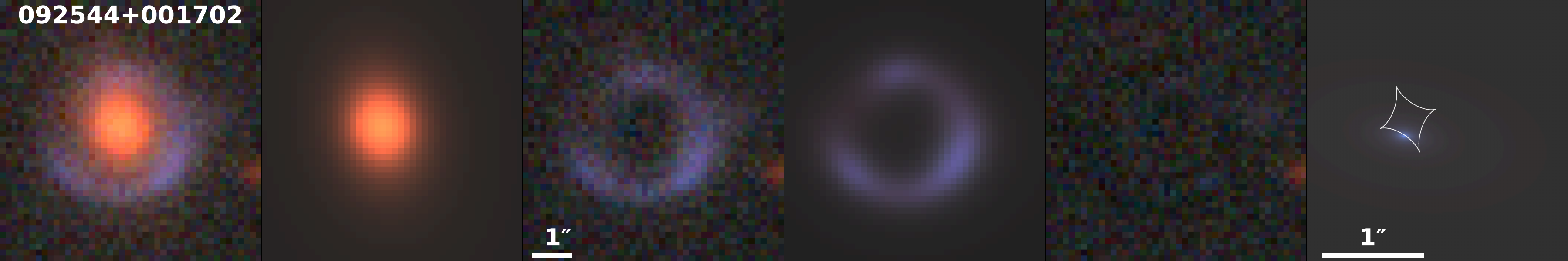}
\includegraphics[width=0.98\textwidth]{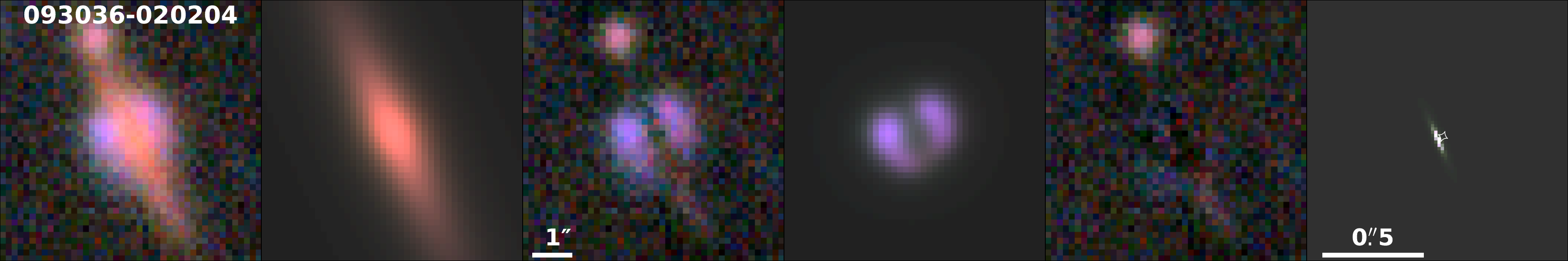}
\includegraphics[width=0.98\textwidth]{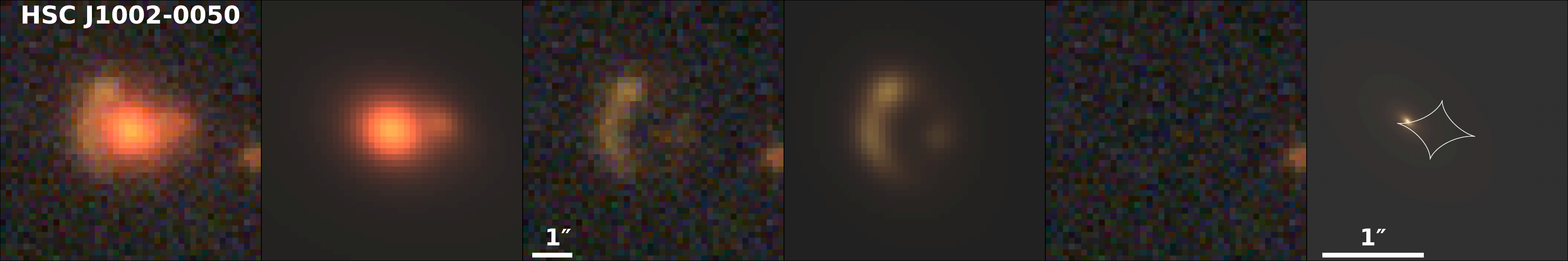}
\includegraphics[width=0.98\textwidth]{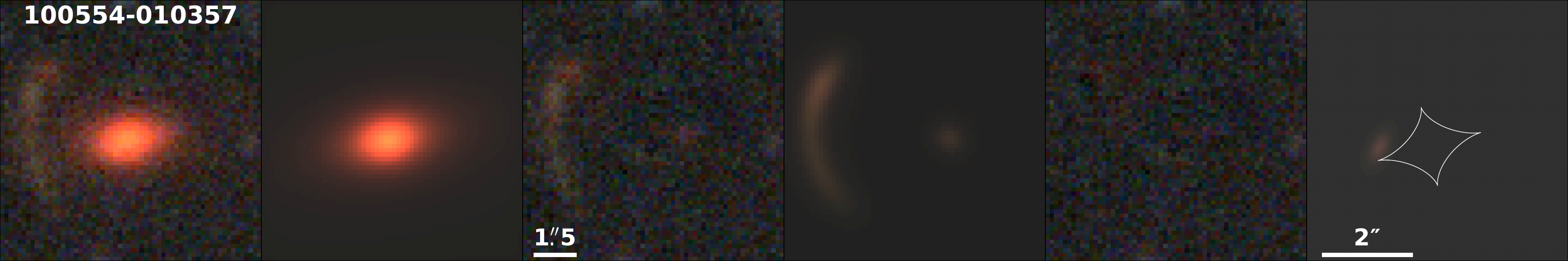}
\includegraphics[width=0.98\textwidth]{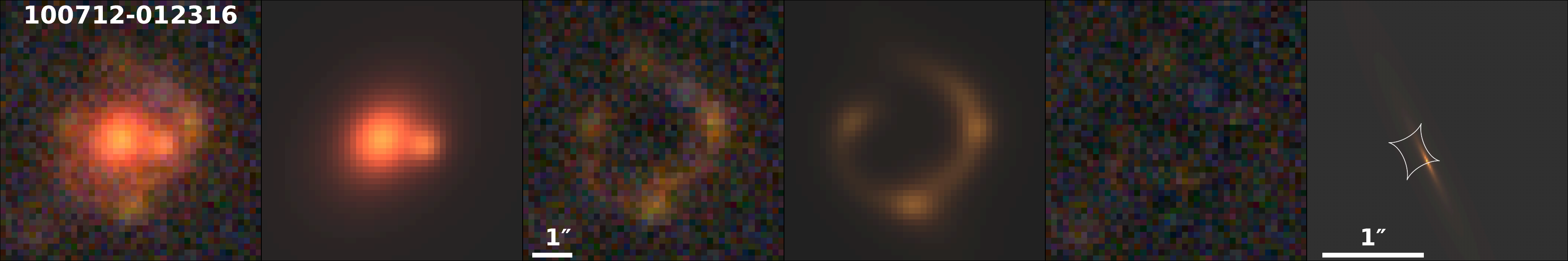}
\includegraphics[width=0.98\textwidth]{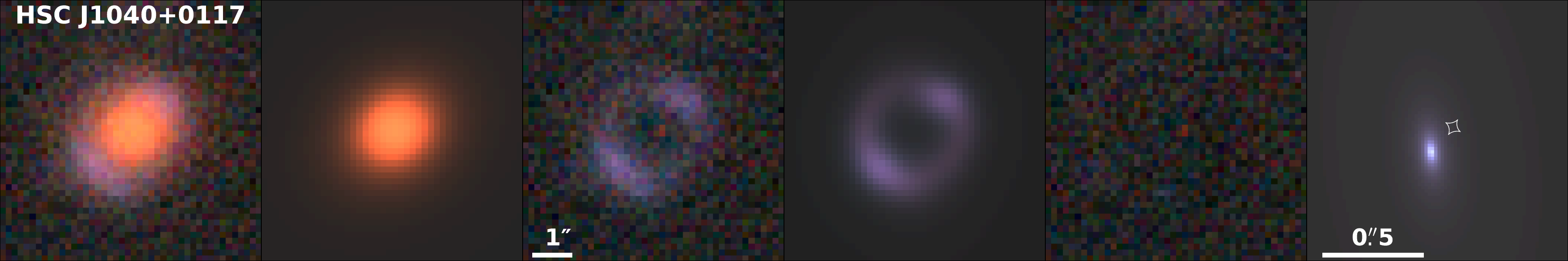}
\includegraphics[width=0.98\textwidth]{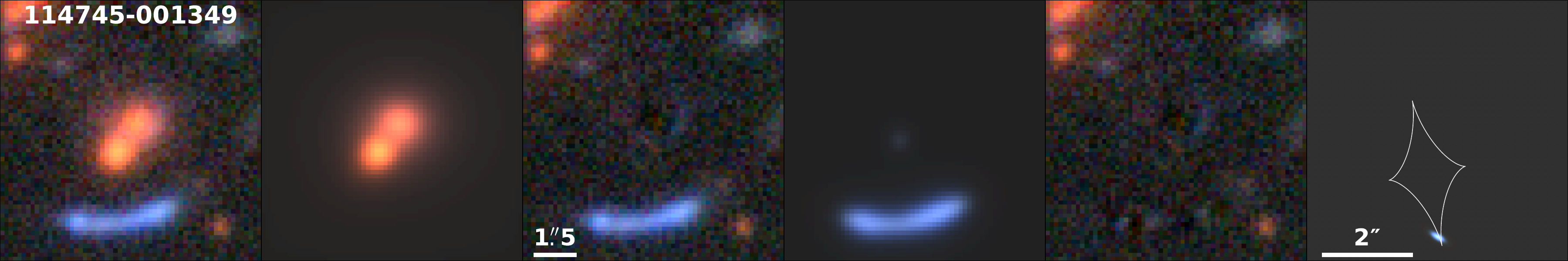}
\includegraphics[width=0.98\textwidth]{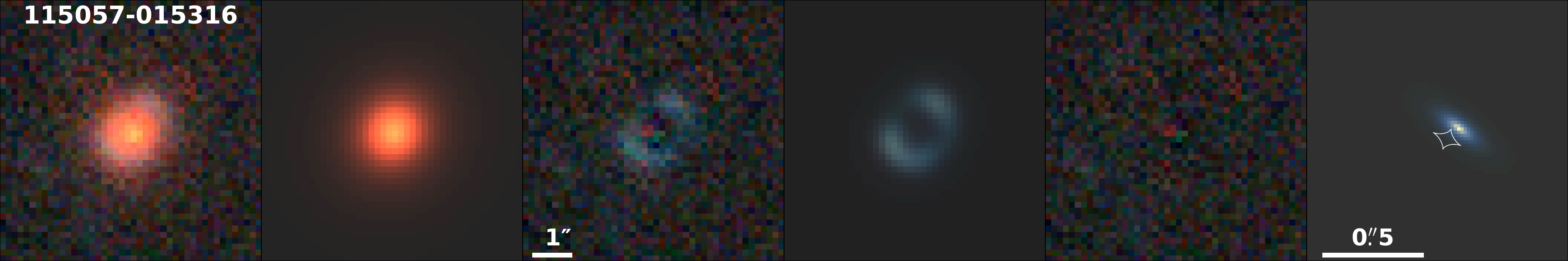}
\caption{\textit{Continued.} The lens galaxy in system 093036$-$020204 exhibits an edge-on disk morphology, while the images in systems HSC\,J1002$-$0050, 100554$-$010357, and 100712$-$012316 display an unusual orange color.}
\label{fig:continued_figure1_3}
\end{figure*}

\begin{figure*}[htbp]
\ContinuedFloat
\centering
\includegraphics[width=0.98\textwidth]{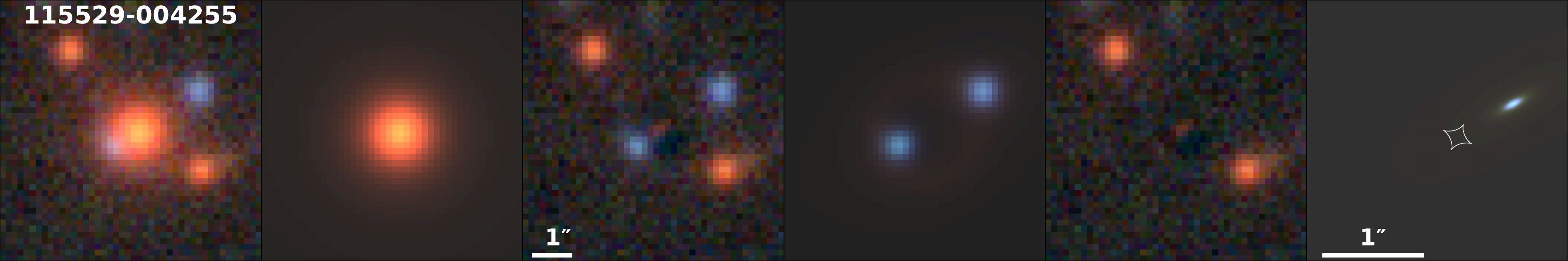}
\includegraphics[width=0.98\textwidth]{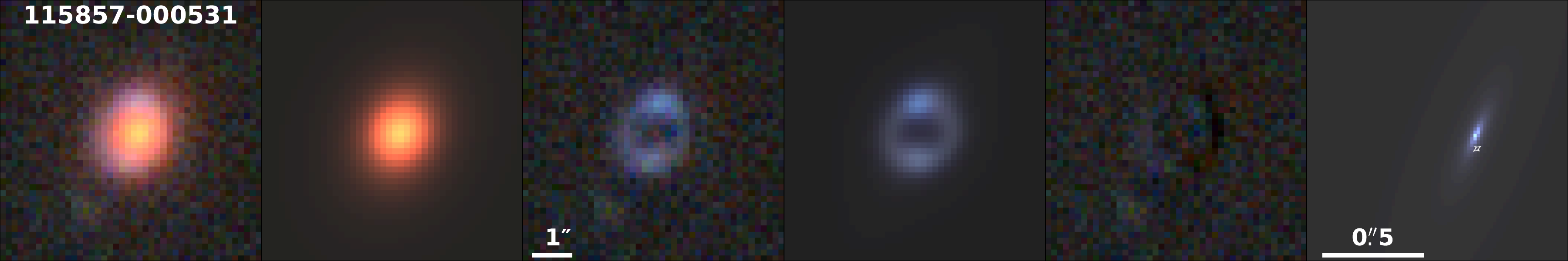}
\includegraphics[width=0.98\textwidth]{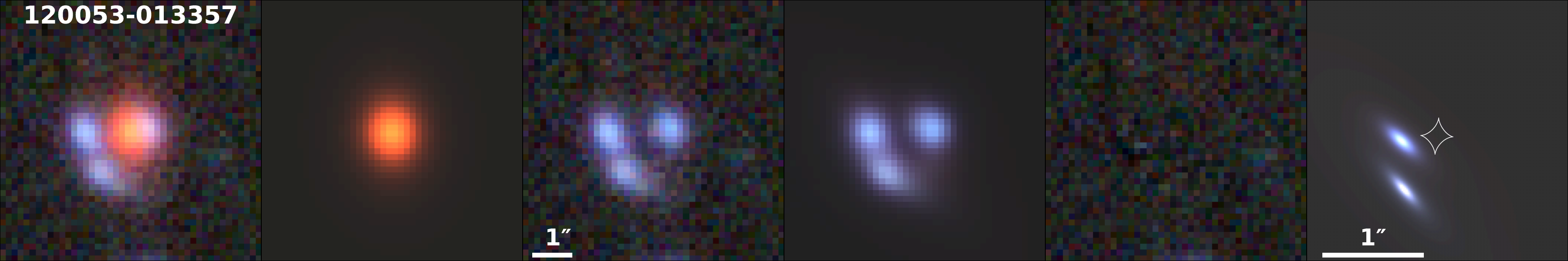}
\includegraphics[width=0.98\textwidth]{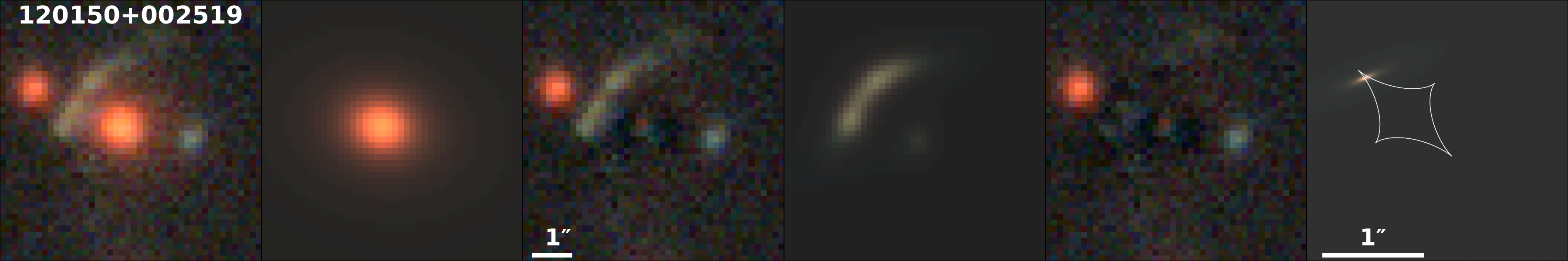}
\includegraphics[width=0.98\textwidth]{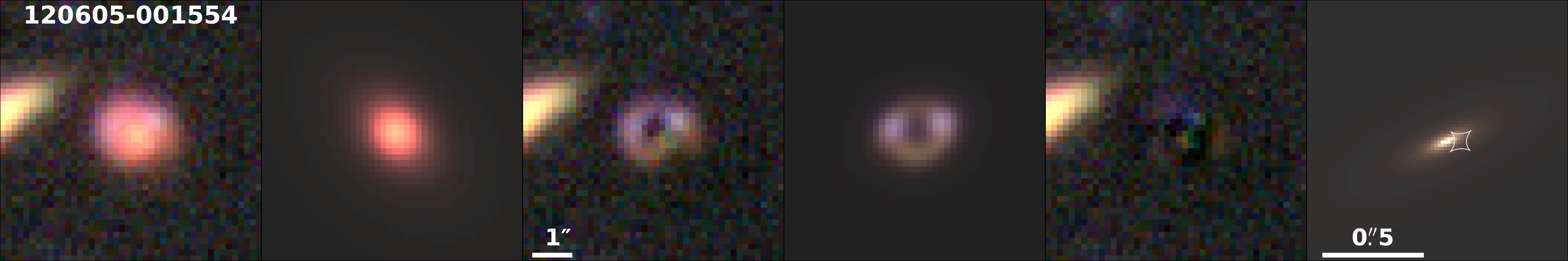}
\includegraphics[width=0.98\textwidth]{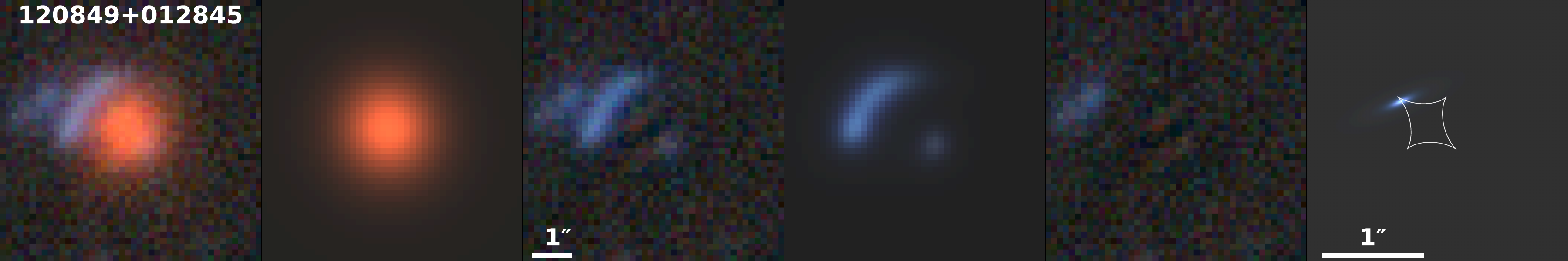}
\includegraphics[width=0.98\textwidth]{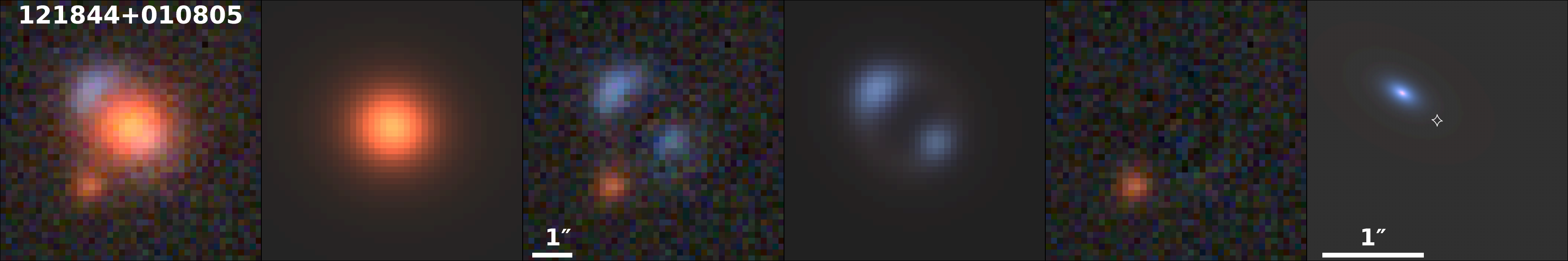}
\includegraphics[width=0.98\textwidth]{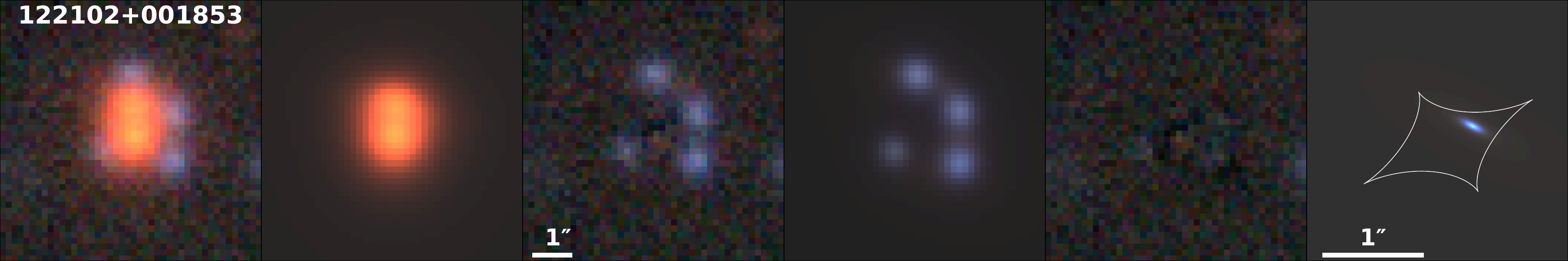}
\makeatletter
\long\def\@makecaption#1#2{%
  \vskip\abovecaptionskip
  {\small\raggedright\noindent\textbf{#1.} #2\par}%
  \vskip\belowcaptionskip
}
\makeatother
\caption{\textit{Continued.}}
\label{fig:continued_figure1_4}
\end{figure*}

\begin{figure*}[htbp]
\ContinuedFloat
\centering
\includegraphics[width=0.98\textwidth]{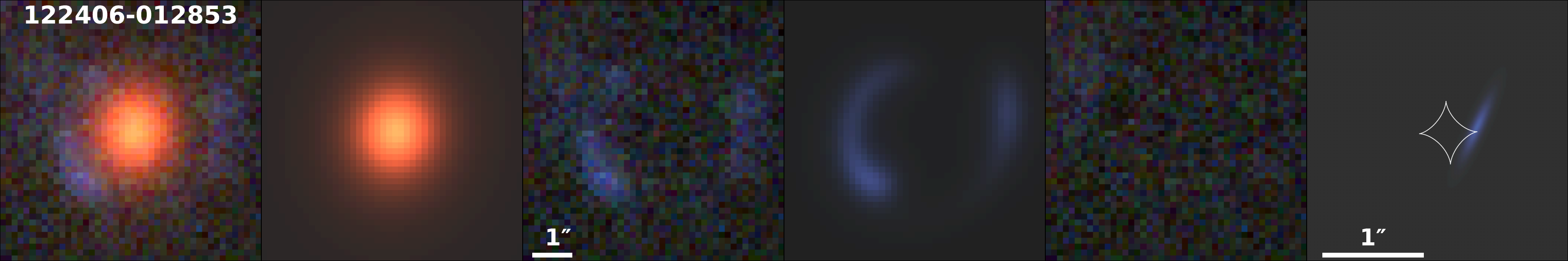}
\includegraphics[width=0.98\textwidth]{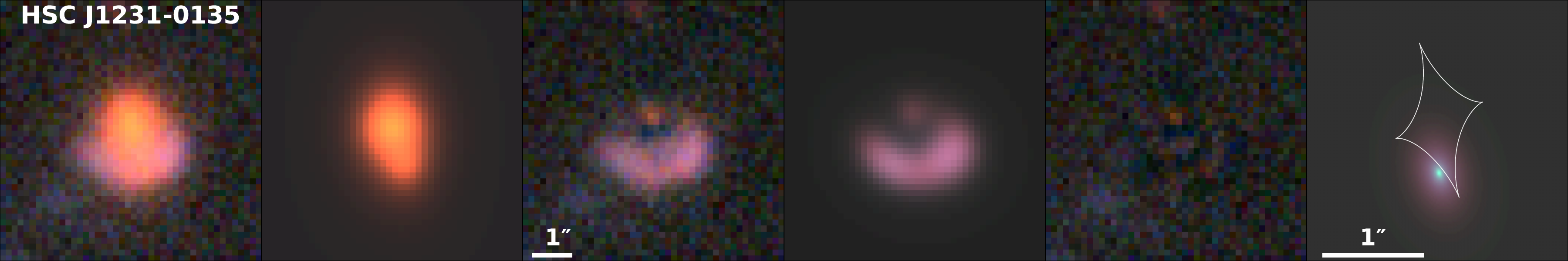}
\includegraphics[width=0.98\textwidth]{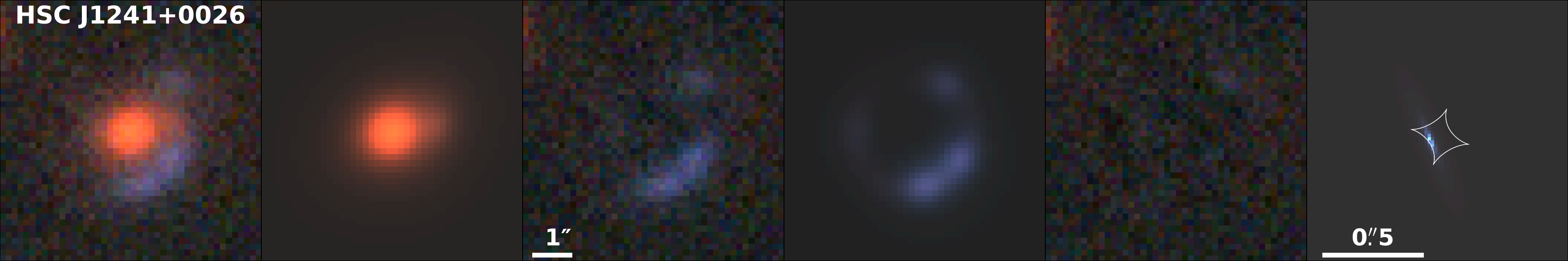}
\includegraphics[width=0.98\textwidth]{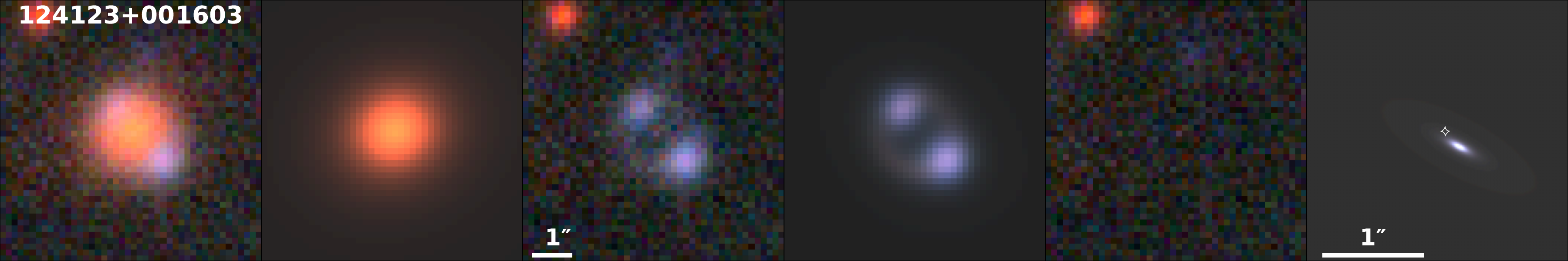}
\includegraphics[width=0.98\textwidth]{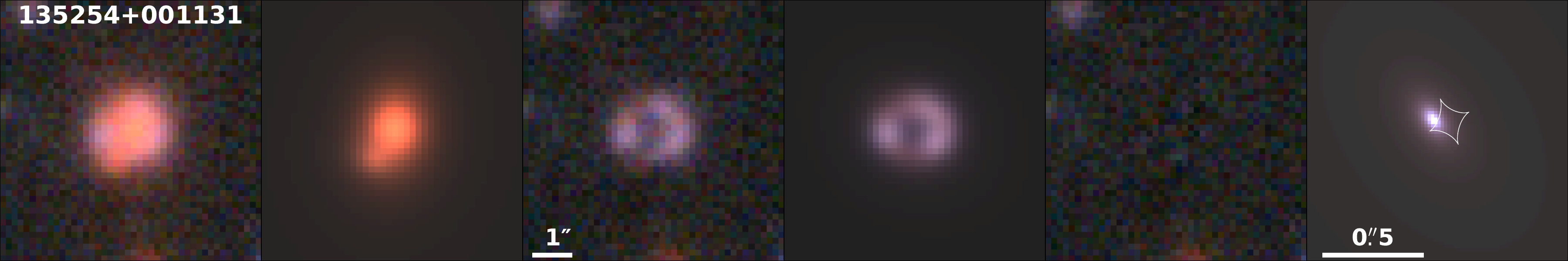}
\includegraphics[width=0.98\textwidth]{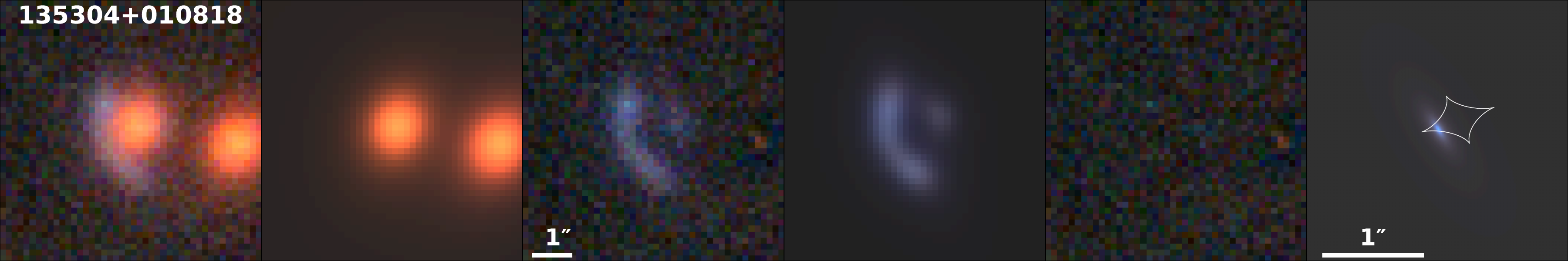}
\includegraphics[width=0.98\textwidth]{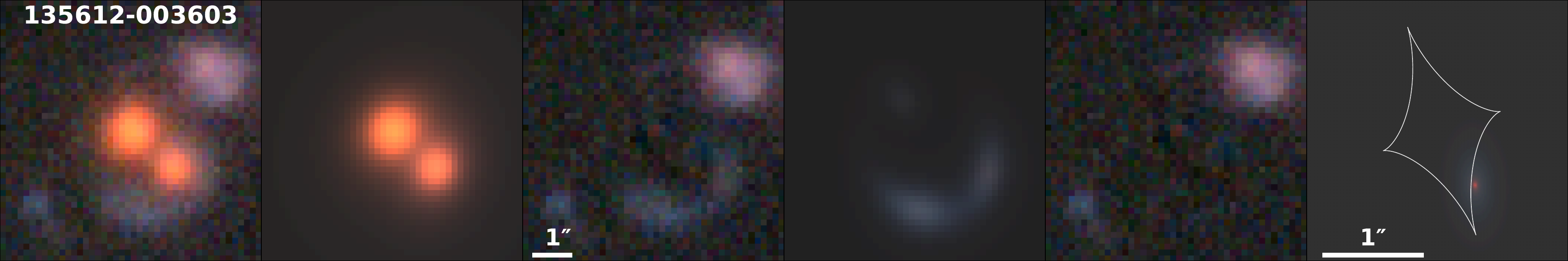}
\includegraphics[width=0.98\textwidth]{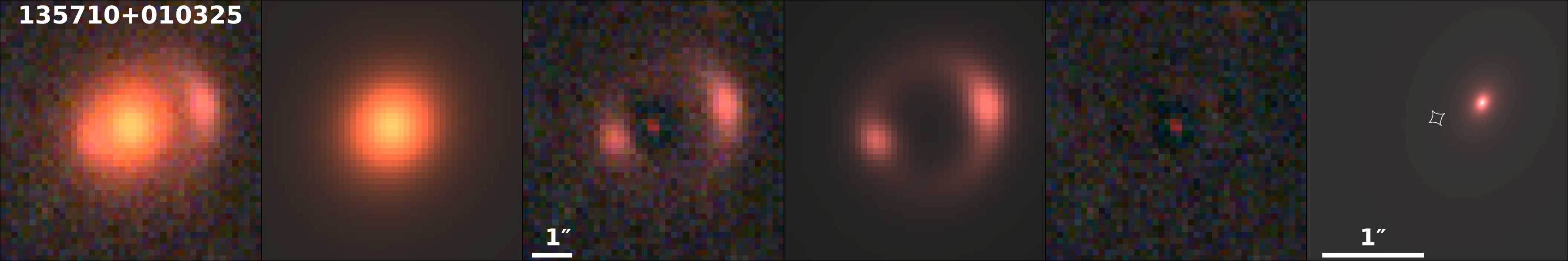}
\caption{\textit{Continued}. System HSC\,J1231$-$0135 exhibits a sub-arc feature, and system 135710$+$010325 is a ETG--ETG lens candidate.}
\label{fig:continued_figure1_5}
\end{figure*}

\begin{figure*}[htbp]
\ContinuedFloat
\centering
\includegraphics[width=0.98\textwidth]{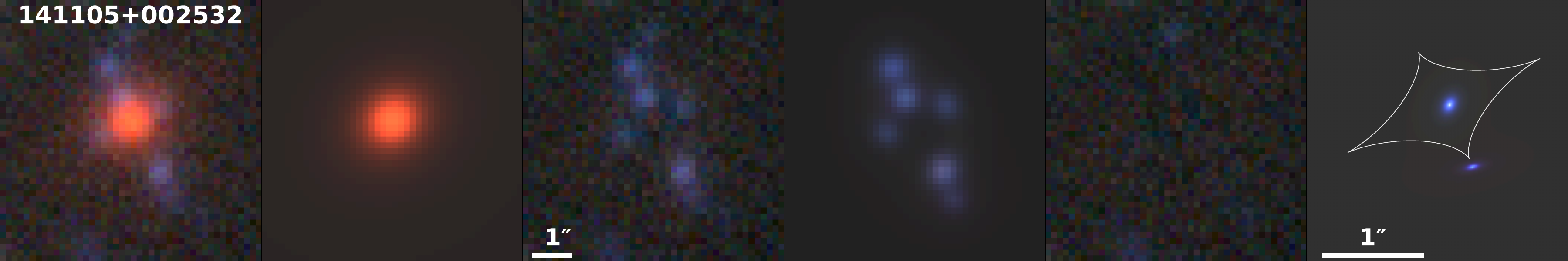}
\includegraphics[width=0.98\textwidth]{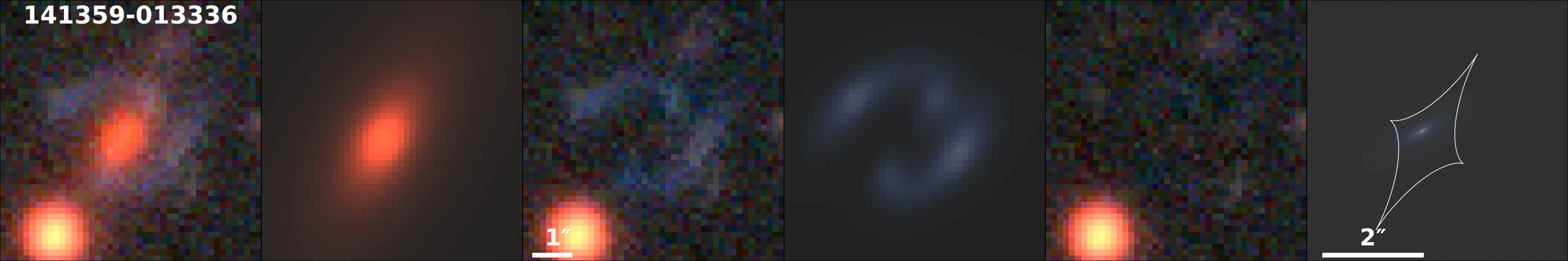}
\includegraphics[width=0.98\textwidth]{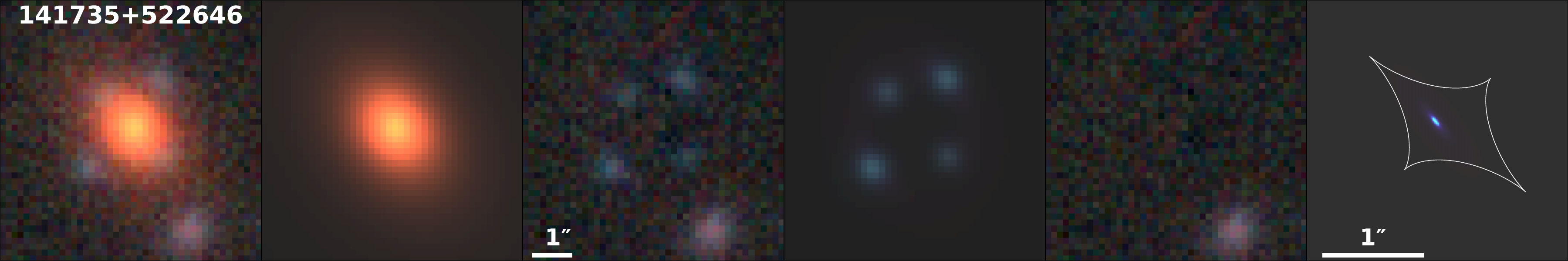}
\includegraphics[width=0.98\textwidth]{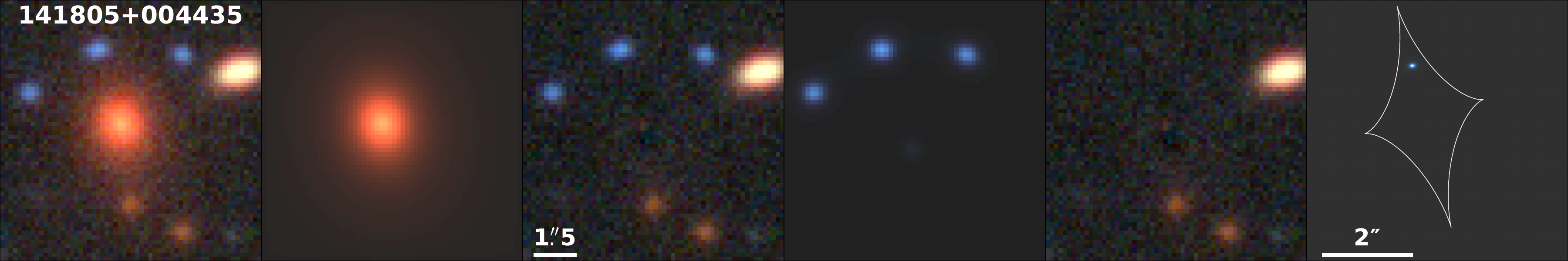}
\includegraphics[width=0.98\textwidth]{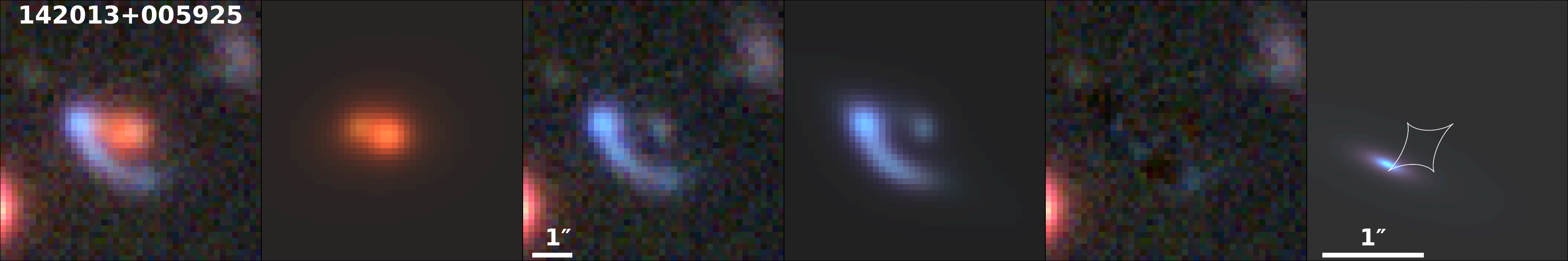}
\includegraphics[width=0.98\textwidth]{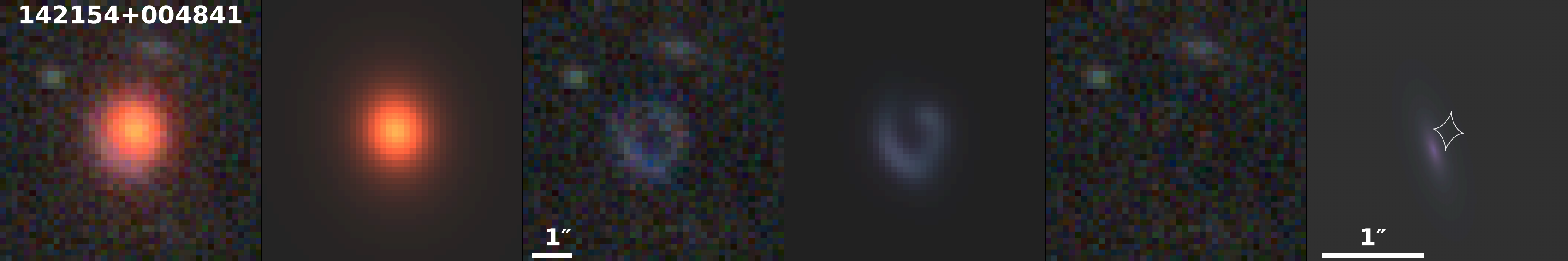}
\includegraphics[width=0.98\textwidth]{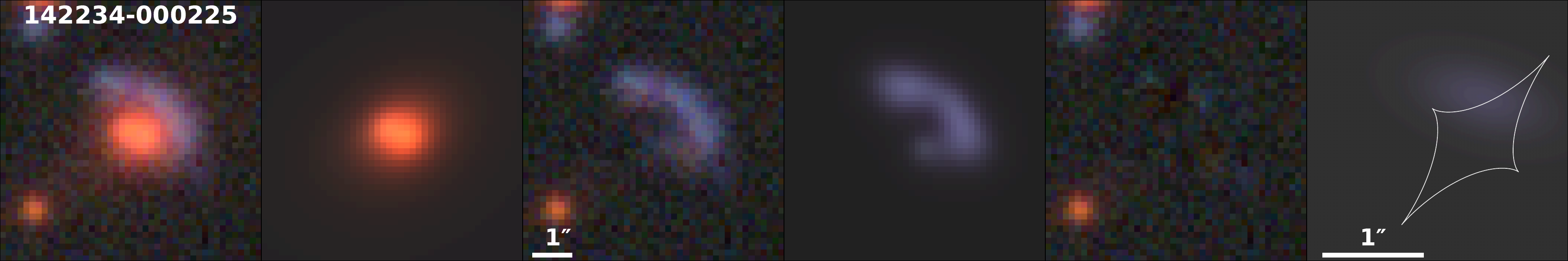}
\includegraphics[width=0.98\textwidth]{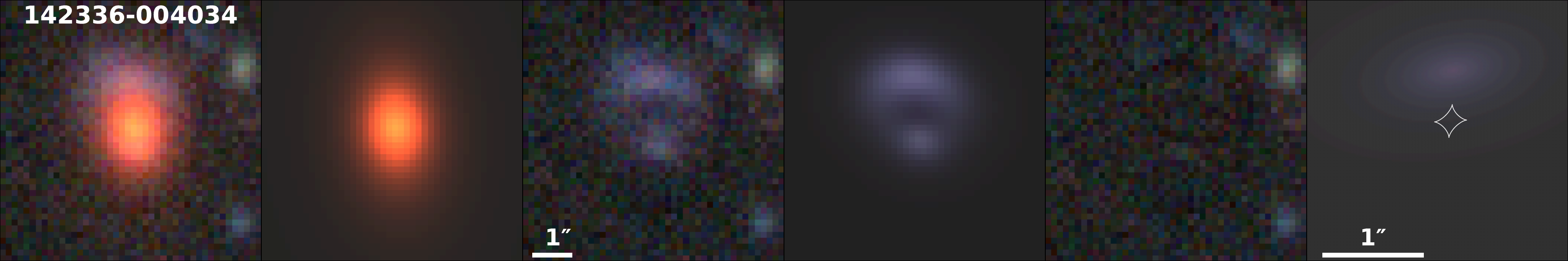}
\caption{\textit{Continued}. System 141105$+$002532 exhibits six blue clumps in a "claymore"-like configuration, well reproduced by a two-component source model.}
\label{fig:continued_figure1_6}
\end{figure*}

\begin{figure*}[htbp]
\ContinuedFloat
\centering
\includegraphics[width=0.98\textwidth]{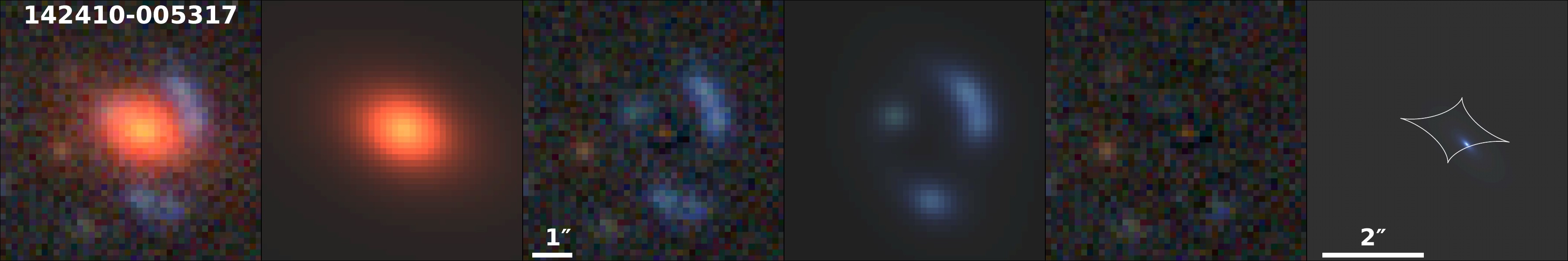}
\includegraphics[width=0.98\textwidth]{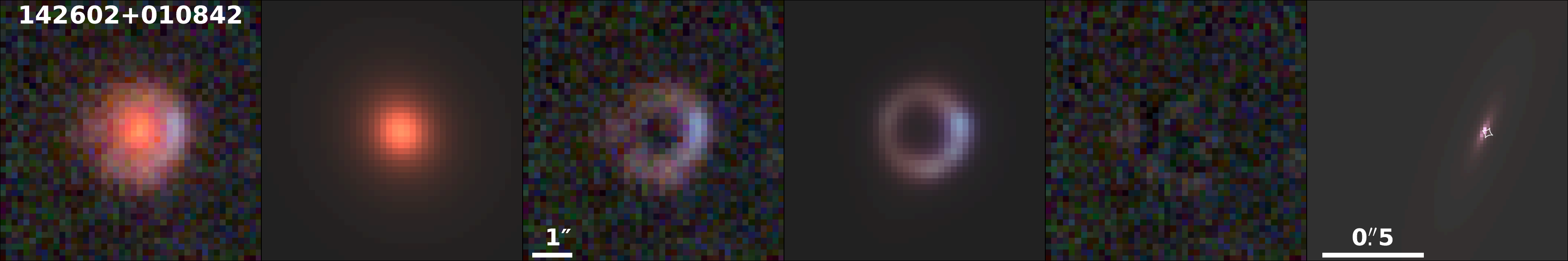}
\includegraphics[width=0.98\textwidth]{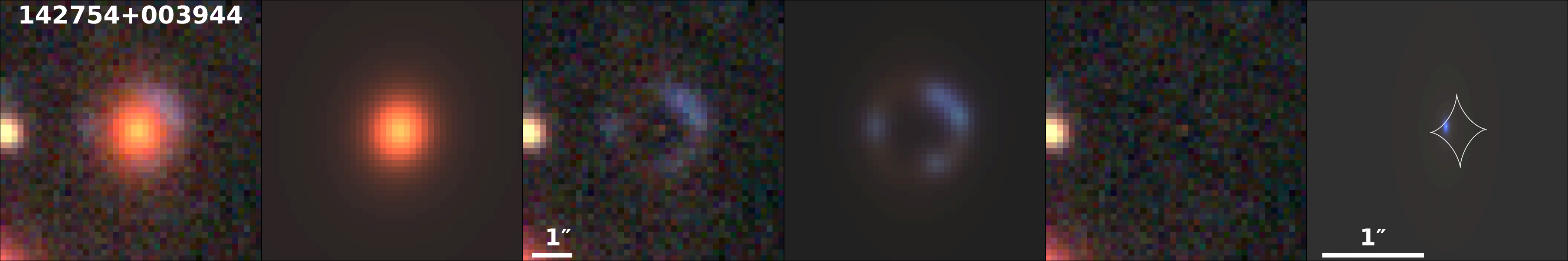}
\includegraphics[width=0.98\textwidth]{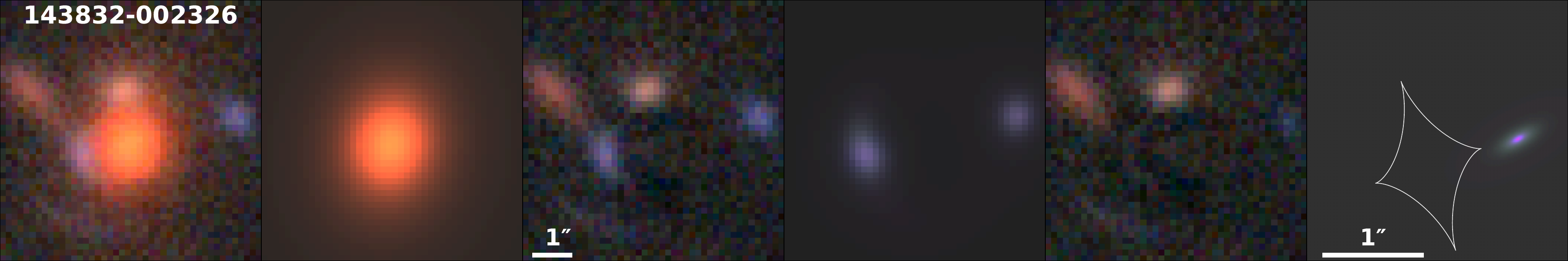}
\includegraphics[width=0.98\textwidth]{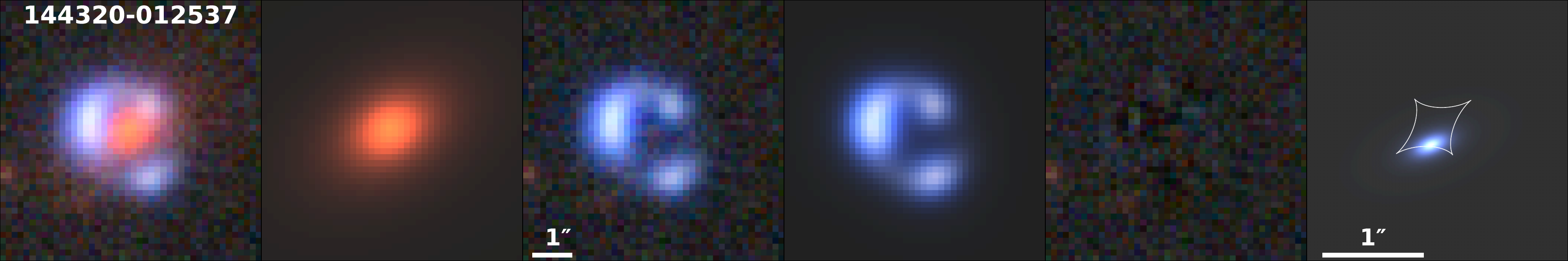}
\includegraphics[width=0.98\textwidth]{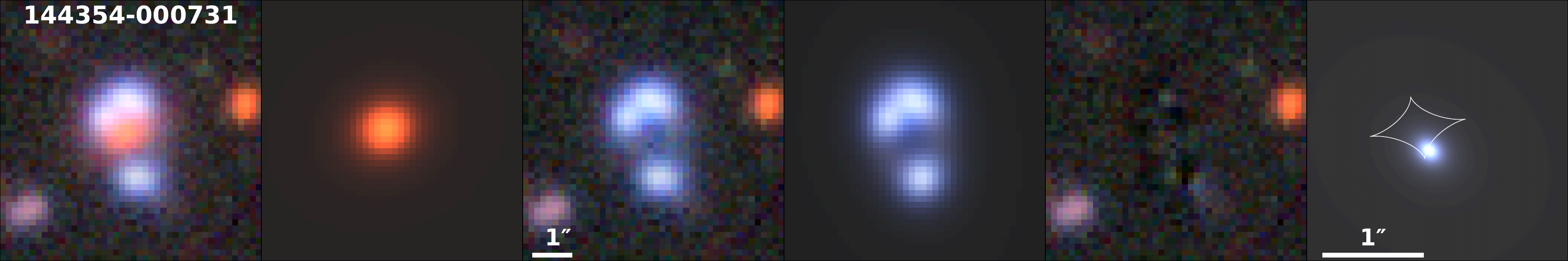}
\includegraphics[width=0.98\textwidth]{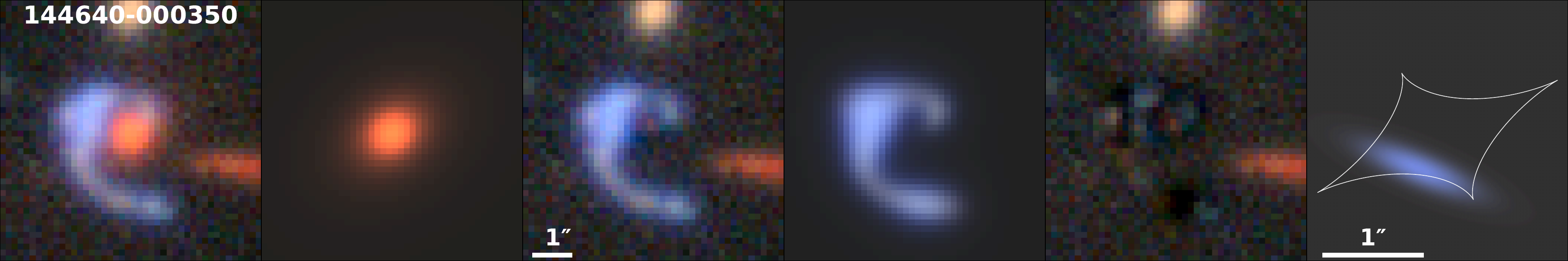}
\includegraphics[width=0.98\textwidth]{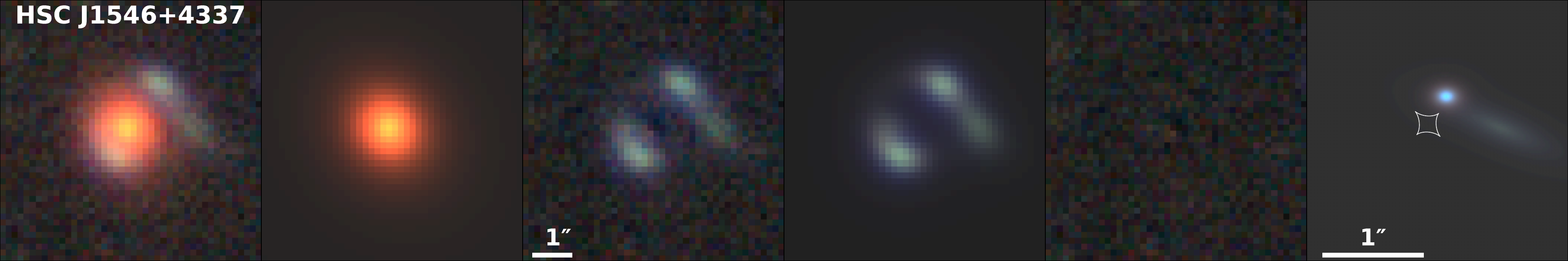}
\makeatletter
\long\def\@makecaption#1#2{%
  \vskip\abovecaptionskip
  {\small\raggedright\noindent\textbf{#1.} #2\par}%
  \vskip\belowcaptionskip
}
\makeatother
\caption{\textit{Continued.}}
\label{fig:continued_figure1_7}
\end{figure*}

\begin{figure*}[htbp]
\ContinuedFloat
\centering
\includegraphics[width=0.98\textwidth]{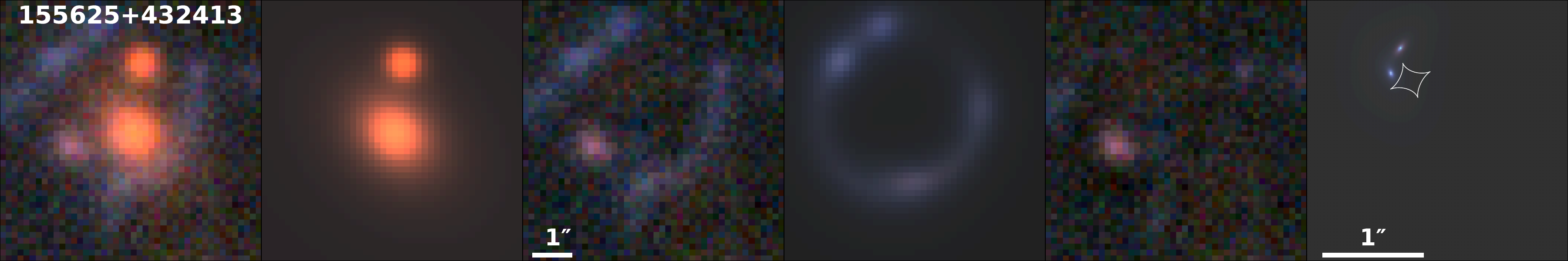}
\includegraphics[width=0.98\textwidth]{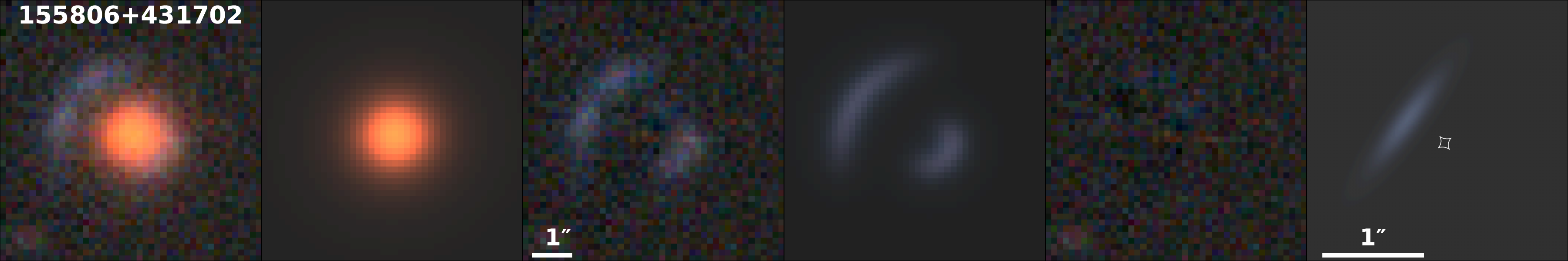}
\includegraphics[width=0.98\textwidth]{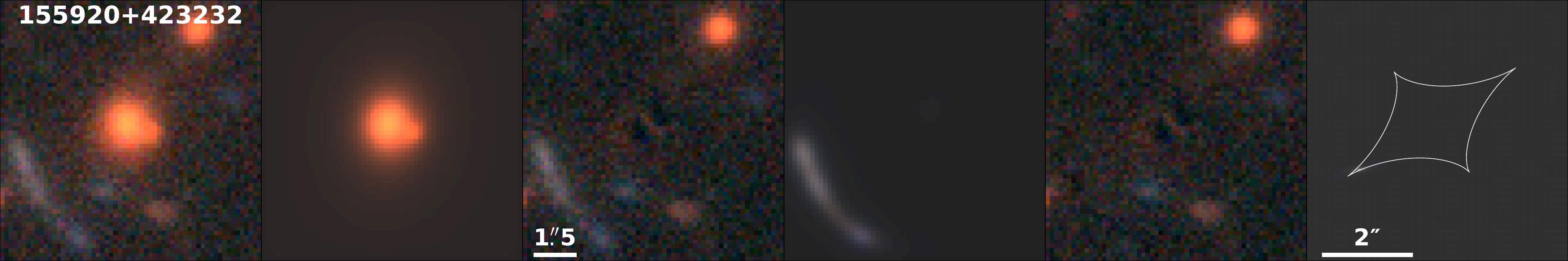}
\includegraphics[width=0.98\textwidth]{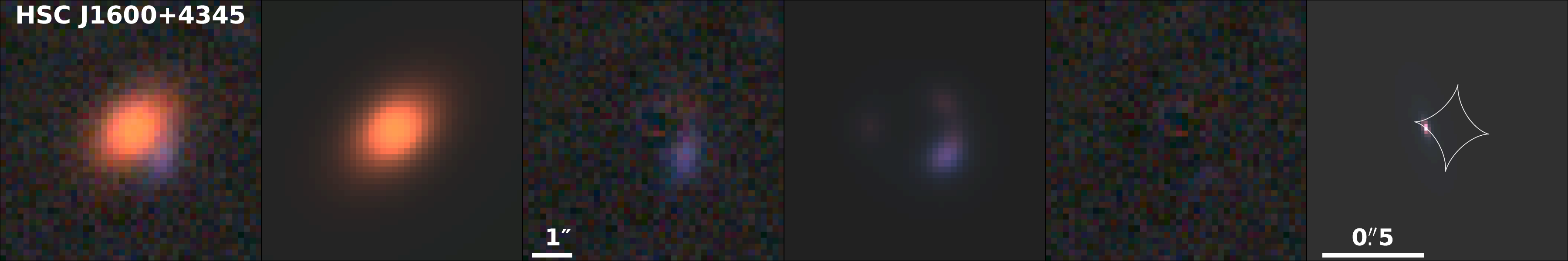}
\includegraphics[width=0.98\textwidth]{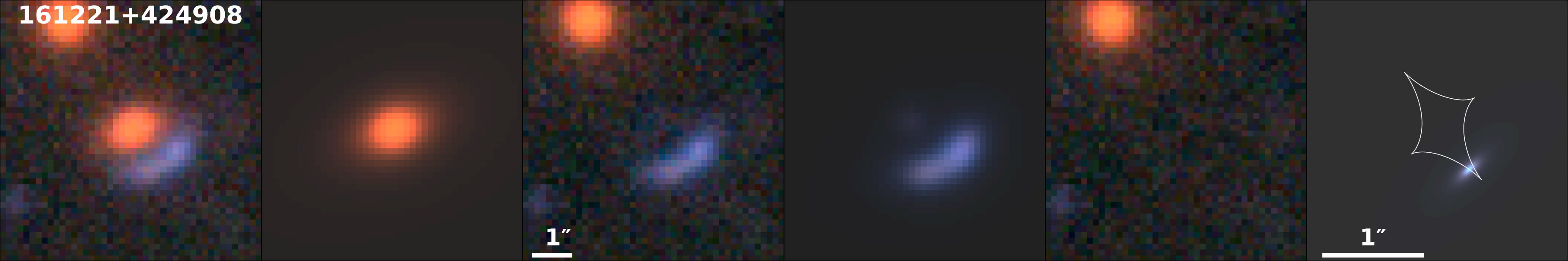}
\includegraphics[width=0.98\textwidth]{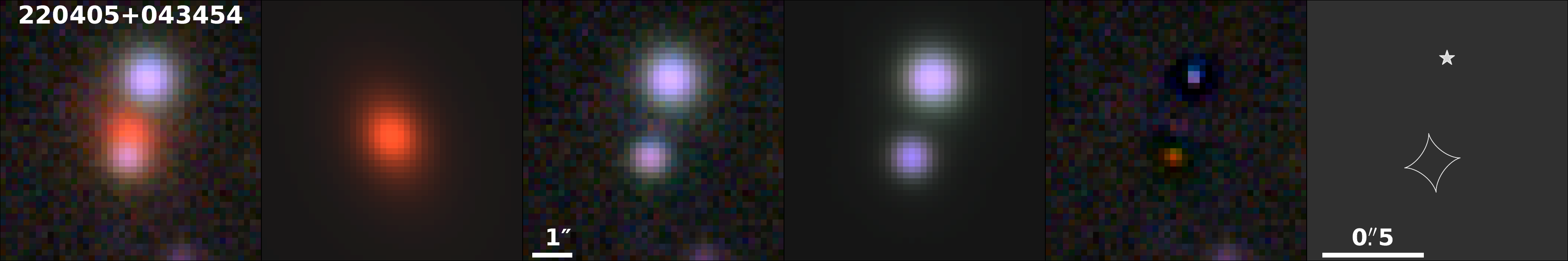}
\includegraphics[width=0.98\textwidth]{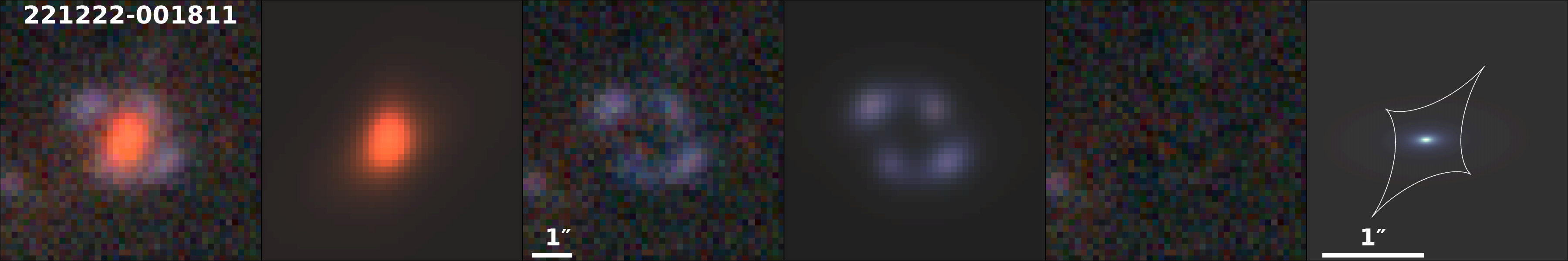}
\includegraphics[width=0.98\textwidth]{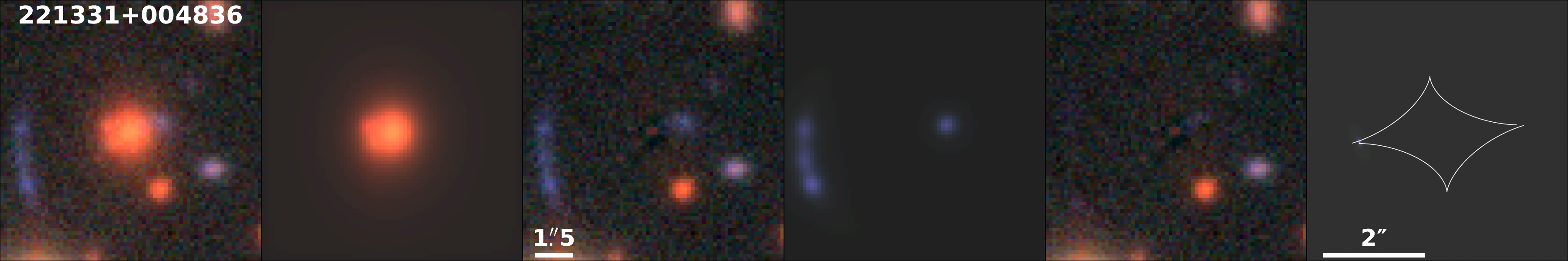}
\caption{\textit{Continued}. For 220405$+$043454, the background source is modeled as a point source, with its reconstructed position marked by an asterisk in the source-plane panel.}
\label{fig:continued_figure1_8}
\end{figure*}

\begin{figure*}[htbp]
\ContinuedFloat
\centering
\includegraphics[width=0.98\textwidth]{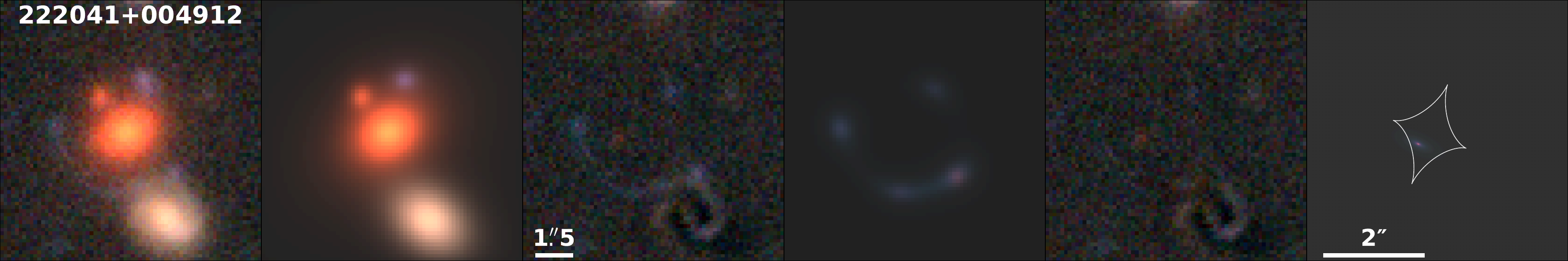}
\includegraphics[width=0.98\textwidth]{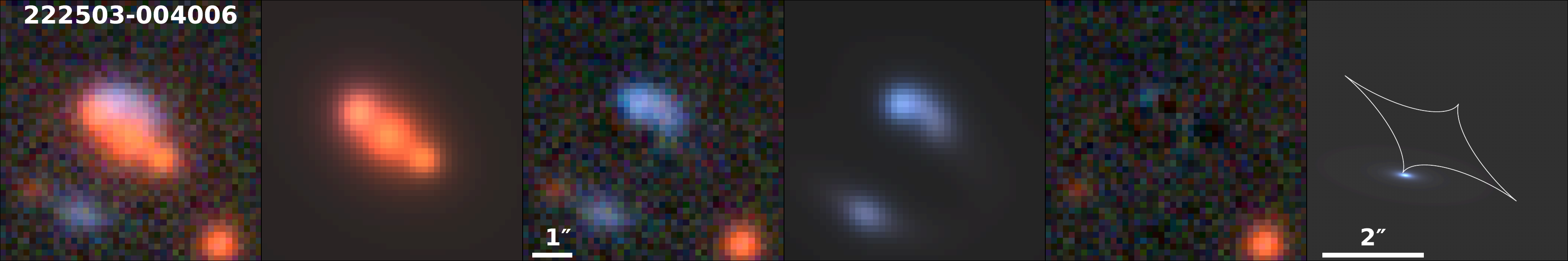}
\includegraphics[width=0.98\textwidth]{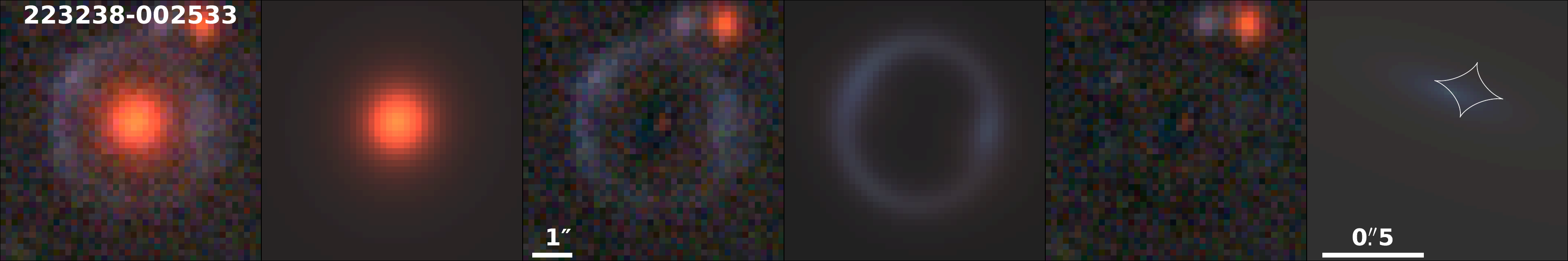}
\includegraphics[width=0.98\textwidth]{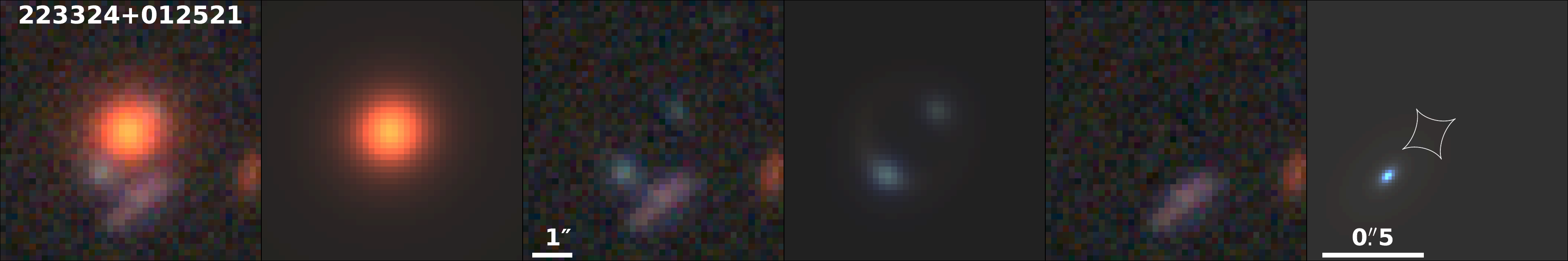}
\includegraphics[width=0.98\textwidth]{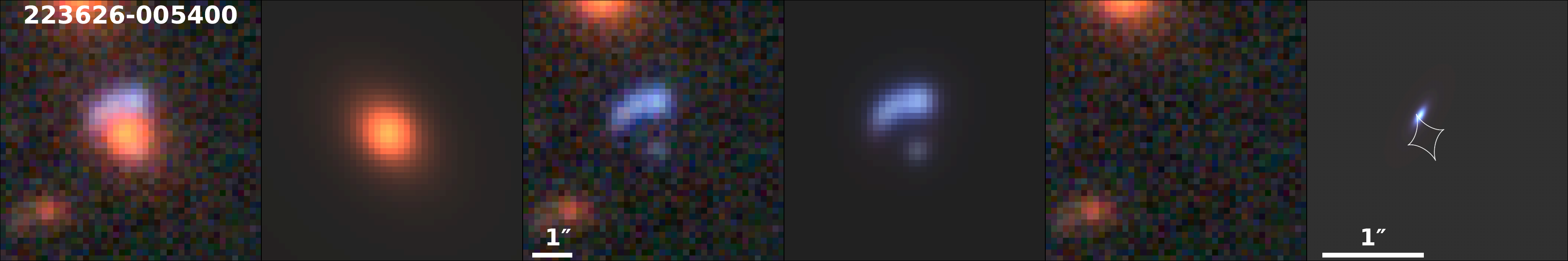}
\includegraphics[width=0.98\textwidth]{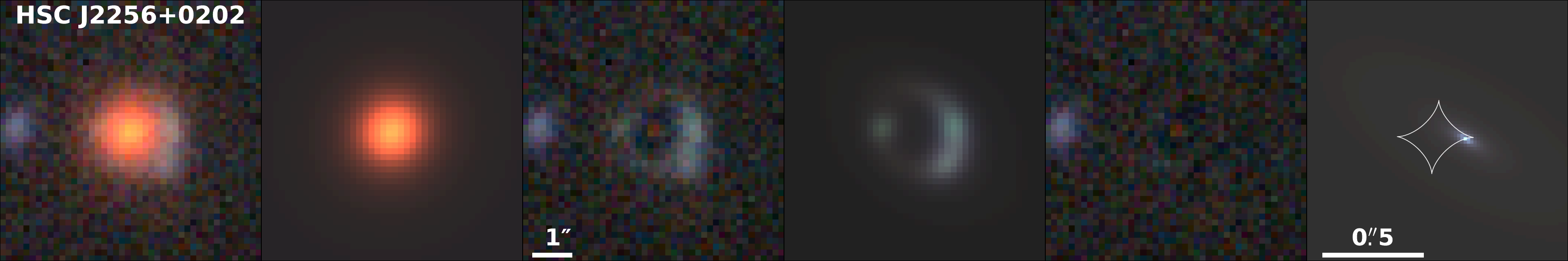}
\includegraphics[width=0.98\textwidth]{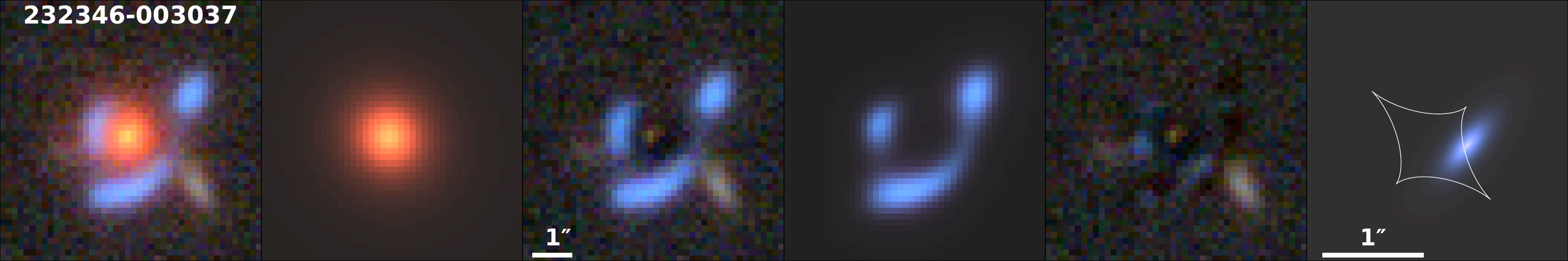}
\includegraphics[width=0.98\textwidth]{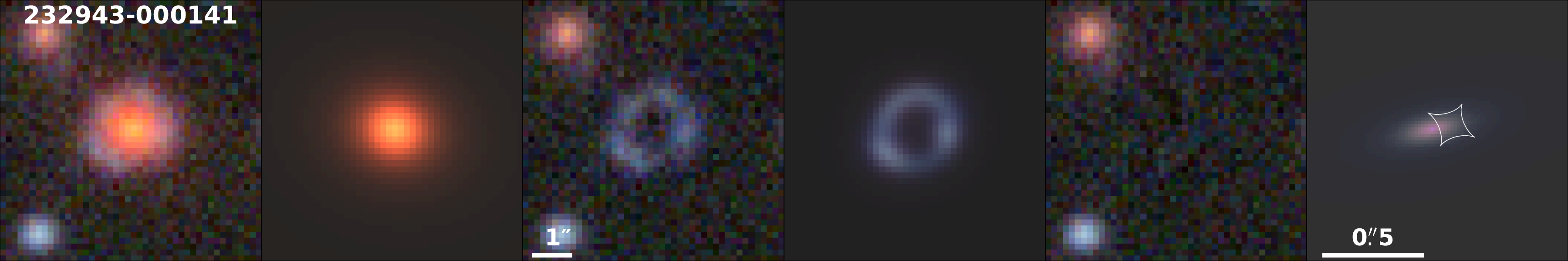}
\caption{\textit{Continued.} The image-plane cutout of 222041$+$004912 is 11\farcs76 $\times$ 11\farcs76, with the light fitting of the neighboring environmental galaxy to the southwest included in the modeling, whose residual map reveals faint double spiral arms indicative of a face-on disk galaxy.}
\label{fig:continued_figure1_9}
\end{figure*}

\section{\label{sec:tables} Summary tables of lens mass, lens light, and source light parameters}

\clearpage
\renewcommand{\thetable}{A\arabic{table}}
\footnotesize
\renewcommand{\arraystretch}{1.35}
\setlength{\tabcolsep}{2pt}
\setlength{\LTleft}{0pt}
\setlength{\LTcapwidth}{\textwidth}
\begin{longtable}{ccccccccc}
\caption{\label{Mass_para}Lens Mass Parameters. \textit{Note.} Columns: (1) system name. (2)--(3) lens and source redshifts, where values marked with $^{\dagger}$ and $^{*}$ are taken from \cite{Jaelani2020} and \cite{sugohi_IV}, respectively. (4)--(6) Einstein radius, axis ratio, and major-axis position angle measured from west. (7)--(8) external shear strength and position angle (east of north). (9) best-fitting stellar velocity dispersion of the lens galaxy and its $1\sigma$ uncertainty. Parameters in Columns (4)--(8) are reported as medians with $1\sigma$ credible intervals derived from MCMC samples.} \\
\hline
Name & $\mathrm{z_L}$ & $\mathrm{z_S}$ & $\theta_{\mathrm{E}}$$(^{\prime\prime})$ & $q_{\mathrm{SIE}}$ & $\phi_{\mathrm{SIE}}(deg)$ & $\gamma$ & $\phi_{\mathrm{gama}}(deg)$ & $\sigma$($\mathrm{km\,s^{-1}}$) \\
\hline
\endfirsthead
\multicolumn{9}{l}{{\bfseries \tablename\ \thetable{} -- continued from previous page}} \\
\hline
Name & $\mathrm{z_L}$ & $\mathrm{z_S}$ & $\theta_{\mathrm{E}}$$(^{\prime\prime})$ & $q_{\mathrm{SIE}}$ & $\phi_{\mathrm{SIE}}(deg)$ & $\gamma$ & $\phi_{\mathrm{gama}}(deg)$ & $\sigma$($\mathrm{km\,s^{-1}}$) \\
\hline
\endhead
\hline
\endlastfoot
085046+003905 & $0.8844 \pm 0.0003$ & $2.83777 \pm 0.00011$ & $1.663^{+0.004}_{-0.003}$ & $0.69^{+0.04}_{-0.02}$ & $79.2^{+0.9}_{-0.9}$ & $0.075^{+0.006}_{-0.004}$ & $3.7^{+1.4}_{-1.8}$ & $307 \pm 43$ \\
083651+003038 & $0.83603 \pm 0.00010$ & $3.0629 \pm 0.0003$ & $1.652^{+0.008}_{-0.007}$ & $0.580^{+0.011}_{-0.013}$ & $51.7^{+0.8}_{-0.7}$ & $0.178^{+0.004}_{-0.004}$ & $17.3^{+1.0}_{-0.8}$ & $232 \pm 30$ \\
090402+031403 & $0.84681 \pm 0.00010$ & $3.3528 \pm 0.0002$ & $0.961^{+0.007}_{-0.009}$ & $0.83^{+0.04}_{-0.04}$ & $144^{+7}_{-7}$ & $0.092^{+0.015}_{-0.015}$ & $164^{+5}_{-4}$ & $193 \pm 24$ \\
090548+004743 & $0.85352 \pm 0.00008$ & $2.91165 \pm 0.00010$ & $1.022^{+0.008}_{-0.008}$ & $0.68^{+0.05}_{-0.04}$ & $120^{+4}_{-4}$ & $0.044^{+0.017}_{-0.018}$ & $115^{+13}_{-13}$ & $252 \pm 30$ \\
HSCJ1104-0052 & $0.87299 \pm 0.00017$ & $2.45593 \pm 0.00019$ & $1.302^{+0.004}_{-0.003}$ & $0.771^{+0.018}_{-0.018}$ & $147^{+2}_{-3}$ & $0.121^{+0.007}_{-0.007}$ & $99.7^{+1.9}_{-1.8}$ & $410 \pm 57$ \\
HSCJ1107-0037 & $0.81443 \pm 0.00010$ & $1.4106 \pm 0.0003$ & $1.370^{+0.019}_{-0.024}$ & $0.44^{+0.04}_{-0.03}$ & $155.9^{+1.9}_{-1.4}$ & $0.27^{+0.02}_{-0.02}$ & $162^{+2}_{-1}$ & $191 \pm 27$ \\
115630-020027 & $0.9983 \pm 0.0002$ & $2.2298 \pm 0.0005$ & $0.654^{+0.008}_{-0.010}$ & $0.82^{+0.04}_{-0.04}$ & $70^{+6}_{-6}$ & $0.076^{+0.012}_{-0.012}$ & $135^{+5}_{-4}$ & $199 \pm 48$ \\
120657-010241 & $0.83004 \pm 0.00015$ & $2.7732 \pm 0.0003$ & $0.68^{+0.02}_{-0.03}$ & $0.82^{+0.06}_{-0.10}$ & $173^{+13}_{-14}$ & $0.11^{+0.04}_{-0.02}$ & $13^{+4}_{-4}$ & $146 \pm 35$ \\
\end{longtable}

\renewcommand{\arraystretch}{1.35}
\setlength{\tabcolsep}{1.5pt}
\setlength{\LTleft}{0pt}
\begin{longtable}{cccccccccc}
\caption{\label{lens_light_para}Lens Light Parameters. \textit{Note.} Columns: (1)--(2) system name and filter. (3)--(9) effective radius, S\'ersic index, axis ratio, position angle (north of west), S\'ersic center offsets from the cutout center (cutout center R.A. and decl. in degrees in parentheses), and apparent AB magnitude, reported as MCMC posterior medians with $1\sigma$ credible intervals. (10) stellar mass of the lens galaxy and its uncertainty, given as the likelihood-weighted mean and standard deviation from \texttt{CIGALE}. Systems 120849$+$012845 and HSC\,J1600$+$4345 adopt concentric $n=4$ + $n=1$ elliptical S\'ersic profiles with shared ellipticities.} \\
\hline
Name & Filter & $R_{\mathrm{eff,lens}}$$(^{\prime\prime})$ & $n_{\mathrm{lens}}$ & $q_{\mathrm{lens}}$ & $\phi_{\mathrm{lens}}(deg)$ & $\Delta x$$(^{\prime\prime})$ & $\Delta y$$(^{\prime\prime})$ & $m_{AB}(mag)$ & $M_*(10^{10} M_\odot)$ \\
\hline
\endfirsthead
\multicolumn{10}{l}{{\bfseries \tablename\ \thetable{} -- continued from previous page}} \\
\hline
Name & Filter & $R_{\mathrm{eff,lens}}$$(^{\prime\prime})$ & $n_{\mathrm{lens}}$ & $q_{\mathrm{lens}}$ & $\phi_{\mathrm{lens}}(deg)$ & $\Delta x$$(^{\prime\prime})$ & $\Delta y$$(^{\prime\prime})$ & $m_{AB}(mag)$ & $M_*(10^{10} M_\odot)$ \\
\hline
\endhead
\hline
\endlastfoot
\hline
 & $g$ & $1.90^{+0.07}_{-0.09}$ &  &  &  &  &  & $23.59^{+0.05}_{-0.04}$ &  \\*
 & $r$ & $0.98^{+0.06}_{-0.06}$ &  &  &  &  &  & $22.14^{+0.05}_{-0.05}$ &  \\*
085046+003905 & $i$ & $0.84^{+0.02}_{-0.03}$ & $4$ & $0.725^{+0.013}_{-0.011}$ & $92.8^{+1.3}_{-1.5}$ & $0.15406^{+0.00018}_{-0.00017}$ & $0.0980^{+0.0002}_{-0.0002}$ & $20.93^{+0.02}_{-0.03}$ & $41 \pm 8$ \\*
 & $z$ & $0.80^{+0.03}_{-0.03}$ &  &  &  & $(132.694211)$ & $(0.651490)$ & $20.11^{+0.02}_{-0.03}$ &  \\*
 & $y$ & $0.84^{+0.03}_{-0.04}$ &  &  &  &  &  & $19.86^{+0.03}_{-0.03}$ &  \\
\hline
 & $g$ & $0.54^{+0.07}_{-0.06}$ &  &  &  &  &  & $24.74^{+0.09}_{-0.08}$ &  \\*
 & $r$ & $0.94^{+0.06}_{-0.06}$ &  &  &  &  &  & $22.37^{+0.05}_{-0.05}$ &  \\*
 & $i$ & $0.88^{+0.05}_{-0.04}$ & $4$ & $0.717^{+0.017}_{-0.015}$ & $88.2^{+1.9}_{-2.2}$ & $0.046^{+0.004}_{-0.004}$ & $0.088^{+0.009}_{-0.009}$ & $20.91^{+0.04}_{-0.03}$ & $57 \pm 11$ \\
\end{longtable}

\renewcommand{\arraystretch}{1.35}
\setlength{\tabcolsep}{2pt}
\setlength{\LTleft}{0pt}
\setlength{\LTcapwidth}{\textwidth}
\begin{longtable}{cccccccccc}
\caption{\label{source_para}Source Light Parameters. \textit{Note.} Columns: (1)--(2) system name and filter. (3)--(9) effective radius, S\'ersic index, axis ratio, position angle (north of west), S\'ersic center offsets relative to the cutout center (cutout-center R.A. and decl. in degrees in parentheses), and apparent AB magnitude, reported as MCMC posterior medians with $1\sigma$ credible intervals. (10) best-fit magnification. The background source in system 220405$+$043454 is modeled as a point source.} \\
\hline
Name & Filter & $R_{\mathrm{eff,source}}$$(^{\prime\prime})$ & $n_{\mathrm{source}}$ & $q_{\mathrm{source}}$ & $\phi_{\mathrm{source}}(deg)$ & $\Delta x$$(^{\prime\prime})$ & $\Delta y$$(^{\prime\prime})$ & $m_{AB}(mag)$ & $\mu$ \\
\hline
\endfirsthead
\multicolumn{10}{l}{{\bfseries \tablename\ \thetable{} -- continued from previous page}} \\
\hline
Name & Filter & $R_{\mathrm{eff,source}}$$(^{\prime\prime})$ & $n_{\mathrm{source}}$ & $q_{\mathrm{source}}$ & $\phi_{\mathrm{source}}(deg)$ & $\Delta x$$(^{\prime\prime})$ & $\Delta y$$(^{\prime\prime})$ & $m_{AB}(mag)$ & $\mu$ \\
\hline
\endhead
\hline
\endlastfoot
\hline
 & $g$ & $0.087^{+0.008}_{-0.012}$ & $5.6^{+0.3}_{-0.3}$ &  &  &  &  & $25.24^{+0.08}_{-0.06}$ & $12.61$ \\*
 & $r$ & $0.064^{+0.009}_{-0.008}$ & $5.2^{+0.3}_{-0.3}$ &  &  &  &  & $25.09^{+0.08}_{-0.06}$ & $13.65$ \\*
085046+003905 & $i$ & $0.216^{+0.009}_{-0.013}$ & $7.89^{+0.08}_{-0.16}$ & $0.51^{+0.04}_{-0.03}$ & $70^{+3}_{-2}$ & $0.114^{+0.003}_{-0.002}$ & $0.343^{+0.005}_{-0.009}$ & $24.62^{+0.07}_{-0.05}$ & $10.50$ \\*
 & $z$ & $0.6690^{+0.0020}_{-0.0048}$ & $7.87^{+0.09}_{-0.17}$ &  &  & $(132.694211)$ & $(0.651490)$ & $24.07^{+0.05}_{-0.04}$ & $8.12$ \\*
 & $y$ & $0.667^{+0.004}_{-0.008}$ & $7.89^{+0.08}_{-0.16}$ &  &  &  &  & $24.03^{+0.06}_{-0.05}$ & $8.14$ \\
\hline
 & $g$ & $0.055^{+0.003}_{-0.003}$ & $0.52^{+0.04}_{-0.02}$ &  &  &  &  & $25.04^{+0.03}_{-0.03}$ & $7.86$ \\*
 & $r$ & $0.054^{+0.003}_{-0.003}$ & $0.54^{+0.07}_{-0.03}$ &  &  &  &  & $24.67^{+0.03}_{-0.03}$ & $7.75$ \\*
083651+003038 & $i$ & $0.056^{+0.004}_{-0.003}$ & $0.54^{+0.05}_{-0.03}$ & $0.243^{+0.017}_{-0.013}$ & $72.0^{+1.3}_{-1.0}$ & $0.022^{+0.006}_{-0.006}$ & $0.735^{+0.010}_{-0.010}$ & $24.57^{+0.03}_{-0.03}$ & $7.89$ \\
\end{longtable}


\bibliography{sample702}{}

@PROCEEDINGS{Toomre1977,
        title = "{Evolution of galaxies and stellar populations}",
    booktitle = {Evolution of Galaxies and Stellar Populations},
         year = 1977,
       editor = {{Tinsley}, Beatrice M. and {Larson}, D. Campbell, Richard B. Gehret},
        month = jan,
       adsurl = {https://ui.adsabs.harvard.edu/abs/1977egsp.conf.....T}
}

@ARTICLE{White1978,
       author = {{White}, S.~D.~M. and {Rees}, M.~J.},
        title = "{Core condensation in heavy halos: a two-stage theory for galaxy formation and clustering.}",
      journal = {\mnras},
         year = 1978,
        month = may,
       volume = {183},
        pages = {341-358},
          doi = {10.1093/mnras/183.3.341},
       adsurl = {https://ui.adsabs.harvard.edu/abs/1978MNRAS.183..341W}
}

@ARTICLE{White1991,
       author = {{White}, Simon D.~M. and {Frenk}, Carlos S.},
        title = "{Galaxy Formation through Hierarchical Clustering}",
      journal = {\apj},
         year = 1991,
        month = sep,
       volume = {379},
        pages = {52},
          doi = {10.1086/170483},
       adsurl = {https://ui.adsabs.harvard.edu/abs/1991ApJ...379...52W}
}

@ARTICLE{Cole1994,
       author = {{Cole}, S. and {Aragon-Salamanca}, A. and {Frenk}, C.~S. and {Navarro}, J.~F. and {Zepf}, S.~E.},
        title = "{A recipe for galaxy formation.}",
      journal = {\mnras},
         year = 1994,
        month = dec,
       volume = {271},
        pages = {781-806},
          doi = {10.1093/mnras/271.4.781},
archivePrefix = {arXiv},
       eprint = {astro-ph/9402001},
 primaryClass = {astro-ph},
       adsurl = {https://ui.adsabs.harvard.edu/abs/1994MNRAS.271..781C}
}

@ARTICLE{Cole2000,
       author = {{Cole}, Shaun and {Lacey}, Cedric G. and {Baugh}, Carlton M. and {Frenk}, Carlos S.},
        title = "{Hierarchical galaxy formation}",
      journal = {\mnras},
         year = 2000,
        month = nov,
       volume = {319},
       number = {1},
        pages = {168-204},
          doi = {10.1046/j.1365-8711.2000.03879.x},
archivePrefix = {arXiv},
       eprint = {astro-ph/0007281},
 primaryClass = {astro-ph},
       adsurl = {https://ui.adsabs.harvard.edu/abs/2000MNRAS.319..168C}
}

@ARTICLE{Ma2004,
       author = {{Ma}, Chung-Pei and {Boylan-Kolchin}, Michael},
        title = "{Are Halos of Collisionless Cold Dark Matter Collisionless?}",
      journal = {\prl},
         year = 2004,
        month = jul,
       volume = {93},
       number = {2},
          eid = {021301},
        pages = {021301},
          doi = {10.1103/PhysRevLett.93.021301},
archivePrefix = {arXiv},
       eprint = {astro-ph/0403102},
 primaryClass = {astro-ph},
       adsurl = {https://ui.adsabs.harvard.edu/abs/2004PhRvL..93b1301M}
}

@ARTICLE{Naab2009,
       author = {{Naab}, Thorsten and {Johansson}, Peter H. and {Ostriker}, Jeremiah P.},
        title = "{Minor Mergers and the Size Evolution of Elliptical Galaxies}",
      journal = {\apjl},
         year = 2009,
        month = jul,
       volume = {699},
       number = {2},
        pages = {L178-L182},
          doi = {10.1088/0004-637X/699/2/L178},
archivePrefix = {arXiv},
       eprint = {0903.1636},
 primaryClass = {astro-ph.CO},
       adsurl = {https://ui.adsabs.harvard.edu/abs/2009ApJ...699L.178N}
}

@ARTICLE{Duffy2010,
       author = {{Duffy}, Alan R. and {Schaye}, Joop and {Kay}, Scott T. and {Dalla Vecchia}, Claudio and {Battye}, Richard A. and {Booth}, C.~M.},
        title = "{Impact of baryon physics on dark matter structures: a detailed simulation study of halo density profiles}",
      journal = {\mnras},
         year = 2010,
        month = jul,
       volume = {405},
       number = {4},
        pages = {2161-2178},
          doi = {10.1111/j.1365-2966.2010.16613.x},
archivePrefix = {arXiv},
       eprint = {1001.3447},
 primaryClass = {astro-ph.CO},
       adsurl = {https://ui.adsabs.harvard.edu/abs/2010MNRAS.405.2161D}
}

@ARTICLE{Velliscig2014,
       author = {{Velliscig}, Marco and {van Daalen}, Marcel P. and {Schaye}, Joop and {McCarthy}, Ian G. and {Cacciato}, Marcello and {Le Brun}, Amandine M.~C. and {Dalla Vecchia}, Claudio},
        title = "{The impact of galaxy formation on the total mass, mass profile and abundance of haloes}",
      journal = {\mnras},
         year = 2014,
        month = aug,
       volume = {442},
       number = {3},
        pages = {2641-2658},
          doi = {10.1093/mnras/stu1044},
archivePrefix = {arXiv},
       eprint = {1402.4461},
 primaryClass = {astro-ph.CO},
       adsurl = {https://ui.adsabs.harvard.edu/abs/2014MNRAS.442.2641V}
}

@ARTICLE{Remus2017,
       author = {{Remus}, Rhea-Silvia and {Dolag}, Klaus and {Naab}, Thorsten and {Burkert}, Andreas and {Hirschmann}, Michaela and {Hoffmann}, Tadziu L. and {Johansson}, Peter H.},
        title = "{The co-evolution of total density profiles and central dark matter fractions in simulated early-type galaxies}",
      journal = {\mnras},
         year = 2017,
        month = jan,
       volume = {464},
       number = {3},
        pages = {3742-3756},
          doi = {10.1093/mnras/stw2594},
archivePrefix = {arXiv},
       eprint = {1603.01619},
 primaryClass = {astro-ph.GA},
       adsurl = {https://ui.adsabs.harvard.edu/abs/2017MNRAS.464.3742R}
}

@ARTICLE{Peirani2019,
       author = {{Peirani}, S{\'e}bastien and {Sonnenfeld}, Alessandro and {Gavazzi}, Rapha{\"e}l and {Oguri}, Masamune and {Dubois}, Yohan and {Silk}, Joe and {Pichon}, Christophe and {Devriendt}, Julien and {Kaviraj}, Sugata},
        title = "{Total density profile of massive early-type galaxies in HORIZON-AGN simulation: impact of AGN feedback and comparison with observations}",
      journal = {\mnras},
         year = 2019,
        month = mar,
       volume = {483},
       number = {4},
        pages = {4615-4627},
          doi = {10.1093/mnras/sty3475},
archivePrefix = {arXiv},
       eprint = {1801.09754},
 primaryClass = {astro-ph.GA},
       adsurl = {https://ui.adsabs.harvard.edu/abs/2019MNRAS.483.4615P}
}

@ARTICLE{Wang2019,
       author = {{Wang}, Yunchong and {Vogelsberger}, Mark and {Xu}, Dandan and {Shen}, Xuejian and {Mao}, Shude and {Barnes}, David and {Li}, Hui and {Marinacci}, Federico and {Torrey}, Paul and {Springel}, Volker and {Hernquist}, Lars},
        title = "{Early-type galaxy density profiles from IllustrisTNG - II. Evolutionary trend of the total density profile}",
      journal = {\mnras},
         year = 2019,
        month = oct,
       volume = {490},
       number = {4},
        pages = {5722-5738},
          doi = {10.1093/mnras/stz2907},
archivePrefix = {arXiv},
       eprint = {1910.08552},
 primaryClass = {astro-ph.GA},
       adsurl = {https://ui.adsabs.harvard.edu/abs/2019MNRAS.490.5722W}
}

@ARTICLE{Ruff2011,
       author = {{Ruff}, Andrea J. and {Gavazzi}, Rapha{\"e}l and {Marshall}, Philip J. and {Treu}, Tommaso and {Auger}, Matthew W. and {Brault}, Florence},
        title = "{The SL2S Galaxy-scale Lens Sample. II. Cosmic Evolution of Dark and Luminous Mass in Early-type Galaxies}",
      journal = {\apj},
         year = 2011,
        month = feb,
       volume = {727},
       number = {2},
          eid = {96},
        pages = {96},
          doi = {10.1088/0004-637X/727/2/96},
archivePrefix = {arXiv},
       eprint = {1008.3167},
 primaryClass = {astro-ph.CO},
       adsurl = {https://ui.adsabs.harvard.edu/abs/2011ApJ...727...96R}
}

@ARTICLE{Bolton2012,
       author = {{Bolton}, Adam S. and {Brownstein}, Joel R. and {Kochanek}, Christopher S. and {Shu}, Yiping and {Schlegel}, David J. and {Eisenstein}, Daniel J. and {Wake}, David A. and {Connolly}, Natalia and {Maraston}, Claudia and {Arneson}, Ryan A. and {Weaver}, Benjamin A.},
        title = "{The BOSS Emission-Line Lens Survey. II. Investigating Mass-density Profile Evolution in the SLACS+BELLS Strong Gravitational Lens Sample}",
      journal = {\apj},
         year = 2012,
        month = sep,
       volume = {757},
       number = {1},
          eid = {82},
        pages = {82},
          doi = {10.1088/0004-637X/757/1/82},
archivePrefix = {arXiv},
       eprint = {1201.2988},
 primaryClass = {astro-ph.CO},
       adsurl = {https://ui.adsabs.harvard.edu/abs/2012ApJ...757...82B}
}

@ARTICLE{Etherington2023,
       author = {{Etherington}, Amy and {Nightingale}, James W. and {Massey}, Richard and {Robertson}, Andrew and {Cao}, XiaoYue and {Amvrosiadis}, Aristeidis and {Cole}, Shaun and {Frenk}, Carlos S. and {He}, Qiuhan and {Lagattuta}, David J. and {Lange}, Samuel and {Li}, Ran},
        title = "{Beyond the bulge-halo conspiracy? Density profiles of early-type galaxies from extended-source strong lensing}",
      journal = {\mnras},
         year = 2023,
        month = jun,
       volume = {521},
       number = {4},
        pages = {6005-6018},
          doi = {10.1093/mnras/stad582},
archivePrefix = {arXiv},
       eprint = {2207.04070},
 primaryClass = {astro-ph.GA},
       adsurl = {https://ui.adsabs.harvard.edu/abs/2023MNRAS.521.6005E}
}

@ARTICLE{Poci2017,
       author = {{Poci}, Adriano and {Cappellari}, Michele and {McDermid}, Richard M.},
        title = "{Systematic trends in total-mass profiles from dynamical models of early-type galaxies}",
      journal = {\mnras},
         year = 2017,
        month = may,
       volume = {467},
       number = {2},
        pages = {1397-1413},
          doi = {10.1093/mnras/stx101},
archivePrefix = {arXiv},
       eprint = {1612.05805},
 primaryClass = {astro-ph.GA},
       adsurl = {https://ui.adsabs.harvard.edu/abs/2017MNRAS.467.1397P}
}

@ARTICLE{Kaviraj2017,
       author = {{Kaviraj}, S. and {Laigle}, C. and {Kimm}, T. and {Devriendt}, J.~E.~G. and {Dubois}, Y. and {Pichon}, C. and {Slyz}, A. and {Chisari}, E. and {Peirani}, S.},
        title = "{The Horizon-AGN simulation: evolution of galaxy properties over cosmic time}",
      journal = {\mnras},
         year = 2017,
        month = jun,
       volume = {467},
       number = {4},
        pages = {4739-4752},
          doi = {10.1093/mnras/stx126},
archivePrefix = {arXiv},
       eprint = {1605.09379},
 primaryClass = {astro-ph.GA},
       adsurl = {https://ui.adsabs.harvard.edu/abs/2017MNRAS.467.4739K}
}

@ARTICLE{Pillepich2018,
       author = {{Pillepich}, Annalisa and {Springel}, Volker and {Nelson}, Dylan and {Genel}, Shy and {Naiman}, Jill and {Pakmor}, R{\"u}diger and {Hernquist}, Lars and {Torrey}, Paul and {Vogelsberger}, Mark and {Weinberger}, Rainer and {Marinacci}, Federico},
        title = "{Simulating galaxy formation with the IllustrisTNG model}",
      journal = {\mnras},
         year = 2018,
        month = jan,
       volume = {473},
       number = {3},
        pages = {4077-4106},
          doi = {10.1093/mnras/stx2656},
archivePrefix = {arXiv},
       eprint = {1703.02970},
 primaryClass = {astro-ph.GA},
       adsurl = {https://ui.adsabs.harvard.edu/abs/2018MNRAS.473.4077P}
}

@ARTICLE{Derkenne2021,
       author = {{Derkenne}, Caro and {McDermid}, Richard M. and {Poci}, Adriano and {Remus}, Rhea-Silvia and {J{\o}rgensen}, Inger and {Emsellem}, Eric},
        title = "{Total mass density slopes of early-type galaxies using Jeans dynamical modelling at redshifts 0.29 < z < 0.55}",
      journal = {\mnras},
         year = 2021,
        month = sep,
       volume = {506},
       number = {3},
        pages = {3691-3716},
          doi = {10.1093/mnras/stab1996},
archivePrefix = {arXiv},
       eprint = {2107.05206},
 primaryClass = {astro-ph.GA},
       adsurl = {https://ui.adsabs.harvard.edu/abs/2021MNRAS.506.3691D}
}

@ARTICLE{Derkenne2023,
       author = {{Derkenne}, Caro and {McDermid}, Richard M. and {Poci}, Adriano and {Mendel}, J. Trevor and {D'Eugenio}, Francesco and {Jeon}, Seyoung and {Remus}, Rhea-Silvia and {Bellstedt}, Sabine and {Battisti}, Andrew J. and {Bland-Hawthorn}, Joss and {Ferr{\'e}-Mateu}, Anna and {Foster}, Caroline and {Harborne}, K.~E. and {Lagos}, Claudia D.~P. and {Peng}, Yingjie and {Sharda}, Piyush and {Sharma}, Gauri and {Sweet}, Sarah and {Tran}, Kim-Vy H. and {Valenzuela}, Lucas M. and {Vaughan}, Sam and {Wisnioski}, Emily and {Yi}, Sukyoung K.},
        title = "{The MAGPI Survey: impact of environment on the total internal mass distribution of galaxies in the last 5 Gyr}",
      journal = {\mnras},
         year = 2023,
        month = jul,
       volume = {522},
       number = {3},
        pages = {3602-3626},
          doi = {10.1093/mnras/stad1079},
archivePrefix = {arXiv},
       eprint = {2306.09630},
 primaryClass = {astro-ph.GA},
       adsurl = {https://ui.adsabs.harvard.edu/abs/2023MNRAS.522.3602D}
}

@ARTICLE{Treu2004,
       author = {{Treu}, Tommaso and {Koopmans}, L{\'e}on V.~E.},
        title = "{Massive Dark Matter Halos and Evolution of Early-Type Galaxies to z \raisebox{-0.5ex}\textasciitilde 1}",
      journal = {\apj},
         year = 2004,
        month = aug,
       volume = {611},
       number = {2},
        pages = {739-760},
          doi = {10.1086/422245},
archivePrefix = {arXiv},
       eprint = {astro-ph/0401373},
 primaryClass = {astro-ph},
       adsurl = {https://ui.adsabs.harvard.edu/abs/2004ApJ...611..739T}
}

@ARTICLE{Sahu2024,
       author = {{Sahu}, Nandini and {Tran}, Kim-Vy and {Suyu}, Sherry H. and {Shajib}, Anowar J. and {Ertl}, Sebastian and {Kacprzak}, Glenn G. and {Glazebrook}, Karl and {Jones}, Tucker and {G.~C.}, Keerthi Vasan and {Barone}, Tania M. and {Baker}, A. Makai and {Skobe}, Hannah and {Derkenne}, Caro and {Lewis}, Geraint F. and {Sweet}, Sarah M. and {Lopez}, Sebastian},
        title = "{AGEL: Is the Conflict Real? Investigating Galaxy Evolution Models Using Strong Lensing at 0.3 < z < 0.9}",
      journal = {\apj},
         year = 2024,
        month = jul,
       volume = {970},
       number = {1},
          eid = {86},
        pages = {86},
          doi = {10.3847/1538-4357/ad4ce3},
archivePrefix = {arXiv},
       eprint = {2405.15427},
 primaryClass = {astro-ph.GA},
       adsurl = {https://ui.adsabs.harvard.edu/abs/2024ApJ...970...86S}
}

@ARTICLE{Huang2025,
       author = {{Huang}, Xiaosheng and {Alvarez-Garcia}, David and {Ubeda}, Monica and {Bhamre}, Vikram and {Xu}, Sean and {Baltasar}, S. and {Ratier-Werbin}, N. and {Urcelay}, F. and {Agarwal}, S. and {Cikota}, A. and {Hsu}, Y. and {Lin}, E. and {Schlegel}, D.~J. and {Silver}, E. and {Storfer}, C.~J. and {Tamargo-Arizmendi}, M.},
        title = "{DESI Strong Lens Foundry V: A Sample of HST-Observed Strong Lenses Modeled with GIGA-Lens}",
      journal = {arXiv e-prints},
         year = 2025,
        month = dec,
          eid = {arXiv:2512.07823},
        pages = {arXiv:2512.07823},
          doi = {10.48550/arXiv.2512.07823},
archivePrefix = {arXiv},
       eprint = {2512.07823},
 primaryClass = {astro-ph.CO},
       adsurl = {https://ui.adsabs.harvard.edu/abs/2025arXiv251207823H}
}

@ARTICLE{DES,
       author = {{Dark Energy Survey Collaboration} and {Abbott}, T. and {Abdalla}, F.~B. and {Aleksi{\'c}}, J. and {Allam}, S. and {Amara}, A. and {Bacon}, D. and {Balbinot}, E. and {Banerji}, M. and {Bechtol}, K. and {Benoit-L{\'e}vy}, A. and {Bernstein}, G.~M. and {Bertin}, E. and {Blazek}, J. and {Bonnett}, C. and {Bridle}, S. and {Brooks}, D. and {Brunner}, R.~J. and {Buckley-Geer}, E. and {Burke}, D.~L. and {Caminha}, G.~B. and {Capozzi}, D. and {Carlsen}, J. and {Carnero-Rosell}, A. and {Carollo}, M. and {Carrasco-Kind}, M. and {Carretero}, J. and {Castander}, F.~J. and {Clerkin}, L. and {Collett}, T. and {Conselice}, C. and {Crocce}, M. and {Cunha}, C.~E. and {D'Andrea}, C.~B. and {da Costa}, L.~N. and {Davis}, T.~M. and {Desai}, S. and {Diehl}, H.~T. and {Dietrich}, J.~P. and {Dodelson}, S. and {Doel}, P. and {Drlica-Wagner}, A. and {Estrada}, J. and {Etherington}, J. and {Evrard}, A.~E. and {Fabbri}, J. and {Finley}, D.~A. and {Flaugher}, B. and {Foley}, R.~J. and {Fosalba}, P. and {Frieman}, J. and {Garc{\'\i}a-Bellido}, J. and {Gaztanaga}, E. and {Gerdes}, D.~W. and {Giannantonio}, T. and {Goldstein}, D.~A. and {Gruen}, D. and {Gruendl}, R.~A. and {Guarnieri}, P. and {Gutierrez}, G. and {Hartley}, W. and {Honscheid}, K. and {Jain}, B. and {James}, D.~J. and {Jeltema}, T. and {Jouvel}, S. and {Kessler}, R. and {King}, A. and {Kirk}, D. and {Kron}, R. and {Kuehn}, K. and {Kuropatkin}, N. and {Lahav}, O. and {Li}, T.~S. and {Lima}, M. and {Lin}, H. and {Maia}, M.~A.~G. and {Makler}, M. and {Manera}, M. and {Maraston}, C. and {Marshall}, J.~L. and {Martini}, P. and {McMahon}, R.~G. and {Melchior}, P. and {Merson}, A. and {Miller}, C.~J. and {Miquel}, R. and {Mohr}, J.~J. and {Morice-Atkinson}, X. and {Naidoo}, K. and {Neilsen}, E. and {Nichol}, R.~C. and {Nord}, B. and {Ogando}, R. and {Ostrovski}, F. and {Palmese}, A. and {Papadopoulos}, A. and {Peiris}, H.~V. and {Peoples}, J. and {Percival}, W.~J. and {Plazas}, A.~A. and {Reed}, S.~L. and {Refregier}, A. and {Romer}, A.~K. and {Roodman}, A. and {Ross}, A. and {Rozo}, E. and {Rykoff}, E.~S. and {Sadeh}, I. and {Sako}, M. and {S{\'a}nchez}, C. and {Sanchez}, E. and {Santiago}, B. and {Scarpine}, V. and {Schubnell}, M. and {Sevilla-Noarbe}, I. and {Sheldon}, E. and {Smith}, M. and {Smith}, R.~C. and {Soares-Santos}, M. and {Sobreira}, F. and {Soumagnac}, M. and {Suchyta}, E. and {Sullivan}, M. and {Swanson}, M. and {Tarle}, G. and {Thaler}, J. and {Thomas}, D. and {Thomas}, R.~C. and {Tucker}, D. and {Vieira}, J.~D. and {Vikram}, V. and {Walker}, A.~R. and {Wechsler}, R.~H. and {Weller}, J. and {Wester}, W. and {Whiteway}, L. and {Wilcox}, H. and {Yanny}, B. and {Zhang}, Y. and {Zuntz}, J.},
        title = "{The Dark Energy Survey: more than dark energy - an overview}",
      journal = {\mnras},
         year = 2016,
        month = aug,
       volume = {460},
       number = {2},
        pages = {1270-1299},
          doi = {10.1093/mnras/stw641},
archivePrefix = {arXiv},
       eprint = {1601.00329},
 primaryClass = {astro-ph.CO},
       adsurl = {https://ui.adsabs.harvard.edu/abs/2016MNRAS.460.1270D}
}

@ARTICLE{HSC,
       author = {{Aihara}, Hiroaki and {Arimoto}, Nobuo and {Armstrong}, Robert and {Arnouts}, St{\'e}phane and {Bahcall}, Neta A. and {Bickerton}, Steven and {Bosch}, James and {Bundy}, Kevin and {Capak}, Peter L. and {Chan}, James H.~H. and {Chiba}, Masashi and {Coupon}, Jean and {Egami}, Eiichi and {Enoki}, Motohiro and {Finet}, Francois and {Fujimori}, Hiroki and {Fujimoto}, Seiji and {Furusawa}, Hisanori and {Furusawa}, Junko and {Goto}, Tomotsugu and {Goulding}, Andy and {Greco}, Johnny P. and {Greene}, Jenny E. and {Gunn}, James E. and {Hamana}, Takashi and {Harikane}, Yuichi and {Hashimoto}, Yasuhiro and {Hattori}, Takashi and {Hayashi}, Masao and {Hayashi}, Yusuke and {He{\l}miniak}, Krzysztof G. and {Higuchi}, Ryo and {Hikage}, Chiaki and {Ho}, Paul T.~P. and {Hsieh}, Bau-Ching and {Huang}, Kuiyun and {Huang}, Song and {Ikeda}, Hiroyuki and {Imanishi}, Masatoshi and {Inoue}, Akio K. and {Iwasawa}, Kazushi and {Iwata}, Ikuru and {Jaelani}, Anton T. and {Jian}, Hung-Yu and {Kamata}, Yukiko and {Karoji}, Hiroshi and {Kashikawa}, Nobunari and {Katayama}, Nobuhiko and {Kawanomoto}, Satoshi and {Kayo}, Issha and {Koda}, Jin and {Koike}, Michitaro and {Kojima}, Takashi and {Komiyama}, Yutaka and {Konno}, Akira and {Koshida}, Shintaro and {Koyama}, Yusei and {Kusakabe}, Haruka and {Leauthaud}, Alexie and {Lee}, Chien-Hsiu and {Lin}, Lihwai and {Lin}, Yen-Ting and {Lupton}, Robert H. and {Mandelbaum}, Rachel and {Matsuoka}, Yoshiki and {Medezinski}, Elinor and {Mineo}, Sogo and {Miyama}, Shoken and {Miyatake}, Hironao and {Miyazaki}, Satoshi and {Momose}, Rieko and {More}, Anupreeta and {More}, Surhud and {Moritani}, Yuki and {Moriya}, Takashi J. and {Morokuma}, Tomoki and {Mukae}, Shiro and {Murata}, Ryoma and {Murayama}, Hitoshi and {Nagao}, Tohru and {Nakata}, Fumiaki and {Niida}, Mana and {Niikura}, Hiroko and {Nishizawa}, Atsushi J. and {Obuchi}, Yoshiyuki and {Oguri}, Masamune and {Oishi}, Yukie and {Okabe}, Nobuhiro and {Okamoto}, Sakurako and {Okura}, Yuki and {Ono}, Yoshiaki and {Onodera}, Masato and {Onoue}, Masafusa and {Osato}, Ken and {Ouchi}, Masami and {Price}, Paul A. and {Pyo}, Tae-Soo and {Sako}, Masao and {Sawicki}, Marcin and {Shibuya}, Takatoshi and {Shimasaku}, Kazuhiro and {Shimono}, Atsushi and {Shirasaki}, Masato and {Silverman}, John D. and {Simet}, Melanie and {Speagle}, Joshua and {Spergel}, David N. and {Strauss}, Michael A. and {Sugahara}, Yuma and {Sugiyama}, Naoshi and {Suto}, Yasushi and {Suyu}, Sherry H. and {Suzuki}, Nao and {Tait}, Philip J. and {Takada}, Masahiro and {Takata}, Tadafumi and {Tamura}, Naoyuki and {Tanaka}, Manobu M. and {Tanaka}, Masaomi and {Tanaka}, Masayuki and {Tanaka}, Yoko and {Terai}, Tsuyoshi and {Terashima}, Yuichi and {Toba}, Yoshiki and {Tominaga}, Nozomu and {Toshikawa}, Jun and {Turner}, Edwin L. and {Uchida}, Tomohisa and {Uchiyama}, Hisakazu and {Umetsu}, Keiichi and {Uraguchi}, Fumihiro and {Urata}, Yuji and {Usuda}, Tomonori and {Utsumi}, Yousuke and {Wang}, Shiang-Yu and {Wang}, Wei-Hao and {Wong}, Kenneth C. and {Yabe}, Kiyoto and {Yamada}, Yoshihiko and {Yamanoi}, Hitomi and {Yasuda}, Naoki and {Yeh}, Sherry and {Yonehara}, Atsunori and {Yuma}, Suraphong},
        title = "{The Hyper Suprime-Cam SSP Survey: Overview and survey design}",
      journal = {\pasj},
         year = 2018,
        month = jan,
       volume = {70},
          eid = {S4},
        pages = {S4},
          doi = {10.1093/pasj/psx066},
archivePrefix = {arXiv},
       eprint = {1704.05858},
 primaryClass = {astro-ph.IM},
       adsurl = {https://ui.adsabs.harvard.edu/abs/2018PASJ...70S...4A}
}

@INPROCEEDINGS{DECaLS,
       author = {{Schlegel}, David and {Beechert}, Jacqueline and {Burleigh}, Kaylan and {Dey}, Arjun and {Findlay}, Joseph and {Herrera}, David and {Juneau}, Stephanie and {Landriau}, Martin and {Lang}, Dustin and {Meisner}, Aaron M. and {Moustakas}, John and {Myers}, Adam D. and {Schlafly}, Edward and {Valdes}, Frank and {Weaver}, Benjamin A. and {Yang}, Jinyi and {Yeche}, Christophe},
        title = "{Overview of the DESI Legacy Imaging Surveys}",
    booktitle = {American Astronomical Society Meeting Abstracts \#233},
         year = 2019,
       series = {American Astronomical Society Meeting Abstracts},
       volume = {233},
        month = jan,
          eid = {380.03},
        pages = {380.03},
       adsurl = {https://ui.adsabs.harvard.edu/abs/2019AAS...23338003S}
}

@ARTICLE{Pan-STARRS,
       author = {{Chambers}, K.~C. and {Magnier}, E.~A. and {Metcalfe}, N. and {Flewelling}, H.~A. and {Huber}, M.~E. and {Waters}, C.~Z. and {Denneau}, L. and {Draper}, P.~W. and {Farrow}, D. and {Finkbeiner}, D.~P. and {Holmberg}, C. and {Koppenhoefer}, J. and {Price}, P.~A. and {Rest}, A. and {Saglia}, R.~P. and {Schlafly}, E.~F. and {Smartt}, S.~J. and {Sweeney}, W. and {Wainscoat}, R.~J. and {Burgett}, W.~S. and {Chastel}, S. and {Grav}, T. and {Heasley}, J.~N. and {Hodapp}, K.~W. and {Jedicke}, R. and {Kaiser}, N. and {Kudritzki}, R.-P. and {Luppino}, G.~A. and {Lupton}, R.~H. and {Monet}, D.~G. and {Morgan}, J.~S. and {Onaka}, P.~M. and {Shiao}, B. and {Stubbs}, C.~W. and {Tonry}, J.~L. and {White}, R. and {Ba{\~n}ados}, E. and {Bell}, E.~F. and {Bender}, R. and {Bernard}, E.~J. and {Boegner}, M. and {Boffi}, F. and {Botticella}, M.~T. and {Calamida}, A. and {Casertano}, S. and {Chen}, W.-P. and {Chen}, X. and {Cole}, S. and {Deacon}, N. and {Frenk}, C. and {Fitzsimmons}, A. and {Gezari}, S. and {Gibbs}, V. and {Goessl}, C. and {Goggia}, T. and {Gourgue}, R. and {Goldman}, B. and {Grant}, P. and {Grebel}, E.~K. and {Hambly}, N.~C. and {Hasinger}, G. and {Heavens}, A.~F. and {Heckman}, T.~M. and {Henderson}, R. and {Henning}, T. and {Holman}, M. and {Hopp}, U. and {Ip}, W.-H. and {Isani}, S. and {Jackson}, M. and {Keyes}, C.~D. and {Koekemoer}, A.~M. and {Kotak}, R. and {Le}, D. and {Liska}, D. and {Long}, K.~S. and {Lucey}, J.~R. and {Liu}, M. and {Martin}, N.~F. and {Masci}, G. and {McLean}, B. and {Mindel}, E. and {Misra}, P. and {Morganson}, E. and {Murphy}, D.~N.~A. and {Obaika}, A. and {Narayan}, G. and {Nieto-Santisteban}, M.~A. and {Norberg}, P. and {Peacock}, J.~A. and {Pier}, E.~A. and {Postman}, M. and {Primak}, N. and {Rae}, C. and {Rai}, A. and {Riess}, A. and {Riffeser}, A. and {Rix}, H.~W. and {R{\"o}ser}, S. and {Russel}, R. and {Rutz}, L. and {Schilbach}, E. and {Schultz}, A.~S.~B. and {Scolnic}, D. and {Strolger}, L. and {Szalay}, A. and {Seitz}, S. and {Small}, E. and {Smith}, K.~W. and {Soderblom}, D.~R. and {Taylor}, P. and {Thomson}, R. and {Taylor}, A.~N. and {Thakar}, A.~R. and {Thiel}, J. and {Thilker}, D. and {Unger}, D. and {Urata}, Y. and {Valenti}, J. and {Wagner}, J. and {Walder}, T. and {Walter}, F. and {Watters}, S.~P. and {Werner}, S. and {Wood-Vasey}, W.~M. and {Wyse}, R.},
        title = "{The Pan-STARRS1 Surveys}",
      journal = {arXiv e-prints},
         year = 2016,
        month = dec,
          eid = {arXiv:1612.05560},
        pages = {arXiv:1612.05560},
          doi = {10.48550/arXiv.1612.05560},
archivePrefix = {arXiv},
       eprint = {1612.05560},
 primaryClass = {astro-ph.IM},
       adsurl = {https://ui.adsabs.harvard.edu/abs/2016arXiv161205560C}
}

@ARTICLE{Euclid,
       author = {{Euclid Collaboration} and {Mellier}, Y. and {Abdurro'uf} and {Acevedo Barroso}, J.~A. and {Ach{\'u}carro}, A. and {Adamek}, J. and {Adam}, R. and {Addison}, G.~E. and {Aghanim}, N. and {Aguena}, M. and {Ajani}, V. and {Akrami}, Y. and {Al-Bahlawan}, A. and {Alavi}, A. and {Albuquerque}, I.~S. and {Alestas}, G. and {Alguero}, G. and {Allaoui}, A. and {Allen}, S.~W. and {Allevato}, V. and {Alonso-Tetilla}, A.~V. and {Altieri}, B. and {Alvarez-Candal}, A. and {Alvi}, S. and {Amara}, A. and {Amendola}, L. and {Amiaux}, J. and {Andika}, I.~T. and {Andreon}, S. and {Andrews}, A. and {Angora}, G. and {Angulo}, R.~E. and {Annibali}, F. and {Anselmi}, A. and {Anselmi}, S. and {Arcari}, S. and {Archidiacono}, M. and {Aric{\`o}}, G. and {Arnaud}, M. and {Arnouts}, S. and {Asgari}, M. and {Asorey}, J. and {Atayde}, L. and {Atek}, H. and {Atrio-Barandela}, F. and {Aubert}, M. and {Aubourg}, E. and {Auphan}, T. and {Auricchio}, N. and {Aussel}, B. and {Aussel}, H. and {Avelino}, P.~P. and {Avgoustidis}, A. and {Avila}, S. and {Awan}, S. and {Azzollini}, R. and {Baccigalupi}, C. and {Bachelet}, E. and {Bacon}, D. and {Baes}, M. and {Bagley}, M.~B. and {Bahr-Kalus}, B. and {Balaguera-Antolinez}, A. and {Balbinot}, E. and {Balcells}, M. and {Baldi}, M. and {Baldry}, I. and {Balestra}, A. and {Ballardini}, M. and {Ballester}, O. and {Balogh}, M. and {Ba{\~n}ados}, E. and {Barbier}, R. and {Bardelli}, S. and {Baron}, M. and {Barreiro}, T. and {Barrena}, R. and {Barriere}, J. -C. and {Barros}, B.~J. and {Barthelemy}, A. and {Bartolo}, N. and {Basset}, A. and {Battaglia}, P. and {Battisti}, A.~J. and {Baugh}, C.~M. and {Baumont}, L. and {Bazzanini}, L. and {Beaulieu}, J. -P. and {Beckmann}, V. and {Belikov}, A.~N. and {Bel}, J. and {Bellagamba}, F. and {Bella}, M. and {Bellini}, E. and {Benabed}, K. and {Bender}, R. and {Benevento}, G. and {Bennett}, C.~L. and {Benson}, K. and {Bergamini}, P. and {Bermejo-Climent}, J.~R. and {Bernardeau}, F. and {Bertacca}, D. and {Berthe}, M. and {Berthier}, J. and {Bethermin}, M. and {Beutler}, F. and {Bevillon}, C. and {Bhargava}, S. and {Bhatawdekar}, R. and {Bianchi}, D. and {Bisigello}, L. and {Biviano}, A. and {Blake}, R.~P. and {Blanchard}, A. and {Blazek}, J. and {Blot}, L. and {Bosco}, A. and {Bodendorf}, C. and {Boenke}, T. and {B{\"o}hringer}, H. and {Boldrini}, P. and {Bolzonella}, M. and {Bonchi}, A. and {Bonici}, M. and {Bonino}, D. and {Bonino}, L. and {Bonvin}, C. and {Bon}, W. and {Booth}, J.~T. and {Borgani}, S. and {Borlaff}, A.~S. and {Borsato}, E. and {Bose}, B. and {Botticella}, M.~T. and {Boucaud}, A. and {Bouche}, F. and {Boucher}, J.~S. and {Boutigny}, D. and {Bouvard}, T. and {Bouwens}, R. and {Bouy}, H. and {Bowler}, R.~A.~A. and {Bozza}, V. and {Bozzo}, E. and {Branchini}, E. and {Brando}, G. and {Brau-Nogue}, S. and {Brekke}, P. and {Bremer}, M.~N. and {Brescia}, M. and {Breton}, M. -A. and {Brinchmann}, J. and {Brinckmann}, T. and {Brockley-Blatt}, C. and {Brodwin}, M. and {Brouard}, L. and {Brown}, M.~L. and {Bruton}, S. and {Bucko}, J. and {Buddelmeijer}, H. and {Buenadicha}, G. and {Buitrago}, F. and {Burger}, P. and {Burigana}, C. and {Busillo}, V. and {Busonero}, D. and {Cabanac}, R. and {Cabayol-Garcia}, L. and {Cagliari}, M.~S. and {Caillat}, A. and {Caillat}, L. and {Calabrese}, M. and {Calabro}, A. and {Calderone}, G. and {Calura}, F. and {Camacho Quevedo}, B. and {Camera}, S. and {Campos}, L. and {Ca{\~n}as-Herrera}, G. and {Candini}, G.~P. and {Cantiello}, M. and {Capobianco}, V. and {Cappellaro}, E. and {Cappelluti}, N. and {Cappi}, A. and {Caputi}, K.~I. and {Cara}, C. and {Carbone}, C. and {Cardone}, V.~F. and {Carella}, E. and {Carlberg}, R.~G. and {Carle}, M. and {Carminati}, L. and {Caro}, F. and {Carrasco}, J.~M. and {Carretero}, J. and {Carrilho}, P. and {Carron Duque}, J. and {Carry}, B.},
        title = "{Euclid: I. Overview of the Euclid mission}",
      journal = {\aap},
         year = 2025,
        month = may,
       volume = {697},
          eid = {A1},
        pages = {A1},
          doi = {10.1051/0004-6361/202450810},
archivePrefix = {arXiv},
       eprint = {2405.13491},
 primaryClass = {astro-ph.CO},
       adsurl = {https://ui.adsabs.harvard.edu/abs/2025A&A...697A...1E}
}

@ARTICLE{SDSS,
       author = {{York}, Donald G. and {Adelman}, J. and {Anderson}, Jr., John E. and {Anderson}, Scott F. and {Annis}, James and {Bahcall}, Neta A. and {Bakken}, J.~A. and {Barkhouser}, Robert and {Bastian}, Steven and {Berman}, Eileen and {Boroski}, William N. and {Bracker}, Steve and {Briegel}, Charlie and {Briggs}, John W. and {Brinkmann}, J. and {Brunner}, Robert and {Burles}, Scott and {Carey}, Larry and {Carr}, Michael A. and {Castander}, Francisco J. and {Chen}, Bing and {Colestock}, Patrick L. and {Connolly}, A.~J. and {Crocker}, J.~H. and {Csabai}, Istv{\'a}n and {Czarapata}, Paul C. and {Davis}, John Eric and {Doi}, Mamoru and {Dombeck}, Tom and {Eisenstein}, Daniel and {Ellman}, Nancy and {Elms}, Brian R. and {Evans}, Michael L. and {Fan}, Xiaohui and {Federwitz}, Glenn R. and {Fiscelli}, Larry and {Friedman}, Scott and {Frieman}, Joshua A. and {Fukugita}, Masataka and {Gillespie}, Bruce and {Gunn}, James E. and {Gurbani}, Vijay K. and {de Haas}, Ernst and {Haldeman}, Merle and {Harris}, Frederick H. and {Hayes}, J. and {Heckman}, Timothy M. and {Hennessy}, G.~S. and {Hindsley}, Robert B. and {Holm}, Scott and {Holmgren}, Donald J. and {Huang}, Chi-hao and {Hull}, Charles and {Husby}, Don and {Ichikawa}, Shin-Ichi and {Ichikawa}, Takashi and {Ivezi{\'c}}, {\v{Z}}eljko and {Kent}, Stephen and {Kim}, Rita S.~J. and {Kinney}, E. and {Klaene}, Mark and {Kleinman}, A.~N. and {Kleinman}, S. and {Knapp}, G.~R. and {Korienek}, John and {Kron}, Richard G. and {Kunszt}, Peter Z. and {Lamb}, D.~Q. and {Lee}, B. and {Leger}, R. French and {Limmongkol}, Siriluk and {Lindenmeyer}, Carl and {Long}, Daniel C. and {Loomis}, Craig and {Loveday}, Jon and {Lucinio}, Rich and {Lupton}, Robert H. and {MacKinnon}, Bryan and {Mannery}, Edward J. and {Mantsch}, P.~M. and {Margon}, Bruce and {McGehee}, Peregrine and {McKay}, Timothy A. and {Meiksin}, Avery and {Merelli}, Aronne and {Monet}, David G. and {Munn}, Jeffrey A. and {Narayanan}, Vijay K. and {Nash}, Thomas and {Neilsen}, Eric and {Neswold}, Rich and {Newberg}, Heidi Jo and {Nichol}, R.~C. and {Nicinski}, Tom and {Nonino}, Mario and {Okada}, Norio and {Okamura}, Sadanori and {Ostriker}, Jeremiah P. and {Owen}, Russell and {Pauls}, A. George and {Peoples}, John and {Peterson}, R.~L. and {Petravick}, Donald and {Pier}, Jeffrey R. and {Pope}, Adrian and {Pordes}, Ruth and {Prosapio}, Angela and {Rechenmacher}, Ron and {Quinn}, Thomas R. and {Richards}, Gordon T. and {Richmond}, Michael W. and {Rivetta}, Claudio H. and {Rockosi}, Constance M. and {Ruthmansdorfer}, Kurt and {Sandford}, Dale and {Schlegel}, David J. and {Schneider}, Donald P. and {Sekiguchi}, Maki and {Sergey}, Gary and {Shimasaku}, Kazuhiro and {Siegmund}, Walter A. and {Smee}, Stephen and {Smith}, J. Allyn and {Snedden}, S. and {Stone}, R. and {Stoughton}, Chris and {Strauss}, Michael A. and {Stubbs}, Christopher and {SubbaRao}, Mark and {Szalay}, Alexander S. and {Szapudi}, Istvan and {Szokoly}, Gyula P. and {Thakar}, Anirudda R. and {Tremonti}, Christy and {Tucker}, Douglas L. and {Uomoto}, Alan and {Vanden Berk}, Dan and {Vogeley}, Michael S. and {Waddell}, Patrick and {Wang}, Shu-i. and {Watanabe}, Masaru and {Weinberg}, David H. and {Yanny}, Brian and {Yasuda}, Naoki and {SDSS Collaboration}},
        title = "{The Sloan Digital Sky Survey: Technical Summary}",
      journal = {\aj},
         year = 2000,
        month = sep,
       volume = {120},
       number = {3},
        pages = {1579-1587},
          doi = {10.1086/301513},
archivePrefix = {arXiv},
       eprint = {astro-ph/0006396},
 primaryClass = {astro-ph},
       adsurl = {https://ui.adsabs.harvard.edu/abs/2000AJ....120.1579Y}
}

@ARTICLE{DESI,
       author = {{DESI Collaboration} and {Aghamousa}, Amir and {Aguilar}, Jessica and {Ahlen}, Steve and {Alam}, Shadab and {Allen}, Lori E. and {Allende Prieto}, Carlos and {Annis}, James and {Bailey}, Stephen and {Balland}, Christophe and {Ballester}, Otger and {Baltay}, Charles and {Beaufore}, Lucas and {Bebek}, Chris and {Beers}, Timothy C. and {Bell}, Eric F. and {Bernal}, Jos{\'e} Luis and {Besuner}, Robert and {Beutler}, Florian and {Blake}, Chris and {Bleuler}, Hannes and {Blomqvist}, Michael and {Blum}, Robert and {Bolton}, Adam S. and {Briceno}, Cesar and {Brooks}, David and {Brownstein}, Joel R. and {Buckley-Geer}, Elizabeth and {Burden}, Angela and {Burtin}, Etienne and {Busca}, Nicolas G. and {Cahn}, Robert N. and {Cai}, Yan-Chuan and {Cardiel-Sas}, Laia and {Carlberg}, Raymond G. and {Carton}, Pierre-Henri and {Casas}, Ricard and {Castander}, Francisco J. and {Cervantes-Cota}, Jorge L. and {Claybaugh}, Todd M. and {Close}, Madeline and {Coker}, Carl T. and {Cole}, Shaun and {Comparat}, Johan and {Cooper}, Andrew P. and {Cousinou}, M. -C. and {Crocce}, Martin and {Cuby}, Jean-Gabriel and {Cunningham}, Daniel P. and {Davis}, Tamara M. and {Dawson}, Kyle S. and {de la Macorra}, Axel and {De Vicente}, Juan and {Delubac}, Timoth{\'e}e and {Derwent}, Mark and {Dey}, Arjun and {Dhungana}, Govinda and {Ding}, Zhejie and {Doel}, Peter and {Duan}, Yutong T. and {Ealet}, Anne and {Edelstein}, Jerry and {Eftekharzadeh}, Sarah and {Eisenstein}, Daniel J. and {Elliott}, Ann and {Escoffier}, St{\'e}phanie and {Evatt}, Matthew and {Fagrelius}, Parker and {Fan}, Xiaohui and {Fanning}, Kevin and {Farahi}, Arya and {Farihi}, Jay and {Favole}, Ginevra and {Feng}, Yu and {Fernandez}, Enrique and {Findlay}, Joseph R. and {Finkbeiner}, Douglas P. and {Fitzpatrick}, Michael J. and {Flaugher}, Brenna and {Flender}, Samuel and {Font-Ribera}, Andreu and {Forero-Romero}, Jaime E. and {Fosalba}, Pablo and {Frenk}, Carlos S. and {Fumagalli}, Michele and {Gaensicke}, Boris T. and {Gallo}, Giuseppe and {Garcia-Bellido}, Juan and {Gaztanaga}, Enrique and {Pietro Gentile Fusillo}, Nicola and {Gerard}, Terry and {Gershkovich}, Irena and {Giannantonio}, Tommaso and {Gillet}, Denis and {Gonzalez-de-Rivera}, Guillermo and {Gonzalez-Perez}, Violeta and {Gott}, Shelby and {Graur}, Or and {Gutierrez}, Gaston and {Guy}, Julien and {Habib}, Salman and {Heetderks}, Henry and {Heetderks}, Ian and {Heitmann}, Katrin and {Hellwing}, Wojciech A. and {Herrera}, David A. and {Ho}, Shirley and {Holland}, Stephen and {Honscheid}, Klaus and {Huff}, Eric and {Hutchinson}, Timothy A. and {Huterer}, Dragan and {Hwang}, Ho Seong and {Illa Laguna}, Joseph Maria and {Ishikawa}, Yuzo and {Jacobs}, Dianna and {Jeffrey}, Niall and {Jelinsky}, Patrick and {Jennings}, Elise and {Jiang}, Linhua and {Jimenez}, Jorge and {Johnson}, Jennifer and {Joyce}, Richard and {Jullo}, Eric and {Juneau}, St{\'e}phanie and {Kama}, Sami and {Karcher}, Armin and {Karkar}, Sonia and {Kehoe}, Robert and {Kennamer}, Noble and {Kent}, Stephen and {Kilbinger}, Martin and {Kim}, Alex G. and {Kirkby}, David and {Kisner}, Theodore and {Kitanidis}, Ellie and {Kneib}, Jean-Paul and {Koposov}, Sergey and {Kovacs}, Eve and {Koyama}, Kazuya and {Kremin}, Anthony and {Kron}, Richard and {Kronig}, Luzius and {Kueter-Young}, Andrea and {Lacey}, Cedric G. and {Lafever}, Robin and {Lahav}, Ofer and {Lambert}, Andrew and {Lampton}, Michael and {Landriau}, Martin and {Lang}, Dustin and {Lauer}, Tod R. and {Le Goff}, Jean-Marc and {Le Guillou}, Laurent and {Le Van Suu}, Auguste and {Lee}, Jae Hyeon and {Lee}, Su-Jeong and {Leitner}, Daniela and {Lesser}, Michael and {Levi}, Michael E. and {L'Huillier}, Benjamin and {Li}, Baojiu and {Liang}, Ming and {Lin}, Huan and {Linder}, Eric and {Loebman}, Sarah R. and {Luki{\'c}}, Zarija and {Ma}, Jun and {MacCrann}, Niall and {Magneville}, Christophe and {Makarem}, Laleh and {Manera}, Marc and {Manser}, Christopher J. and {Marshall}, Robert and {Martini}, Paul and {Massey}, Richard and {Matheson}, Thomas and {McCauley}, Jeremy and {McDonald}, Patrick and {McGreer}, Ian D. and {Meisner}, Aaron and {Metcalfe}, Nigel and {Miller}, Timothy N. and {Miquel}, Ramon and {Moustakas}, John and {Myers}, Adam and {Naik}, Milind and {Newman}, Jeffrey A. and {Nichol}, Robert C. and {Nicola}, Andrina and {Nicolati da Costa}, Luiz and {Nie}, Jundan and {Niz}, Gustavo and {Norberg}, Peder and {Nord}, Brian and {Norman}, Dara and {Nugent}, Peter and {O'Brien}, Thomas and {Oh}, Minji and {Olsen}, Knut A.~G.},
        title = "{The DESI Experiment Part I: Science,Targeting, and Survey Design}",
      journal = {arXiv e-prints},
         year = 2016,
        month = oct,
          eid = {arXiv:1611.00036},
        pages = {arXiv:1611.00036},
          doi = {10.48550/arXiv.1611.00036},
archivePrefix = {arXiv},
       eprint = {1611.00036},
 primaryClass = {astro-ph.IM},
       adsurl = {https://ui.adsabs.harvard.edu/abs/2016arXiv161100036D}
}

@ARTICLE{DESI_DR1,
       author = {{DESI Collaboration} and {Abdul Karim}, M. and {Adame}, A.~G. and {Aguado}, D. and {Aguilar}, J. and {Ahlen}, S. and {Alam}, S. and {Aldering}, G. and {Alexander}, D.~M. and {Alfarsy}, R. and {Allen}, L. and {Allende Prieto}, C. and {Alves}, O. and {Anand}, A. and {Andrade}, U. and {Armengaud}, E. and {Avila}, S. and {Aviles}, A. and {Awan}, H. and {Bailey}, S. and {Baleato Lizancos}, A. and {Ballester}, O. and {Bault}, A. and {Bautista}, J. and {Bean}, R. and {Behera}, J. and {BenZvi}, S. and {Beraldo e Silva}, L. and {Bermejo-Climent}, J.~R. and {Beutler}, F. and {Bianchi}, D. and {Blake}, C. and {Blum}, R. and {Bolton}, A.~S. and {Bonici}, M. and {Brieden}, S. and {Brodzeller}, A. and {Brooks}, D. and {Buckley-Geer}, E. and {Burtin}, E. and {Bystr{\"o}m}, A. and {Canning}, R. and {Carnero Rosell}, A. and {Carr}, A. and {Carrilho}, P. and {Casas}, L. and {Castander}, F.~J. and {Cereskaite}, R. and {Cervantes-Cota}, J.~L. and {Chaussidon}, E. and {Chaves-Montero}, J. and {Chen}, S. and {Chen}, X. and {Circosta}, C. and {Claybaugh}, T. and {Cole}, S. and {Cooper}, A.~P. and {Cousinou}, M.-C. and {Cuceu}, A. and {Davis}, T.~M. and {Dawson}, K.~S. and {de Belsunce}, R. and {de la Cruz}, R. and {de la Macorra}, A. and {de Mattia}, A. and {Deiosso}, N. and {Della Costa}, J. and {Demina}, R. and {Demirbozan}, U. and {DeRose}, J. and {Dey}, A. and {Dey}, B. and {Ding}, J. and {Ding}, Z. and {Doel}, P. and {Douglass}, K. and {Dowicz}, M. and {Ebina}, H. and {Edelstein}, J. and {Eisenstein}, D.~J. and {Elbers}, W. and {Emas}, N. and {Escoffier}, S. and {Fagrelius}, P. and {Fan}, X. and {Fanning}, K. and {Favole}, G. and {Fawcett}, V.~A. and {Fern{\'a}ndez-Garc{\'\i}a}, E. and {Ferraro}, S. and {Findlay}, N. and {Font-Ribera}, A. and {Forero-Romero}, J.~E. and {Forero-S{\'a}nchez}, D. and {Frenk}, C.~S. and {G{\"a}nsicke}, B.~T. and {Galbany}, L. and {Garc{\'\i}a-Bellido}, J. and {Garcia-Quintero}, C. and {Garrison}, L.~H. and {Gazta{\~n}aga}, E. and {Gil-Mar{\'\i}n}, H. and {Gloudemans}, A. and {Gnedin}, O.~Y. and {Gontcho A Gontcho}, S. and {Gonzalez}, D. and {Gonzalez-Morales}, A.~X. and {Gonzalez-Perez}, V. and {Gordon}, C. and {Graur}, O. and {Green}, D. and {Gruen}, D. and {Gsponer}, R. and {Guandalin}, C. and {Gutierrez}, G. and {Guy}, J. and {Hahn}, C. and {Han}, J.~J. and {Han}, J. and {He}, S. and {Herrera-Alcantar}, H.~K. and {Heydenreich}, S. and {Honscheid}, K. and {Hou}, J. and {Howlett}, C. and {Huterer}, D. and {Ir{\v{s}}i{\v{c}}}, V. and {Ishak}, M. and {Jacques}, A. and {Jiang}, L. and {Jimenez}, J. and {Jing}, Y.~P. and {Joachimi}, B. and {Joudaki}, S. and {Joyce}, R. and {Jullo}, E. and {Juneau}, S. and {Kara{\c{c}}ayl{\i}}, N.~G. and {Karim}, T. and {Kehoe}, R. and {Kent}, S. and {Khederlarian}, A. and {Kirkby}, D. and {Kisner}, T. and {Kitaura}, F.-S. and {Kizhuprakkat}, N. and {Kong}, H. and {Koposov}, S.~E. and {Kremin}, A. and {Krolewski}, A. and {Lahav}, O. and {Lai}, Y. and {Lamman}, C. and {Lan}, T.-W. and {Landriau}, M. and {Lang}, D. and {Lange}, J.~U. and {Lasker}, J. and {Le Goff}, J.~M. and {Le Guillou}, L. and {Leauthaud}, A. and {Levi}, M.~E. and {Li}, S. and {Li}, T.~S. and {Liu}, W. and {Lodha}, K. and {Lokken}, M. and {Luo}, Y. and {Magneville}, C. and {Manera}, M. and {Manser}, C.~J. and {Margala}, D. and {Martini}, P. and {Maus}, M. and {McCullough}, J. and {McDonald}, P. and {Medina}, G.~E. and {Medina-Varela}, L. and {Meisner}, A. and {Mena-Fern{\'a}ndez}, J. and {Menegas}, A. and {Meneses-Rizo}, J. and {Mezcua}, M. and {Miquel}, R. and {Montero-Camacho}, P. and {Moon}, J. and {Moustakas}, J. and {Mu{\~n}oz-Guti{\'e}rrez}, A. and {Mu noz-Santos}, D. and {Myers}, A.~D. and {Myles}, J. and {Nadathur}, S. and {Najita}, J. and {Napolitano}, L. and {Newman}, J.~A. and {Nikakhtar}, F. and {Nikutta}, R. and {Niz}, G. and {Noriega}, H.~E. and {Nugent}, P.},
        title = "{Data Release 1 of the Dark Energy Spectroscopic Instrument}",
      journal = {\aj},
         year = 2026,
        month = may,
       volume = {171},
       number = {5},
          eid = {285},
        pages = {285},
          doi = {10.3847/1538-3881/ae4c43},
archivePrefix = {arXiv},
       eprint = {2503.14745},
 primaryClass = {astro-ph.CO},
       adsurl = {https://ui.adsabs.harvard.edu/abs/2026AJ....171..285D}
}

@ARTICLE{sugohi_I,
       author = {{Sonnenfeld}, Alessandro and {Chan}, James H.~H. and {Shu}, Yiping and {More}, Anupreeta and {Oguri}, Masamune and {Suyu}, Sherry H. and {Wong}, Kenneth C. and {Lee}, Chien-Hsiu and {Coupon}, Jean and {Yonehara}, Atsunori and {Bolton}, Adam S. and {Jaelani}, Anton T. and {Tanaka}, Masayuki and {Miyazaki}, Satoshi and {Komiyama}, Yutaka},
        title = "{Survey of Gravitationally-lensed Objects in HSC Imaging (SuGOHI). I. Automatic search for galaxy-scale strong lenses}",
      journal = {\pasj},
         year = 2018,
        month = jan,
       volume = {70},
          eid = {S29},
        pages = {S29},
          doi = {10.1093/pasj/psx062},
archivePrefix = {arXiv},
       eprint = {1704.01585},
 primaryClass = {astro-ph.GA},
       adsurl = {https://ui.adsabs.harvard.edu/abs/2018PASJ...70S..29S}
}

@ARTICLE{sugohi_II,
       author = {{Wong}, Kenneth C. and {Sonnenfeld}, Alessandro and {Chan}, James H.~H. and {Rusu}, Cristian E. and {Tanaka}, Masayuki and {Jaelani}, Anton T. and {Lee}, Chien-Hsiu and {More}, Anupreeta and {Oguri}, Masamune and {Suyu}, Sherry H. and {Komiyama}, Yutaka},
        title = "{Survey of Gravitationally Lensed Objects in HSC Imaging (SuGOHI). II. Environments and Line-of-Sight Structure of Strong Gravitational Lens Galaxies to z {\ensuremath{\sim}} 0.8}",
      journal = {\apj},
         year = 2018,
        month = nov,
       volume = {867},
       number = {2},
          eid = {107},
        pages = {107},
          doi = {10.3847/1538-4357/aae381},
archivePrefix = {arXiv},
       eprint = {1809.07341},
 primaryClass = {astro-ph.GA},
       adsurl = {https://ui.adsabs.harvard.edu/abs/2018ApJ...867..107W}
}

@ARTICLE{sugohi_IV,
       author = {{Chan}, James H.~H. and {Suyu}, Sherry H. and {Sonnenfeld}, Alessandro and {Jaelani}, Anton T. and {More}, Anupreeta and {Yonehara}, Atsunori and {Kubota}, Yuriko and {Coupon}, Jean and {Lee}, Chien-Hsiu and {Oguri}, Masamune and {Rusu}, Cristian E. and {Wong}, Kenneth C.},
        title = "{Survey of Gravitationally lensed Objects in HSC Imaging (SuGOHI). IV. Lensed quasar search in the HSC survey}",
      journal = {\aap},
         year = 2020,
        month = apr,
       volume = {636},
          eid = {A87},
        pages = {A87},
          doi = {10.1051/0004-6361/201937030},
archivePrefix = {arXiv},
       eprint = {1911.02587},
 primaryClass = {astro-ph.GA},
       adsurl = {https://ui.adsabs.harvard.edu/abs/2020A&A...636A..87C}
}

@ARTICLE{sugohi_V,
       author = {{Jaelani}, Anton T. and {More}, Anupreeta and {Oguri}, Masamune and {Sonnenfeld}, Alessandro and {Suyu}, Sherry H. and {Rusu}, Cristian E. and {Wong}, Kenneth C. and {Chan}, James H.~H. and {Kayo}, Issha and {Lee}, Chien-Hsiu and {Chao}, Dani C. -Y. and {Coupon}, Jean and {Inoue}, Kaiki T. and {Futamase}, Toshifumi},
        title = "{Survey of Gravitationally lensed Objects in HSC Imaging (SuGOHI) - V. Group-to-cluster scale lens search from the HSC-SSP Survey}",
      journal = {\mnras},
         year = 2020,
        month = jun,
       volume = {495},
       number = {1},
        pages = {1291-1310},
          doi = {10.1093/mnras/staa1062},
archivePrefix = {arXiv},
       eprint = {2002.01611},
 primaryClass = {astro-ph.GA},
       adsurl = {https://ui.adsabs.harvard.edu/abs/2020MNRAS.495.1291J}
}

@ARTICLE{sugohi_VI,
       author = {{Sonnenfeld}, Alessandro and {Verma}, Aprajita and {More}, Anupreeta and {Baeten}, Elisabeth and {Macmillan}, Christine and {Wong}, Kenneth C. and {Chan}, James H.~H. and {Jaelani}, Anton T. and {Lee}, Chien-Hsiu and {Oguri}, Masamune and {Rusu}, Cristian E. and {Veldthuis}, Marten and {Trouille}, Laura and {Marshall}, Philip J. and {Hutchings}, Roger and {Allen}, Campbell and {O'Donnell}, James and {Cornen}, Claude and {Davis}, Christopher P. and {McMaster}, Adam and {Lintott}, Chris and {Miller}, Grant},
        title = "{Survey of Gravitationally-lensed Objects in HSC Imaging (SuGOHI). VI. Crowdsourced lens finding with Space Warps}",
      journal = {\aap},
         year = 2020,
        month = oct,
       volume = {642},
          eid = {A148},
        pages = {A148},
          doi = {10.1051/0004-6361/202038067},
archivePrefix = {arXiv},
       eprint = {2004.00634},
 primaryClass = {astro-ph.IM},
       adsurl = {https://ui.adsabs.harvard.edu/abs/2020A&A...642A.148S}
}

@ARTICLE{sugohi_VII,
       author = {{Jaelani}, Anton T. and {Rusu}, Cristian E. and {Kayo}, Issha and {More}, Anupreeta and {Sonnenfeld}, Alessandro and {Silverman}, John D. and {Schramm}, Malte and {Anguita}, Timo and {Inada}, Naohisa and {Kondo}, Daichi and {Schechter}, Paul L. and {Lee}, Khee-Gan and {Oguri}, Masamune and {Chan}, James H.~H. and {Wong}, Kenneth C. and {Inoue}, Kaiki T.},
        title = "{Survey of Gravitationally Lensed Objects in HSC Imaging (SuGOHI) - VII. Discovery and confirmation of three strongly lensed quasars}",
      journal = {\mnras},
         year = 2021,
        month = mar,
       volume = {502},
       number = {1},
        pages = {1487-1493},
          doi = {10.1093/mnras/stab145},
archivePrefix = {arXiv},
       eprint = {2006.16584},
 primaryClass = {astro-ph.GA},
       adsurl = {https://ui.adsabs.harvard.edu/abs/2021MNRAS.502.1487J}
}

@ARTICLE{sugohi_VIII,
       author = {{Wong}, Kenneth C. and {Chan}, James H.~H. and {Chao}, Dani C. -Y. and {Jaelani}, Anton T. and {Kayo}, Issha and {Lee}, Chien-Hsiu and {More}, Anupreeta and {Oguri}, Masamune},
        title = "{Survey of Gravitationally lensed objects in HSC Imaging (SuGOHI). VIII. New galaxy-scale lenses from the HSC SSP}",
      journal = {\pasj},
         year = 2022,
        month = oct,
       volume = {74},
       number = {5},
        pages = {1209-1219},
          doi = {10.1093/pasj/psac065},
archivePrefix = {arXiv},
       eprint = {2203.12805},
 primaryClass = {astro-ph.GA},
       adsurl = {https://ui.adsabs.harvard.edu/abs/2022PASJ...74.1209W}
}

@ARTICLE{sugohi_IX,
       author = {{Chan}, James H.~H. and {Wong}, Kenneth C. and {Ding}, Xuheng and {Chao}, Dani and {Chiu}, I. -Non and {Jaelani}, Anton T. and {Kayo}, Issha and {More}, Anupreeta and {Oguri}, Masamune and {Suyu}, Sherry H.},
        title = "{Survey of gravitationally lensed objects in HSC imaging (SuGOHI) - IX. Discovery of strongly lensed quasar candidates}",
      journal = {\mnras},
         year = 2024,
        month = jan,
       volume = {527},
       number = {3},
        pages = {6253-6275},
          doi = {10.1093/mnras/stad2953},
archivePrefix = {arXiv},
       eprint = {2304.05425},
 primaryClass = {astro-ph.GA},
       adsurl = {https://ui.adsabs.harvard.edu/abs/2024MNRAS.527.6253C}
}

@ARTICLE{sugohi_X,
       author = {{Jaelani}, Anton T. and {More}, Anupreeta and {Wong}, Kenneth C. and {Inoue}, Kaiki T. and {Chao}, Dani C. -Y. and {Premadi}, Premana W. and {Ca{\~n}ameras}, Raoul},
        title = "{Survey of gravitationally lensed objects in HSC imaging (SuGOHI) - X. Strong lens finding in the HSC-SSP using convolutional neural networks}",
      journal = {\mnras},
         year = 2024,
        month = dec,
       volume = {535},
       number = {2},
        pages = {1625-1639},
          doi = {10.1093/mnras/stae2442},
archivePrefix = {arXiv},
       eprint = {2312.07333},
 primaryClass = {astro-ph.GA},
       adsurl = {https://ui.adsabs.harvard.edu/abs/2024MNRAS.535.1625J}
}

@ARTICLE{holismokesII,
       author = {{Ca{\~n}ameras}, R. and {Schuldt}, S. and {Suyu}, S.~H. and {Taubenberger}, S. and {Meinhardt}, T. and {Leal-Taix{\'e}}, L. and {Lemon}, C. and {Rojas}, K. and {Savary}, E.},
        title = "{HOLISMOKES. II. Identifying galaxy-scale strong gravitational lenses in Pan-STARRS using convolutional neural networks}",
      journal = {\aap},
         year = 2020,
        month = dec,
       volume = {644},
          eid = {A163},
        pages = {A163},
          doi = {10.1051/0004-6361/202038219},
archivePrefix = {arXiv},
       eprint = {2004.13048},
 primaryClass = {astro-ph.GA},
       adsurl = {https://ui.adsabs.harvard.edu/abs/2020A&A...644A.163C}
}

@ARTICLE{holismokesVI,
       author = {{Ca{\~n}ameras}, R. and {Schuldt}, S. and {Shu}, Y. and {Suyu}, S.~H. and {Taubenberger}, S. and {Meinhardt}, T. and {Leal-Taix{\'e}}, L. and {Chao}, D.~C. -Y. and {Inoue}, K.~T. and {Jaelani}, A.~T. and {More}, A.},
        title = "{HOLISMOKES. VI. New galaxy-scale strong lens candidates from the HSC-SSP imaging survey}",
      journal = {\aap},
         year = 2021,
        month = sep,
       volume = {653},
          eid = {L6},
        pages = {L6},
          doi = {10.1051/0004-6361/202141758},
archivePrefix = {arXiv},
       eprint = {2107.07829},
 primaryClass = {astro-ph.GA},
       adsurl = {https://ui.adsabs.harvard.edu/abs/2021A&A...653L...6C}
}

@ARTICLE{holismokesVIII,
       author = {{Shu}, Yiping and {Ca{\~n}ameras}, Raoul and {Schuldt}, Stefan and {Suyu}, Sherry H. and {Taubenberger}, Stefan and {Inoue}, Kaiki Taro and {Jaelani}, Anton T.},
        title = "{HOLISMOKES. VIII. High-redshift, strong-lens search in the Hyper Suprime-Cam Subaru Strategic Program}",
      journal = {\aap},
         year = 2022,
        month = jun,
       volume = {662},
          eid = {A4},
        pages = {A4},
          doi = {10.1051/0004-6361/202243203},
archivePrefix = {arXiv},
       eprint = {2201.11135},
 primaryClass = {astro-ph.GA},
       adsurl = {https://ui.adsabs.harvard.edu/abs/2022A&A...662A...4S}
}

@ARTICLE{holismokesXIII,
       author = {{Schuldt}, S. and {Ca{\~n}ameras}, R. and {Andika}, I.~T. and {Bag}, S. and {Melo}, A. and {Shu}, Y. and {Suyu}, S.~H. and {Taubenberger}, S. and {Grillo}, C.},
        title = "{HOLISMOKES: XIII. Strong-lens candidates at all mass scales and their environments from the Hyper-Suprime Cam and deep learning}",
      journal = {\aap},
         year = 2025,
        month = jan,
       volume = {693},
          eid = {A291},
        pages = {A291},
          doi = {10.1051/0004-6361/202450927},
archivePrefix = {arXiv},
       eprint = {2405.20383},
 primaryClass = {astro-ph.GA},
       adsurl = {https://ui.adsabs.harvard.edu/abs/2025A&A...693A.291S}
}

@ARTICLE{holismokesXVI,
       author = {{Schuldt}, S. and {Ca{\~n}ameras}, R. and {Shu}, Y. and {Andika}, I.~T. and {Bag}, S. and {Grillo}, C. and {Melo}, A. and {Suyu}, S.~H. and {Taubenberger}, S.},
        title = "{HOLISMOKES XVI: Lens search in HSC-PDR3 with a neural network committee and post-processing for false-positive removal}",
      journal = {arXiv e-prints},
         year = 2025,
        month = mar,
          eid = {arXiv:2503.07733},
        pages = {arXiv:2503.07733},
          doi = {10.48550/arXiv.2503.07733},
archivePrefix = {arXiv},
       eprint = {2503.07733},
 primaryClass = {astro-ph.IM},
       adsurl = {https://ui.adsabs.harvard.edu/abs/2025arXiv250307733S}
}

@ARTICLE{Euclid2025B,
       author = {{Euclid Collaboration} and {Rojas}, K. and {Collett}, T.~E. and {Acevedo Barroso}, J.~A. and {Nightingale}, J.~W. and {Stern}, D. and {Moustakas}, L.~A. and {Schuldt}, S. and {Despali}, G. and {Melo}, A. and {Walmsley}, M. and {Ballard}, D.~J. and {Enzi}, W.~J.~R. and {Li}, T. and {Sainz de Murieta}, A. and {Andika}, I.~T. and {Cl{\'e}ment}, B. and {Courbin}, F. and {Ecker}, L.~R. and {Gavazzi}, R. and {Jackson}, N. and {Kov{\'a}cs}, A. and {Matavulj}, P. and {Meneghetti}, M. and {Serjeant}, S. and {Sluse}, D. and {Tortora}, C. and {Verma}, A. and {Marchetti}, L. and {O'Riordan}, C.~M. and {McCarthy}, K. and {Suyu}, S.~H. and {Metcalf}, R.~B. and {Aghanim}, N. and {Altieri}, B. and {Amara}, A. and {Andreon}, S. and {Auricchio}, N. and {Aussel}, H. and {Baccigalupi}, C. and {Baldi}, M. and {Balestra}, A. and {Bardelli}, S. and {Battaglia}, P. and {Bender}, R. and {Biviano}, A. and {Bonchi}, A. and {Branchini}, E. and {Brescia}, M. and {Brinchmann}, J. and {Camera}, S. and {Ca{\~n}as-Herrera}, G. and {Capobianco}, V. and {Carbone}, C. and {Cardone}, V.~F. and {Carretero}, J. and {Casas}, S. and {Castellano}, M. and {Castignani}, G. and {Cavuoti}, S. and {Chambers}, K.~C. and {Cimatti}, A. and {Colodro-Conde}, C. and {Congedo}, G. and {Conselice}, C.~J. and {Conversi}, L. and {Copin}, Y. and {Courtois}, H.~M. and {Cropper}, M. and {Da Silva}, A. and {Degaudenzi}, H. and {De Lucia}, G. and {Di Giorgio}, A.~M. and {Dolding}, C. and {Dole}, H. and {Dubath}, F. and {Dupac}, X. and {Escoffier}, S. and {Fabricius}, M. and {Farina}, M. and {Farinelli}, R. and {Faustini}, F. and {Ferriol}, S. and {Finelli}, F. and {Fotopoulou}, S. and {Frailis}, M. and {Franceschi}, E. and {Galeotta}, S. and {George}, K. and {Gillard}, W. and {Gillis}, B. and {Giocoli}, C. and {G{\'o}mez-Alvarez}, P. and {Gracia-Carpio}, J. and {Granett}, B.~R. and {Grazian}, A. and {Grupp}, F. and {Guzzo}, L. and {Gwyn}, S. and {Haugan}, S.~V.~H. and {Holmes}, W. and {Hook}, I.~M. and {Hormuth}, F. and {Hornstrup}, A. and {Hudelot}, P. and {Jahnke}, K. and {Jhabvala}, M. and {Keih{\"a}nen}, E. and {Kermiche}, S. and {Kiessling}, A. and {Kubik}, B. and {Kuijken}, K. and {K{\"u}mmel}, M. and {Kunz}, M. and {Kurki-Suonio}, H. and {Le Boulc'h}, Q. and {Le Brun}, A.~M.~C. and {Le Mignant}, D. and {Liebing}, P. and {Ligori}, S. and {Lilje}, P.~B. and {Lindholm}, V. and {Lloro}, I. and {Mainetti}, G. and {Maino}, D. and {Maiorano}, E. and {Mansutti}, O. and {Marcin}, S. and {Marggraf}, O. and {Martinelli}, M. and {Martinet}, N. and {Marulli}, F. and {Massey}, R. and {Maurogordato}, S. and {McCracken}, H.~J. and {Medinaceli}, E. and {Mei}, S. and {Melchior}, M. and {Mellier}, Y. and {Merlin}, E. and {Meylan}, G. and {Mora}, A. and {Moresco}, M. and {Moscardini}, L. and {Nakajima}, R. and {Neissner}, C. and {Nichol}, R.~C. and {Niemi}, S.-M. and {Padilla}, C. and {Paltani}, S. and {Pasian}, F. and {Pedersen}, K. and {Percival}, W.~J. and {Pettorino}, V. and {Pires}, S. and {Polenta}, G. and {Poncet}, M. and {Popa}, L.~A. and {Pozzetti}, L. and {Raison}, F. and {Rebolo}, R. and {Renzi}, A. and {Rhodes}, J. and {Riccio}, G. and {Romelli}, E. and {Roncarelli}, M. and {Saglia}, R. and {Sakr}, Z. and {S{\'a}nchez}, A.~G. and {Sapone}, D. and {Sartoris}, B. and {Schewtschenko}, J.~A. and {Schirmer}, M. and {Schneider}, P. and {Schrabback}, T. and {Secroun}, A. and {Seidel}, G. and {Seiffert}, M. and {Serrano}, S. and {Simon}, P. and {Sirignano}, C. and {Sirri}, G. and {Stanco}, L. and {Steinwagner}, J. and {Tallada-Cresp{\'\i}}, P. and {Taylor}, A.~N. and {Tereno}, I. and {Toft}, S. and {Toledo-Moreo}, R. and {Torradeflot}, F. and {Tutusaus}, I. and {Valenziano}, L. and {Valiviita}, J. and {Vassallo}, T. and {Verdoes Kleijn}, G. and {Veropalumbo}, A. and {Wang}, Y. and {Weller}, J. and {Zacchei}, A. and {Zamorani}, G.},
        title = "{Euclid Quick Data Release (Q1) The Strong Lensing Discovery Engine B -- Early strong lens candidates from visual inspection of high velocity dispersion galaxies}",
      journal = {arXiv e-prints},
         year = 2025,
        month = mar,
          eid = {arXiv:2503.15325},
        pages = {arXiv:2503.15325},
          doi = {10.48550/arXiv.2503.15325},
archivePrefix = {arXiv},
       eprint = {2503.15325},
 primaryClass = {astro-ph.GA},
       adsurl = {https://ui.adsabs.harvard.edu/abs/2025arXiv250315325E}
}

@ARTICLE{Nightingale2025,
       author = {{Nightingale}, James W. and {Mahler}, Guillaume and {McCleary}, Jacqueline and {He}, Qiuhan and {Hogg}, Natalie B. and {Amvrosiadis}, Aristeidis and {Gozaliasl}, Ghassem and {Mercier}, Wilfried and {Scognamiglio}, Diana and {Berman}, Edward and {Leroy}, Gavin and {Liu}, Daizhong and {Massey}, Richard J. and {Shuntov}, Marko and {von Wietersheim-Kramsta}, Maximilian and {Franco}, Maximilien and {Paquereau}, Louise and {Ilbert}, Olivier and {Allen}, Natalie and {Toft}, Sune and {Akins}, Hollis B. and {Casey}, Caitlin M. and {Kartaltepe}, Jeyhan S. and {Koekemoer}, Anton M. and {McCracken}, Henry Joy and {Rhodes}, Jason D. and {Robertson}, Brant E. and {Drakos}, Nicole E. and {Faisst}, Andreas L. and {Jin}, Shuowen},
        title = "{The COSMOS-Web Lens Survey (COWLS) I: discovery of >100 high redshift strong lenses in contiguous JWST imaging}",
      journal = {\mnras},
         year = 2025,
        month = oct,
       volume = {543},
       number = {1},
        pages = {203-222},
          doi = {10.1093/mnras/staf1253},
archivePrefix = {arXiv},
       eprint = {2503.08777},
 primaryClass = {astro-ph.GA},
       adsurl = {https://ui.adsabs.harvard.edu/abs/2025MNRAS.543..203N}
}

@ARTICLE{Bolton2006,
       author = {{Bolton}, Adam S. and {Burles}, Scott and {Koopmans}, L{\'e}on V.~E. and {Treu}, Tommaso and {Moustakas}, Leonidas A.},
        title = "{The Sloan Lens ACS Survey. I. A Large Spectroscopically Selected Sample of Massive Early-Type Lens Galaxies}",
      journal = {\apj},
         year = 2006,
        month = feb,
       volume = {638},
       number = {2},
        pages = {703-724},
          doi = {10.1086/498884},
archivePrefix = {arXiv},
       eprint = {astro-ph/0511453},
 primaryClass = {astro-ph},
       adsurl = {https://ui.adsabs.harvard.edu/abs/2006ApJ...638..703B}
}

@BOOK{Sersic1968,
       author = {{Sersic}, Jose Luis},
        title = "{Atlas de Galaxias Australes}",
         year = 1968,
       adsurl = {https://ui.adsabs.harvard.edu/abs/1968adga.book.....S}
}
\bibliographystyle{aasjournalv7.1}



\end{document}